\documentclass[]{aa}  

\usepackage{float}
\usepackage{graphicx}
\usepackage[varg]{txfonts}
\usepackage[colorlinks=true,linkcolor=red,anchorcolor=blue,citecolor=blue,filecolor=black,menucolor=black,runcolor=black,urlcolor=blue]{hyperref}
\newcommand{\micron}[0]{\mu m}
\newcommand{\angstrom}{\mbox{\normalfont\AA}}
\usepackage{color}
\usepackage[normalem]{ulem}

\begin{document}

   \title{Stellar kinematics in the nuclear regions of nearby LIRGs with VLT-SINFONI}

   \subtitle{Comparison with gas phases and implications for dynamical mass estimations}

   \author{A. Crespo G\'omez
          \inst{\ref{inst1}}
          \and
          J. Piqueras L\'opez\inst{\ref{inst1}}
          \and
          S. Arribas \inst{\ref{inst1}}
          \and
          M. Pereira-Santaella  \inst{\ref{inst1}}
          \and
          L. Colina \inst{\ref{inst1}}
          \and
          B. Rodr\'iguez del Pino \inst{\ref{inst1}}}

   \institute{
   Centro de Astrobiolog\'ia (CSIC-INTA), Carretera de Ajalvir km. 4, Torrej\'on de Ardoz E28850, Madrid, Spain \label{inst1} 
   \email{acrespo@cab.inta.csic.es}
             }

   \date{Received September 18, 2020; accepted March 26, 2021}

 
  \abstract
  { Nearby luminous infrared galaxies (LIRGs) are often considered to be the local counterpart of the star forming galaxy (SFG) population at z>1. Therefore, local LIRGs are ideal systems with which to perform spatially resolved studies on the physical processes that govern these objects and to validate assumptions made in high-z studies because of a lack of sensitivity and/or spatial resolution.}
   {In this work we analyse the spatially resolved kinematics of the stellar component in the inner r<1-2\,kpc of ten nearby (mean z=0.014) LIRGs, establishing the dynamical state of the stars and estimating their dynamical masses ($M_\mathrm{dyn}$). We compare the stellar kinematics with those for different gas phases, and analyse the relative effects of using different tracers when estimating dynamical masses.}
  {We use seeing-limited SINFONI H- and K-band spectroscopy in combination with ancillary infrared (IR) imaging from various instruments (NICMOS/F160W, NACO/Ks and IRAC/3.6$\mu m$). The stellar kinematics are extracted in both near-IR bands by fitting the continuum emission using \texttt{pPXF}. The velocity maps are then modelled as rotating discs and used to extract the geometrical parameters (i.e. centre, PA, and inclination), which are compared with their photometric counterparts extracted from the near-IR images. We use the stellar and the previously extracted gas velocity and velocity dispersion maps to estimate the dynamical mass using the different tracers.}
   {We find that the different gas phases have similar kinematics, whereas the stellar component is rotating with slightly lower velocities  (i.e. $V_\mathrm{*}$\,$\sim$\,0.8$V_\mathrm{g}$) but in significantly warmer orbits (i.e. $\sigma_\mathrm{*}$\,$\sim$\,2$\sigma_\mathrm{g}$) than the gas phases, resulting in significantly lower $V/\sigma$ for the stars (i.e. $\sim$1.5-2) than for the gas (i.e. $\sim$4-6). These ratios can be understood if the stars are rotating in thick discs while the gas phases are confined in dynamically cooler (i.e. thinner) rotating discs. However, these differences do not lead to significant discrepancies between the dynamical mass estimations based on the stellar and gas kinematics. This result suggests that the gas kinematics can be used to estimate $M_\mathrm{dyn}$ also in z$\sim$2 SFGs, a galaxy population that shares many structural and kinematic properties with local LIRGs.}
   {}


   \maketitle

\section{Introduction}
\label{sec:intro}

Luminous and ultraluminous infrared galaxies (U/LIRGs) are characterised by their high infrared (IR) luminosities (i.e. L$_\mathrm{IR}$>$10^{11}$\,$\mathrm{L}_\odot$ and $10^{12}$\,$\mathrm{L}_\odot$ for LIRGs and ULIRGs, respectively). Their IR light is produced by the dust thermal re-emission of photons originating from massive starbusts and/or active galactic nuclei (AGNs) \citep{Sanders&Ishida+04,Nardini+10}. While LIRGs are mainly isolated galaxies that may have experienced some minor mergers \citep{Larson+16}, ULIRGs are generally major mergers \citep{Clements+96,Dasyra+06}. Although rare in the local Universe, these galaxies are the dominant contributors to the star formation rate density beyond z$\sim$2 \citep{Perez-Gonzalez+05,Lefloch+05,Magnelli+11}. In addition, local LIRGs have been shown to be analogues of high-z IR galaxies \citep{Pope+08,Stacey+10,Diaz-Santos+10,Arribas+12} and are therefore key to our understanding of galaxy evolution through cosmic time.

Low-z U/LIRGs offer an opportunity to study the physical mechanisms occurring in these types of galaxies at high spatial resolution and
signal-to-noise ratio (S/N). In recent years, diverse integral field spectroscopy (IFS), interferometric, and photometric studies have been carried out to characterise their AGN activity \citep{Alonso-Herrero+12,Diaz-Santos+13,U+19}, dust distribution and extinction \citep{Garcia-Marin+09dust,Stierwalt+14,Herrero-Illana+19}, star formation \citep{Arribas+01,Rodriguez-Zaurin+11,Pereira-Santaella+15,PiquerasLopez+16,Larson+20}, and ionisation sources \citep{Monreal-Ibero+06,Monreal-Ibero+10,Inami+13,Colina+15}.

The kinematics studies of these systems have almost entirely focused on their different gas phases (i.e. ionised, hot molecular, and cold molecular) through optical \citep{Arribas+08,Bellocchi+12,Cazzoli+14,Hung+15}, IR \citep{PiquerasLopez+12,Storchi-Bergmann+12,Colina+15,Emonts+17}, and sub-millimetre \citep{Garcia-Burillo+14,Zaragoza-Cardiel+17} wavelengths. These works have revealed velocity distributions varying from isolated rotating-disc patterns to more complex behaviours produced by gravitational interactions and/or disruptive events (e.g. outflows). Indeed the presence of outflows in U/LIRGs is prevalent, if not universal in the ionised \citep{Bellocchi+13,Arribas+14}, atomic \citep{Rupke+13,Cazzoli+16} and molecular \citep{Cicone+14,Pereira-Santaella+18,Herrera-Camus+20} gas phases of U/LIRGs. 

Contrary to the gas phases, stars do not suffer from local disruptive events and therefore should be the preferred probe for gravitational potential. Thus, the stellar kinematics can be taken as reference to evaluate whether or not the gaseous components are virialized and, consequently, whether or not they can be used to infer fundamental galaxy properties like the dynamical mass (i.e. $M_\mathrm{dyn}$). However, the kinematics of the stars is intrinsically complex as the collisionless nature of the stars allows their velocity distribution to deviate from a Gaussian function. Consequently, the stellar kinematics is often represented by elaborate models, such as the Schwarzschild models \citep{Schwarzschild+79} or anisotropic Jeans models \citep{Cappellari+08}, when multiple and high-quality observables (e.g. velocity and velocity dispersion along with higher order moments, surface brightness density, etc.) are available. For systems that are dominated by rotation, as in the stellar discs used in the present work, a rotation-curve-based analysis is commonly considered as a good first-order approximation. In these cases, an asymmetric drift correction is needed to asses the non-negligible velocity dispersion contribution to the observed velocity.

Extracting the stellar kinematics can be challenging as it requires high S/N at the continuum to observe the stellar absorption features. Moreover, the possible presence of AGN emission along with the dust extinction of the inner-most regions can severely modify the continuum shape, hindering the extraction of these stellar absorption bands \citep{Forster-Schreiber+00,Greene+06,Xiao+11,Burtscher+15}. For these reasons, the gas emission lines have been historically used as tracers of the gravitational potential, especially at high z. However, this assumption may not be adequate for high-z star forming galaxies (SFGs), where phenomena associated with the intense SF and AGN activity may disturb the gas phases. Despite the relevance of studying U/LIRGs to our understanding of the dynamics of these systems and the impact of disturbing phenomena in their gas phases, we are still lacking detailed, spatially resolved IFS studies of their stellar kinematics. 

In this work, we present a spatially resolved near-IR IFS SINFONI stellar kinematic analysis of the nuclear regions (r$\sim$2\,kpc) of ten local LIRGs covering different nuclear activity classes and morphological types. These galaxies are part of a larger sample of U/LIRGs that have been the subject of several studies to characterise their physical processes and gas kinematics using photometric \citep{Alonso-Herrero+06}, optical and near-IR IFS \citep{Arribas+04,Bedregal+09,Bellocchi+16}, and sub-mm CO ALMA \citep{Pereira-Santaella+16a,Pereira-Santaella+16b,Pereira-Santaella+18} data. Here, the stellar kinematics are compared with the gas kinematics previously obtained with the same near-IR IFS dataset \citep{PiquerasLopez+12}. Backed with this dataset, we are in the position to analyse whether or not the gas and stellar phases are in similar dynamical states and therefore whether or not the gas kinematics can be used to trace the dynamical mass in these systems. 

The paper is structured as follows. In Section~\ref{sec:sample} we introduce the main properties of the sample, and the observations, data reduction, and ancillary date are described. Section~\ref{sec:analysis} describes the main methods and tools used for both photometric and kinematic analyses. In Section~\ref{sec:results}, we introduce and discuss the main results obtained during the analysis, and, finally, Section~\ref{sec:conclusions} summarises our main results and conclusions. In addition, individual notes for the galaxies of the sample can be found in Appendix~\ref{append:sources_notes}. Figures outlining the photometric and kinematic analysis for the sample are presented in the Appendices~\ref{append:morph} and~\ref{append:kin}, respectively. Appendix~\ref{append:2369} shows the double-Gaussian analysis of Br$\gamma$ for NGC\,2369, and Appendix~\ref{append:13120} presents the kinematic analysis of the ULIRG IRAS\,13120-5453, the results for which were not considered in this study.

\section{Sample, data treatment, and complementary data}
\label{sec:sample}

\subsection{The sample}
\label{subsec:sample}

The sample analysed in this work contains a set of ten LIRGs drawn from a larger sample of local U/LIRGs observed with several IFS facilities \citep[see][]{Arribas+08}. These nearby objects (i.e. mean z=0.014) are within a factor of two in distance (40-80\,Mpc) and cover the lower values of LIRG luminosities, with nine out of the ten objects within 11.10<log($L_{\mathrm{IR}}/\mathrm{L_{\odot}}$)<11.44. The sample encompasses galaxies at different gravitational interaction stages (i.e. five isolated, three interacting and one ongoing merger, and one post-coalescence system; see \citealt{Yuan+10,Bellocchi+16}) and nuclear activity classes (two Seyfert, six \ion{H}{II,} and two `composite' galaxies). In general, these objects present intense star-forming regions (median $\Sigma_{\mathrm{SFR}}$\,$\sim$2\,$\mathrm{M_{\odot}\,yr^{-1}\,kpc^{-2}}$, \citealt{PiquerasLopez+16}) and are highly obscured in their nuclear regions (i.e. $A_{\mathrm{V}}$=15-25, \citealt{PiquerasLopez+13}). Table~\ref{tab:sample} summarises the main properties of these galaxies.

\subsection{Near-IR IFS data}
\label{subsec:ifs}

The integral field spectroscopic data used in this work were obtained with the instrument SINFONI on the VLT \citep{Eisenhauer+03} during the periods 77B, 78B, and 81B (from April 2006 to July 2008). Observations were carried out in the H-band (1.45\,$\mathrm{\micron}$ to 1.85\,$\mathrm{\micron}$) and K-band (1.95\,$\mathrm{\micron}$ to 2.45\,$\mathrm{\micron}$) with scales of 0.125\arcsec$\times$\,0.250\arcsec\,spaxel$^{-1}$. Although the field of view (FoV) of each single exposure is $8\arcsec$\,$\times$\,$8\arcsec$, the dithering pattern adopted during the observations allowed us to achieve FoVs varying from 9\arcsec$\times$\,9$\arcsec$ to 12\arcsec$\times$\,$12\arcsec$, with scales of 0.125\arcsec$\times$\,0.125\arcsec\,spaxel$^{-1}$. 

The spectral resolution for these configurations are $R$\,$\sim$4000 and $R$\,$\sim$3000 for
the H- and K-bands, yielding FWHM$\sim$6.6\,$\angstrom$ and 6.0\,$\angstrom$, respectively. The observations were executed in seeing-limited mode, leading to an average angular resolution in full width at half maximum (FWHM) of $\sim$0.63$\arcsec$ ($\sim$125-0.250\,pc at their redshifts).

The reduction and calibration procedures were carried out using the standard ESO pipeline ESOREX (version 2.0.5) as described in \citet{PiquerasLopez+12} and we refer the reader to that work for further details.

For this study, we treated the original data cubes to remove the so-called `instrumental fingerprint', an effect present in the SINFONI datacubes that modifies the continuum shape at specific spatial and spectral regions \citep[see][for examples]{Neumayer+07,DaSilva+17,Dametto+19}. We used the principal component analysis (PCA) tomography technique described in \citet{Steiner+09} to correct for this instrumental fingerprint. Following the procedure carried out in \citet{Menezes+15}, we first fitted high-order polynomials to the spectra, omitting the emission lines, to trace the shape of the continuum. We then applied the PCA method and created a cube containing the `fingerprint' effect. Finally, we subtracted this cube from the original one, removing the instrumental effect.

In addition, we spatially binned our IFS data using the Voronoi binning code described in \citet{Cappellari+03}. This method creates a tessellation of the FoV by combining adjacent spaxels until an average continuum S/N of 20 is reached (with values ranging from 15 up to 40). We defined the signal as the level of the continuum at 1.6\,$\mathrm{\micron}$ and 2.2\,$\mathrm{\micron}$, whereas the noise was defined as the standard deviation of continuum within the 1.628-1.635\,$\mu m$ and 2.175-2.196\,$\mu m$ spectral ranges, for H- and K-band, respectively.

\subsection{Gas kinematic maps}
\label{subsec:gas_kin}

Besides the stellar kinematics, our near-IR IFS dataset also allows us to extract the gas kinematics based on several emission lines. The velocity and velocity dispersion maps of different gas phases (i.e. ionised, partially ionised, and hot molecular) were extracted and presented by \citet{PiquerasLopez+12} using this same dataset. These gas phases were traced by the Br$\gamma$ $\lambda$2.166\,$\mathrm{\micron}$, [\ion{Fe}{II}] $\lambda$1.644\,$\mathrm{\micron,}$ and H$_2$1-0\,S(1) $\lambda$2.122\,$\mathrm{\micron}$ emission lines, respectively. These lines were chosen as they are the strongest hydrogen recombination and H$_2$ lines in the H- and K-band, whereas the [\ion{Fe}{II}] line is often associated with regions partially ionised by X-rays or shocks \citep{Mouri+00}. As described in \citet{PiquerasLopez+12}, these lines were fitted with single Gaussian profiles that were previously convolved with the SINFONI line spread function (LSF). 
We refer the reader to the original work for further details. These gas kinematics allowed us to compare (see Sect.~\ref{subsec:starsvsgas}) the gas and stellar kinematics with the same spectral and spatial resolutions using a single comprehensive dataset. 

\subsection{Ancillary imaging data}
\label{subsec:phot_data}

Ancillary images from different datasets were used to obtain photometric parameters (e.g. effective radii) that were later used in our analysis of the IFS data. The global stellar distributions were characterised with Spitzer/IRAC/3.6$\mathrm{\micron}$ images (FWHM$\sim$2$\arcsec$; \citealt{Fazio+04}). These images were reduced using the IRAC pipeline v2.0 (2011) and presented in \citet{PereiraSantaella+11}. The inner-most regions (i.e. $\sim$10$\arcsec$) were covered by HST/NICMOS NIC2/F160W and VLT/NACO Ks images (FWHM$\sim$0.15$\arcsec$ and $\sim$0.2$\arcsec$; \citealt{Thompson+98} and \citealt{Rousset+03}, respectively), granting better spatial resolution than our IFS data (i.e. $\sim$0.63$\arcsec$). As the NIC2/F160W image of NGC\,3256 does not cover both nuclei, we used its NIC3/F160W
image (i.e. FWHM$\sim$0.4$\arcsec$ and FoV$\sim$50$\arcsec$). The NICMOS and NACO images were requested through the Mikulski Archive for Space Telescope (MAST; PI: Alonso-Herrero, ID: 10169) and ESO Archive (PI: Escala, ID: 086-B.0901(A)). 

The alignment of the different images with the IFS data was carried as follows. After degrading the images to match the IFS resolution, two-dimensional Gaussian functions were fitted to the brightest galaxy structures (e.g. nuclei, blobs) in the HST and NACO images. These regions were aligned with their counterparts in the H and K continuum maps extracted from the IFS data. For the IRAC images we considered as many field objects as possible to overcome their more limited angular resolution. Finally, we corrected the astrometry of every object by aligning these images with their CO(2-1) and continuum ($\sim$230\,GHz) maps obtained from ALMA data (ID programs 2013.1.00243.S, 2013.1.00271.S and 2017.1.00255.S)

\begin{table*}
\caption{Main properties of the sample. Columns (2), (3), and (4): Right ascension, declination, and redshift from the NASA Extragalactic Database (NED). Columns (5) and (6): Luminosity distance and angular scale from Ned Wright's Cosmology Calculator \citep{Wright+06} assuming $H_{\mathrm{0}}$=70\,km\,s$^{-1}$Mpc$^{-1}$, $\Omega_{\mathrm{M}}=0.3$ and $\Omega_{\mathrm{\Lambda}} =0.7$. Column (7): Infrared luminosities, where $L_{\mathrm{IR}}$(8-1000$\mathrm{\micron}$) was calculated from the IRAS flux densities $f_{12}$, $f_{\mathrm{25}}$, $f_{\mathrm{60}}$ and $f_{\mathrm{100}}$ presented in \citet{Sanders+03}. Column (8): Spectroscopic classification as in Table 1 from \citet{PiquerasLopez+12}. Galaxies classified as composite
present IR emission that can be explained by a combination of AGN activity and star formation. Column (9): Gravitational interaction stage classification as follows: (a) isolated, (b) interacting pair, (c) ongoing merger, and (d) post-coalescence merger. This classification is a simplified version based on the morphology classes presented in \citet{Yuan+10}, when possible, and \citet{Arribas+08}.}            
\centering          
\begin{tabular}{c c c c c c c c c }     
\hline  
 NAME  & RA & Dec & z     & $D_{\mathrm{L}}$  &   Scale     & $\log L_{\mathrm{IR}}$ &  Spect. class  & Interaction stage\\
 &   (J2000)      &     (J2000)    &     & (Mpc) &    (kpc/\arcsec)    &  ($L_{\mathrm{\odot}}$)   &        &    \\
    (1)         &        (2)      &  (3)  & (4)   &  (5)         & (6)         &       (7)   & (8)  & (9) \\ \hline 
NGC\,2369         & $\mathrm{07^h 16^m 37.73^s}$   & $-62\degr20'37.4\arcsec$          & 0.010807 & 46.7 & 0.221    & 11.17      & Composite & a \\ 
NGC\,3110         & $\mathrm{10^h04^m02.11^s}$    & $-06\degr28'29.2\arcsec$     & 0.016858 & 73.1 & 0.343   & 11.34           & Composite  & b \\ 
NGC\,3256         & $\mathrm{10^h27^m51.27^s}$    & $-43\degr54'13.8\arcsec$    & 0.009354 & 40.4 & 0.192   & 11.74    &\ion{H}{II} & c \\ 
ESO\,320-G030     & $\mathrm{11^h53^m11.72^s}$   & $-39\degr07'48.9\arcsec$  & 0.010781 & 46.6 & 0.221   & 11.35       & \ion{H}{II}   & a \\ 
IRAS\,F12115-4656 & $\mathrm{12^h14^m12.84^s}$  & $-47\degr13'43.2\arcsec$     & 0.018489 & 80.3 & 0.375   & 11.10   & \ion{H}{II}    & b \\  
NGC\,5135         & $\mathrm{13^h25^m44.06^s}$   & $-29\degr50'01.2\arcsec$   & 0.013693 & 59.3 & 0.280    & 11.33 &  Sy2     & a \\  
IRAS\,F17138-1017 & $\mathrm{17^h16^m35.79^s}$  & $-10\degr20'39.4\arcsec$     & 0.017335 & 75.2 & 0.352    & 11.42      &\ion{H}{II}  & d  \\  
IC\,4687          & $\mathrm{18^h13^m39.63^s}$   & $-57\degr43'31.3\arcsec$    & 0.017345 & 75.3 & 0.353   & 11.44    & \ion{H}{II}    & b \\  
NGC\,7130         & $\mathrm{21^h48^m19.50^s}$   & $-34\degr57'04.7\arcsec$   & 0.016151  & 70.0 & 0.329  & 11.34   & Sy2 & a \\ 
IC\,5179          & $\mathrm{22^h16^m09.10^s}$  & $-36\degr50'37.4\arcsec$   & 0.011415 & 49.3 & 0.234    & 11.12    & \ion{H}{II}      & a \\ 
\hline           
\end{tabular}
\label{tab:sample}      
\end{table*}

\section{Data analysis}
\label{sec:analysis}

\subsection{Stellar kinematic analysis}
\label{subsec:kin_analysis}

\subsubsection{Velocity and velocity dispersion maps extraction}
\label{subsubsec:ppxf}

\begin{figure*}
\includegraphics[width=\linewidth]{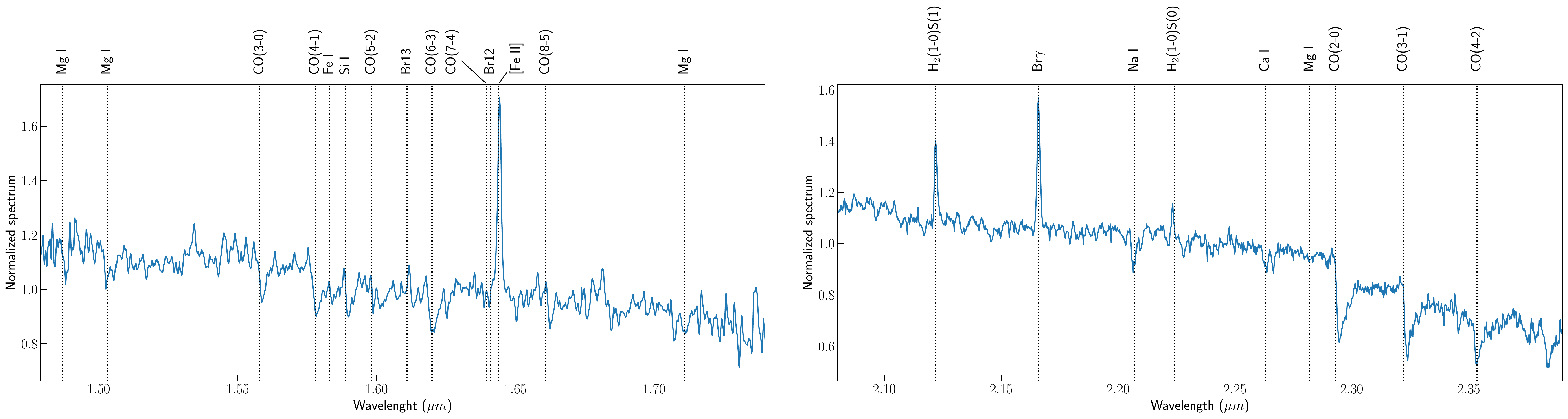}
\caption{Example of SINFONI data. Normalised nuclear spectra of IC5179 from the SINFONI H- and K-band data (left and right panels, respectively). The main emission lines and stellar absorption features have been labelled.}
\label{fig:example_spec}
\end{figure*}

The stellar kinematics maps were derived using the penalised pixel-fitting (\texttt{pPXF}) routine \citep{Cappellari+04,Cappellari+17}. This code fits the continuum with a linear combination of stellar templates, which have been previously convolved with the SINFONI LSF ($\sigma$\,$\sim$50 and 35\,km\,s$^{-1}$ for H- and K-band, respectively), parametrised by Gauss-Hermite polynomials. In this work, we fitted the line-of-sight velocity distribution (LOSVD) with the velocity, velocity dispersion, and the higher order Gauss-Hermite moments h$_\mathrm{3}$ and h$_\mathrm{4}$. However, the h$_\mathrm{3}$ and h$_\mathrm{4}$ moments maps are not considered during the subsequent analysis due to their noise-dominated structure and lack of spatial coherence. In addition, we allowed \texttt{pPXF} to add a first-degree polynomial to account for possible effects of the continuum shape (e.g. AGN continuum emission) and/or template mismatch.

The near-IR absorption features are mainly produced at the atmospheres of evolved giant stars, with a non-negligible contribution of cool AGB stars \citep{Kleinmann+86,Hinkle+95,Dallier+96,Wallace+97,Forster-Schreiber+00,Boker+08,Kotilainen+12,Dametto+14}. In nearby galaxies, the K-band is strongly dominated by the CO(2-0) $\lambda$2.293\,$\mathrm{\micron}$ and CO(3-1) $\lambda$2.312\,$\mathrm{\micron}$ bands, although it also presents absorption lines from atomic species (e.g. \ion{Na}{I} $\lambda$2.207\,$\mathrm{\micron}$, \ion{Ca}{I} $\lambda$2.263\,$\mathrm{\micron}$ and \ion{Mg}{I} $\lambda$2.282\,$\mathrm{\micron}$, see Fig.~\ref{fig:example_spec}). On the contrary, the H-band shows less prominent CO bands, and a larger number of atomic absorption lines (e.g. \ion{Mg}{I} $\lambda$1.487,1.503,1.575,1.711\,$\mathrm{\micron}$, \ion{Fe}{I} $\lambda$1.583\,$\mathrm{\micron}$ and \ion{Si}{I} $\lambda$1.589\,$\mathrm{\micron}$, see Fig.~\ref{fig:example_spec}). We chose the PHOENIX synthetic spectral library presented in \citet{Husser+13}, which covers both near-IR bands with a spectral resolution larger than our IFS data (i.e. $R\mathrm{\sim}$10\,000). As the strength of the CO absorption features is very sensitive to the star surface gravity and effective temperature \citep{Silge&Gebhardt+03}, it is important to include a good representation of giant stars when generating the library of stellar templates. For this reason, we created a subsample of PHOENIX templates within the range of effective temperatures 4000\,K$<$T$<$5200\,K, gravities $\mathrm{1<\log g<4.5}$ and metallicities $\mathrm{-1<\log[Fe/H]<0}$, assuming solar alpha-element abundance (i.e. $\mathrm{log([\alpha/Fe])=0}$. Although this set of kinematic templates may be to some extent limited in terms of properties for a stellar population analysis (e.g. metallicity, alpha-element abundance, etc.), we point out that this is out of the scope of this work. This sample was created mimicking the Gemini Spectral Library \citep{Winge+09}, which is commonly used in near-IR, with the purpose of extracting the stellar kinematics fitting the absorption features present in the continuum. 

We considered the 1.479$-$1.751\,$\mathrm{\micron}$ and 2.176$-$2.356\,$\mathrm{\micron}$ spectral windows during the fitting procedure, granting enough S/N and the same spectral coverage for each target. The emission lines and OH sky residuals were masked. We performed Monte Carlo simulations ---where the standard deviation of the residuals from the subtraction of the best fit and the original spectra were considered as flux noise--- to compute the errors of the \texttt{pPXF} results. In addition, we accounted for the uncertainties associated to the wavelength calibration by varying the wavelength with a normal distribution of half a spectral channel (i.e. $\sim$15\,km\,s$^{-1}$). The kinematic maps resulting from this analysis and their main characteristics are discussed in Sect.~\ref{subsec:vel_maps}. 

In general, we find that the K-band presents larger $\sigma$ uncertainties (i.e. $\sim$40 vs. $\sim$15\,km\,s$^{-1}$ in the H-band) and noisier velocity maps, especially in the outskirts of the FoV where the S/N decreases. Although both near-IR bands have similar S/N per spaxel at the continuum, the stellar continuum fit on the K-band is likely affected by the smaller number of stellar features (where the strong CO bands predominate) and the narrower wavelength range analysed. Even though the H-band is more affected by sky lines and the SINFONI `footprint' (see Sect.~\ref{subsec:ifs}), the wider spectral window analysed in this band along with the larger number and uniformly distributed absorption features provide a better determination of the kinematics.

\subsubsection{Modelling the rotating disc }
\label{subsubsec:DiskFit}

\begin{figure}
\includegraphics[width=\linewidth]{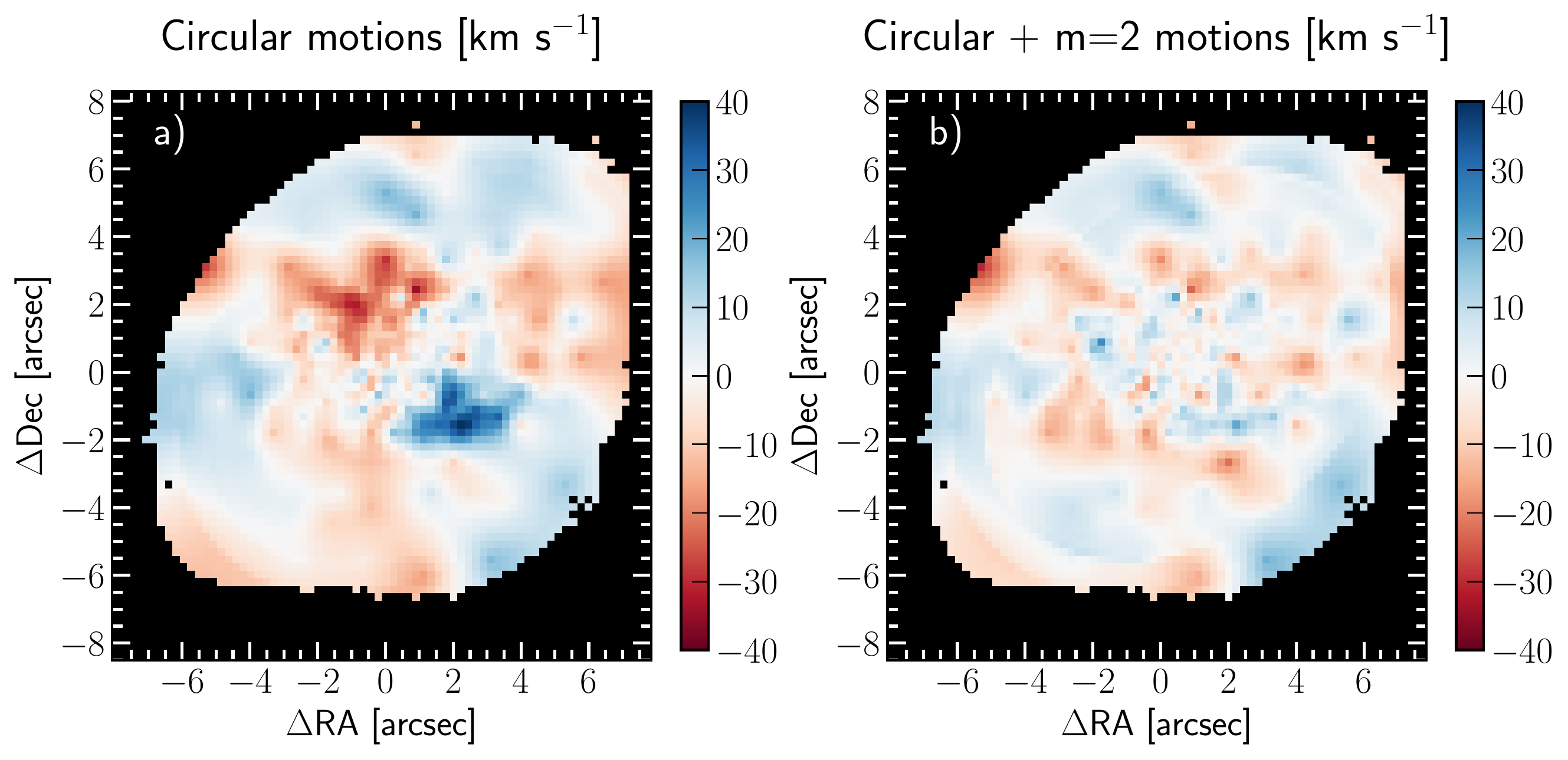}
\caption{Anti-symmetric velocity residuals produced by the bar-like motions. Velocity residuals from the K-band models for ESO\,320-G030 before (left panel) and after (right panel) adding the m=2 velocity component to the rotating disc, which accounts for the non-circular motions. }
\label{fig:example_nonrot}
\end{figure}

After extracting the kinematics maps from the stellar continuum, we modelled the observed stellar velocity maps using \texttt{DiskFit} \citep{SellwoodSpekkens15}. This code uses $\chi^2$ minimisation to find the rotating-disc model (considering possible contributions of bar-like motions) that globally fits the observed velocity map. Before the modelling, we interpolated the velocity maps to avoid artefacts derived from the irregular sampling caused by the Voronoi binning. We used the barycentres from each spatial bin and their velocity values to interpolate a velocity map with the cell size of the original data (i.e. 0.125$\arcsec$). The models for each galaxy were then obtained from these interpolated velocity maps within elliptical apertures ---in which we exclude the external regions with lower S/N--- by fitting a constant position angle (i.e. PA), centre, inclination, and systemic velocity.

In some cases, an additional non-axisymmetric component was required to account for the presence of anti-symmetric residuals produced by bar-like structures (see Appendix~\ref{append:sources_notes} for notes of the individual objects). These structures were modelled as bi-symmetric (m=2) perturbations as:
\begin{equation}
V=V_\mathrm{sys}+\sin{i}\left[V_t\cos{\theta}-V_{2,t}\cos{2\theta_b}\cos{\theta}-V_{2,r}\sin{2\theta_b}\sin{\theta} \right]
,\end{equation}
following Eq. 5 from \citet{SpekkensSellwood07}, where $\theta_b$ is the orientation of the bi-symmetric component with respect to the PA (i.e. $\theta$). Figure~\ref{fig:example_nonrot} shows an example, for ESO\,320-G030, of the anti-symmetric residuals produced by the bar-like distortions that are modelled with the bi-symmetric component. It is important to bear in mind that \texttt{DiskFit} assumes thin rotating discs, although warps are allowed at the outskirts of the velocity map. Therefore, deviations from this co-planarity may contribute to the non-axisymmetric components and/or justify some of the regions with larger residuals. The associated uncertainties to each parameter were computed using the bootstrap utility (i.e. \texttt{Bootlace}) supplied along with the \texttt{DiskFit} code. The models and the extracted parameters for each near-IR band are presented in Sect.~\ref{subsec:results_DiskFit}.

\subsection{Complementary photometric analysis}
\label{subsec:phot_analysis}

\begin{table*}
\caption{Effective and reference radii. Column (2) Effective radius obtained by applying the CoG method to the IRAC/3.6$\mathrm{\mu m}$ images. In brackets, the size of the PSF (FWHM/2) in kpc at the redshift of the target. Column (3): Reference radii, defined as the maximum radii along the major-axis with reliable kinematic data for each object. Column (4): Scale length from the exponential disc models fitted to the IRAC surface brightness profiles (see Sect.~\ref{subsec:Mdyn}). Column (5): Effective radii from the 2MASS K-band images derived using GALFIT, presented in \citet{Bellocchi+13}.$\dagger$: This value was obtained using the CoG using the NACO/Ks image, as there are no IRAC data for this object. }
\centering 
\begin{tabular}{ c c c c| c }
\hline 

NAME & $R_{\mathrm{eff}}$ & $R_{\mathrm{r}}$ & $R_{\mathrm{d}}$  & $R^{\text{2MASS}}_{\mathrm{GALFIT}}$   \\
       & (kpc) & (kpc) & (kpc) & (kpc)  \\
        (1)    &    (2)   &  (3)  & (4)  & (5)   \\ \hline  
NGC\,2369   & 3.72$\pm$0.69 [0.22]   & 0.77   & 2.81$\pm$0.09 & 3.64$\pm$0.35\\
NGC\,3110   &  4.13$\pm$0.41  [0.34]     & 1.20  & 3.37$\pm$0.11   & 3.56$\pm$0.45 \\ 
NGC\,3256   & 1.75$\pm$0.17  [0.19]       & 0.67  & 2.05$\pm$0.07     &  2.66$\pm$0.86 \\
ESO\,320-G030  &  1.18$\pm$0.14 [0.22]     & 0.99 & 2.09$\pm$0.07 &  1.22$\pm$0.20 \\
IRAS\,F12115-4656  & 1.99$\pm$0.10 [0.04]$^\dagger$  & 1.91   & 1.15$\pm$0.02  & 2.97$\pm$0.57  \\
NGC\,5135     & 1.95$\pm$0.53  [0.28]      & 0.98   & 4.11$\pm$0.15    & 3.41$\pm$0.85   \\
IRAS\,F17138-1017   & 1.74$\pm$0.21 [0.35]      & 1.23   & 2.36$\pm$0.13    & 1.78$\pm$0.15  \\ 
IC\,4687      & 1.63$\pm$0.19  [0.35]     &  1.27    & 1.97$\pm$0.10     & 1.75$\pm$0.32\\
NGC\,7130     & 4.22$\pm$0.35  [0.33]      & 0.73      & 3.07$\pm$0.11       & 3.68$\pm$0.51    \\ 
IC\,5179       & 4.54$\pm$0.86  [0.23]     &  1.76    & 3.17$\pm$0.08    & 5.13$\pm$0.66   \\ 
\hline
\end{tabular}
\label{tab:Reff}
\end{table*}

\begin{table*}
\caption{Photometric parameters extracted with \texttt{Photutils}. All the parameters were measured using elliptical $R_{\mathrm{r}}$ apertures whereas their associated uncertainties were computed using Monte Carlo simulations. Columns (2)-(5): Coordinates and errors of the photometric centre determination, assuming ALMA astrometry (see Sect.~\ref{subsec:phot_data}). Column (6): Position angle, defined as the anti-clockwise angle from the north to the semi-major axis. Column (7): Inclination, defined as $\cos{i}=b/a$ where $b/a$ is the axial ratio of the elliptical aperture. Column (8): Inclination extracted from the same HST/NICMOS images in \citet{Bellocchi+13}. ${\mathrm{(\dagger})}$ Values extracted using H$\alpha$ maps. Column (9): Inclination measured with the 2MASS/Ks image \citep{2MASS+06}, obtained from the NASA/IPAC Extragalactic Database (NED).}
\centering      
\begin{tabular}{c c c c c c c | c c}     
\hline  
 NAME  & RA& $\Delta$RA & Dec & $\Delta$Dec  & PA  & $i$ & $i_{\text{Bellocchi+13}}$ & $i_{\text{2MASS}}$\\
&  & (mas)  &   &  (mas) & (deg) &  (deg) &  (deg)   &  (deg)\\
   (1)    &    (2)   &  (3)  & (4)  &  (5) & (6)  & (7) & (8) & (9) \\ \hline 
NGC\,2369        & $\mathrm{07^h 16^m 37.68^s}$ & 59  & $-62\degr20'37.02\arcsec$ & 127 & 170$\pm$1 & 67$\pm$3 & 75$\pm$3 & 63 \\
NGC\,3110         & $\mathrm{10^h04^m02.09^s}$ & 85   & $-06\degr28'29.82\arcsec$ & 86 & 189$\pm$6 & 50$\pm$4 & 65$\pm$3 &  55\\
NGC\,3256         & $\mathrm{10^h27^m51.25^s}$ & 232    & $-43\degr54'14.53\arcsec$ & 322  & 170$\pm$7 & 46$\pm$10 & 45$\pm$10  & 47 \\
ESO\,320-G030     & $\mathrm{11^h53^m11.72^s}$& 145  & $-39\degr07'49.07\arcsec$& 60 & 118$\pm$10 & 46$\pm$5 & 37$\pm$3$^{\mathrm{\dagger}}$ & 61 \\
IRAS F12115-4656 & $\mathrm{12^h14^m12.82^s}$ & 33 & $-47\degr13'42.79\arcsec$& 49  & 119$\pm$4 & 56$\pm$5 & 42$\pm$3$^{\mathrm{\dagger}}$ & 50 \\
NGC\,5135         & $\mathrm{13^h25^m44.02^s}$ & 149   & $-29\degr50'01.20\arcsec$ & 172 & 56$\pm$15 & 31$\pm$3 & 37$\pm$3  & 61 \\
IRAS F17138-1017  & $\mathrm{17^h16^m35.82^s}$ & 121  & $-10\degr20'39.12\arcsec$ & 123 & 198$\pm$5 & 62$\pm$6 & 52$\pm$8  & ... \\
IC\,4687          & $\mathrm{18^h13^m39.74^s}$ & 253   & $-57\degr43'30.84\arcsec$ & 165   & 20$\pm$32 & 27$\pm$7  & 47$\pm$3$^{\mathrm{\dagger}}$  &  50\\
NGC\,7130        & $\mathrm{21^h48^m19.51^s}$ & 103   & $-34\degr57'04.48\arcsec$ & 126 & 161$\pm$13 & 34$\pm$6 & 51$\pm$2 & 28 \\
IC\,5179          & $\mathrm{22^h16^m09.16^s}$ & 278  & $-36\degr50'37.07\arcsec$ & 348 & 29$\pm$6 & 59$\pm$5  & 56$\pm$4$^{\mathrm{\dagger}}$ & 68 \\
\hline           
\end{tabular}
\label{tab:phot_results}      
\end{table*}

\begin{figure*}
\includegraphics[width=\linewidth]{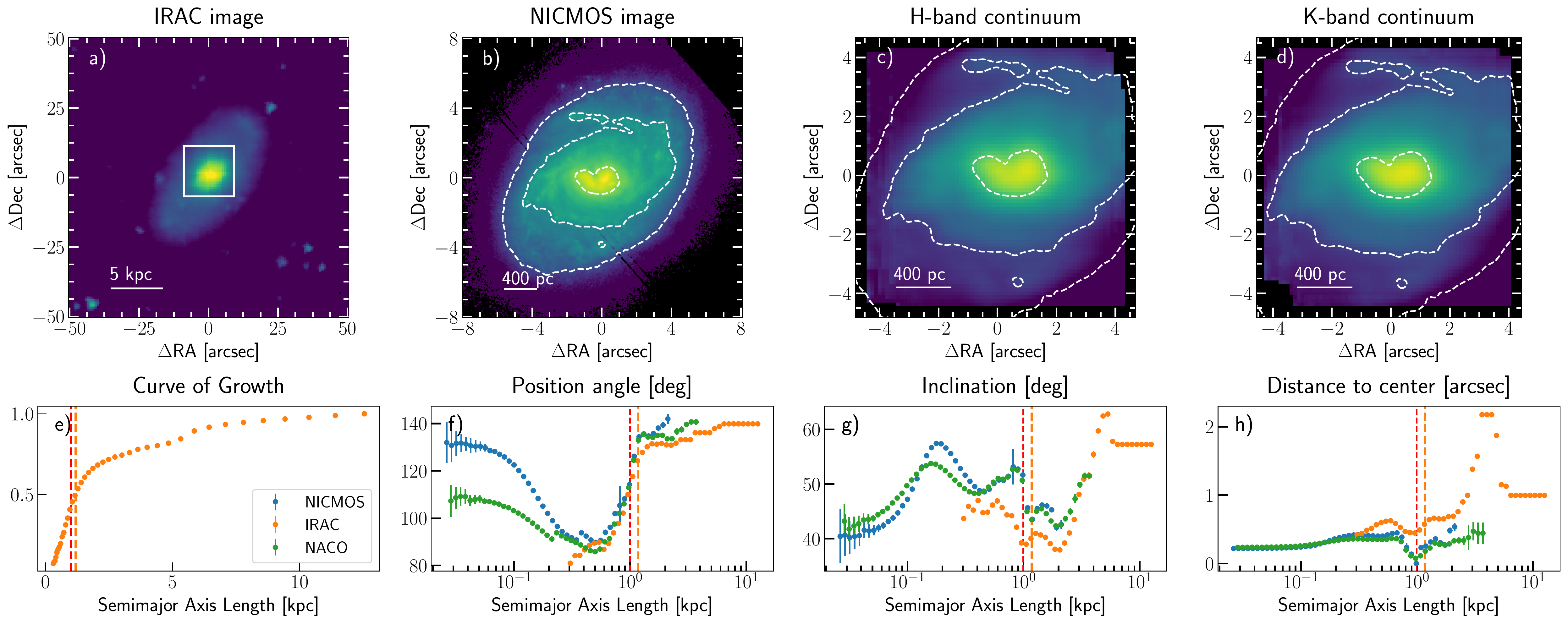}
\caption{Photometric analysis for ESO\,320-G030. Panels (a) and (b): Spitzer/IRAC and HST/NICMOS images used in this work, respectively. The white square in panel (a) represents the FoV displayed on the panel (b). Panels (c) and (d): H- and K-band SINFONI continuum maps used during the alignment (see Sec~\ref{subsec:phot_analysis}). Panel (e): Curve-of-Growth obtained from the IRAC image. Panels (f) and (g): Radial profiles of the PA and inclination with the size of the elliptical aperture. Panel (h): Radial profile of the photo-centre, expressed as the distance from the centre at $R_{\mathrm{r}}$. Vertical dashed lines in the lower panels represent the aperture associated to the effective radius extracted with the CoG method for each image. Red dashed line marks the $R_{\mathrm{r}}$ aperture, at which the parameters are extracted (see Table~\ref{tab:phot_results}).}
\label{fig:example_morph}
\end{figure*}

We made use of the \texttt{Photutils} \citep{Photutils+16} package of Astropy \citep{Astropy+18} to fit increasingly larger elliptical apertures to all the available images (IRAC, NACO and NICMOS). From each aperture, we extracted the photo-centre, PA, and inclination, allowing us to identify potential changes in these parameters with the radial distance and compare them with those obtained from the disc modelling of the stellar kinematics (Sect.~\ref{subsubsec:DiskFit}). We computed the associated uncertainties of these parameters using Monte Carlo simulations, varying the flux of each pixel from the image within a normal distribution in which the background noise level was used to represent the standard deviation.

Figure~\ref{fig:example_morph} shows, as an example, the radial profiles of the photometric parameters of ESO\,320-G030, whereas analogous figures for the rest of the sample can be found in Appendix~\ref{append:morph}. In general, we found similar radial profiles for the different images, with small differences in the PA and inclination (i.e. <$5\degr$, panels `f' and `g'), and in the photo-centre (i.e. <0.3$\arcsec$, panel `h') in the overlapping regions. These radial profiles are affected by the presence of bright regions and nuclear bars/pseudo-bulges along with the loss of S/N and bright asymmetries in the spiral arms, appearing as abrupt changes in the profiles (mostly in the inclination and PA) and/or the deviation of the photo-centre at larger scales (e.g. IC\,5179, NGC\,3110, and NGC\,7130; see Appendix~\ref{append:morph}).

We defined reference radii (i.e. $R_\mathrm{r}$, Table~\ref{tab:Reff}) to extract the results of the photometric parameters and compare them with the values obtained from the kinematics. These radii were defined as the maximum distance along the kinematic major-axis for which we can extract reliable stellar velocity and velocity dispersion values for each galaxy. This definition maximises the area covered simultaneously by the photometric data and the SINFONI FoV and prevents spurious results produced by nuclear structures.

Based on these radii, we obtain the PA, inclination, and distance to the centre and their associated uncertainties, defined as the mean and standard deviation of the radial profiles from all the images within 0.9$
R_\mathrm{r}$ and 1.1$R_\mathrm{r}$. These results are presented in Table~\ref{tab:phot_results}, whereas further comments on the photometric parameters and morphology of individual sources can be found in Appendix~\ref{append:sources_notes}.

\subsubsection{$R_{\mathrm{eff}}$ determinations}
\label{subsubsec:Reff}

The effective radius ($R_{\mathrm{eff}}$) defines the aperture that encloses half of the total light of a system. Considering that the near-IR light is a good proxy of the stellar mass, $R_{\mathrm{eff}}$ can be used to estimate the area that encloses half of the total stellar mass of the galaxy. In principle, the effective radius determination depends on the specific method considered, as well as on the FoV, angular resolution, wavelength range, and depth of the image. In this work, we adopted the curve-of-growth (CoG) method, defining the effective radius as $R_{\mathrm{eff}}\equiv a$, where $a$ is the semi-major axis of the elliptical aperture enclosing half of the light. The associated uncertainties were defined as the mean of the differences with the radii corresponding to the 45\% and 55\% of the total flux in the CoG.

The IRAC/3.6$\mathrm{\mu m}$ images were selected to derive $R_{\mathrm{eff}}$ because their FoV is large enough to cover the whole structure of these nearby galaxies and the near-IR light traces the stellar mass \citep{Gavazzi+96,Zibetti+02} better than the optical light. In addition, IRAC PSF (FWHM/2 ranging $\sim$0.19-0.38\ kpc depending on target distance) is significantly smaller than the derived $R_{\mathrm{eff}}$ values ($>$1 kpc), and so angular resolution has a negligible effect on the $R_{\mathrm{eff}}$ determinations for the targets in our sample (column 2, Table~\ref{tab:Reff}). We note that the SINFONI FoV, and therefore the stellar kinematic maps, do not fully cover the area defined by the $R_{\mathrm{eff}}$ in most of the objects.

Table~\ref{tab:Reff} also presents, for comparison, the effective radii computed by \citet{Bellocchi+13} using GALFIT based on 2MASS K-band images. Omitting the IRAS\,F12115-4656 value, which was extracted from its NACO/Ks image, we observe that the results show good agreement within uncertainties, with a mean difference of $\sim$10\%. Larger differences are observed in NGC\,3256 and NGC\,5135, which is likely due to the larger spatial resolution of the IRAC images, which allows better resolution of their complex and bright internal structure.

\subsubsection{Photometrically derived inclinations }
\label{subsubsec:evolution_morph}

A reliable inclination value is crucial for correction of the observed rotational velocity and thus for determination of the dynamical mass of these objects. Therefore, we compared our derived inclinations with those obtained at different distances from the same HST/NICMOS images by \cite{Bellocchi+13} and from the 2MASS/Ks images \citep{2MASS+06} obtained from the NASA/IPAC Extragalactic Database (NED) listed in Table~\ref{tab:phot_results}. 

Although the inclinations were measured at different radial distances, we observe similar inclinations when using the same HST/NICMOS images (i.e. <sin($i$)/sin($i_{\mathrm{B+13}}$)>=0.97$\pm$0.16, with typical differences smaller than 10$\degr$ for 7 out of 10 galaxies). Similar differences were obtained when we compared with the values extracted from the 2MASS/Ks images (i.e. <sin($i$)/sin($i_{\mathrm{2MASS}}$)>=0.92$\pm$0.17). Only NGC\,5135 displays a difference larger than 20$\degr$. This is likely produced by its complex nuclear structure, as the radial profile of the inclination for this object (Figs.~\ref{fig:morph_5135}) shows values similar to those from 2MASS at larger radii.

These results show that the differences in the inclination produced by using the r<1-2\,kpc regions instead of wider FoVs are <10\%, similar to those obtained when using different apertures and/or isophotes within the same FoV. Therefore, our inclination values are not strongly affected by the aperture assumed to extract them. Despite this, the differences in the inclinations of the most face-on objects imply errors of up to a factor of $\sim$1.5 in sin($i$).

\section{Results and discussion}
\label{sec:results}

 \subsection{Stellar velocity fields and velocity dispersion maps}
\label{subsec:vel_maps}

\begin{figure*}
\includegraphics[width=\linewidth]{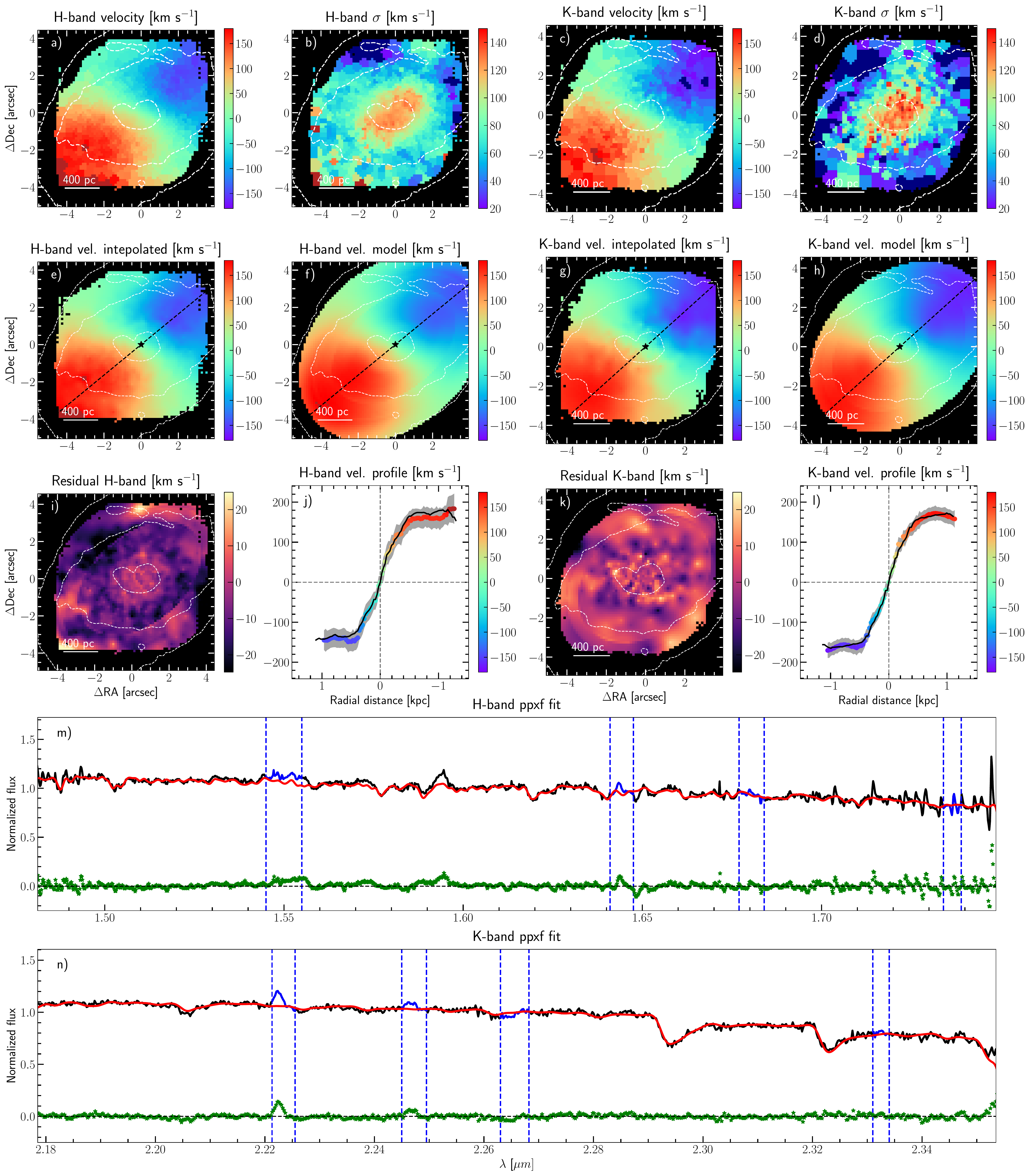} 
\caption{ESO\,320-G030 kinematic results. Panels `a' and `b' (`c' and `d') show the velocity and velocity dispersion maps extracted with \texttt{pPXF} for the H-band (K-band). The HST contours used in Fig.~\ref{fig:example_morph} are over-plotted here for reference. Panels `e' and `f' (`g' and `h') display the interpolated velocity maps and the resulting \texttt{DiskFit} model for the H-band (K-band). The PA and kinematic centre extracted from these models are represented by a dashed line and a star, respectively. The residuals of each model are presented in panels `i' and `k'. The observed velocity profiles extracted along the major axis on the interpolated velocity maps are shown in panels `j' and `l' as coloured dots, whereas the grey areas delimit the 1$\sigma$ confidence interval from their \texttt{pPXF} errors. In these panels, the modelled rotation curves are represented as black lines. Panel `m' (`n') shows, as an example, the \texttt{pPXF} fit of the spectra of the H-band (K-band) kinematic centre. Black and red lines represent the normalised spectra and their best fit, with their residuals plotted as green dots. Blue dashed lines delimit the wavelength ranges excluded during the analysis because of the presence of emission lines (e.g. [\ion{Fe}{II}] $\lambda$1.664\,$\mathrm{\micron}$, Br11 $\lambda$1.681\,$\mathrm{\micron}$, H$_2$\,1-0\,S(0) $\lambda$2.224\,$\mathrm{\micron}$, etc.), atmospheric lines, or strong residuals from the SINFONI `footprint' (see Sect.~\ref{subsec:ifs}).}
\label{fig:example_kin}
\end{figure*}

Figure~\ref{fig:example_kin} shows the velocity and velocity dispersion maps obtained for ESO\,320-G030. Analogous figures for the rest of the sample can be found in Appendix~\ref{append:kin}. In general, our galaxies display velocity maps (panels `a' and `c') with clear anti-symmetric patterns consistent with rotating discs. We observe similar orientation and amplitudes in both near-IR bands, although the velocity maps seem to be smoother in the H-band than in the K-band. Besides the general rotating-disc structure, few objects present distinctive signatures. Non-axisymmetric motions identified by the so-called `S' pattern, usually associated with the presence of bar-like structures (e.g. \citealt{Emsellem+06,Riffel&Storchi+11,Busch+17}), are clearly noticeable in ESO\,320-G030. Moreover, NGC\,2369 and IRAS\,F17138-1017 present velocity structures in the
inner $\sim$1\,kpc that deviate from their global rotating-disc patterns. In contrast with the rotating-disc structure found in most of the sample, the velocity distribution observed in NGC\,3256 reveals a complex pattern, presumably produced by its merger nature. See Appendix~\ref{append:sources_notes} for notes on individual sources.

The velocity dispersion maps (panels `b' and `d') show, in general, flat or slightly decreasing radial distributions with central values in the range $\sim$80$-$110\,km\,s$^{-1}$, in agreement with values obtained at the nuclear regions (i.e. r<2\,kpc) of spiral galaxies in the literature \citep{Bottema+92,deZeeuw+02,Batcheldor+05,Falcon-Barroso+17,Mogotsi+19}. At r$\sim$\,$R_\mathrm{r}$, the velocity dispersion is typically $\sim$80\,km\,s$^{-1}$, whereas at larger radii the values become unreliable as they are likely overestimated due to the beam-smearing inside the large Voronoi bins and the low S/N at the continuum. Although the whole sample shows velocity maps compatible with rotating discs, only ESO\,320-G030 presents a $\sigma$ map with the expected nuclear-peak structure with large central values (i.e. $\sim$125\,km\,s$^{-1}$) radially decreasing until $\sim$50\,km\,s$^{-1}$ in agreement with those values obtained in \citet{Cazzoli+14}. 
In IRAS\,F17138-1017 and NGC\,5135, the $\sigma$ maps display areas with lower values ($\sim$60\,km\,s$^{-1}$) than their neighbouring regions, spatially correlating with the intense SF blobs studied in \citet{PiquerasLopez+16}. Decrements of the stellar dispersion in the nuclear region and/or ring-like structures have previously been observed in spiral galaxies \citep{Marquez+03,Falcon-Barroso+06,Riffel+11,Busch+17}. The most likely scenario supported by numerical simulations \citep{Wozniak+03} is that these stellar dispersion drops are produced by young stars that were born from kinematically cold gas, inheriting its low-$\sigma$ values. The spatial correlation between our low-$\sigma$ regions and the SF blobs is consistent with this hypothesis.

\subsection{Rotating-disc models and stellar velocity profiles}
\label{subsec:results_DiskFit}

\begin{figure}
    \includegraphics[width=\columnwidth]{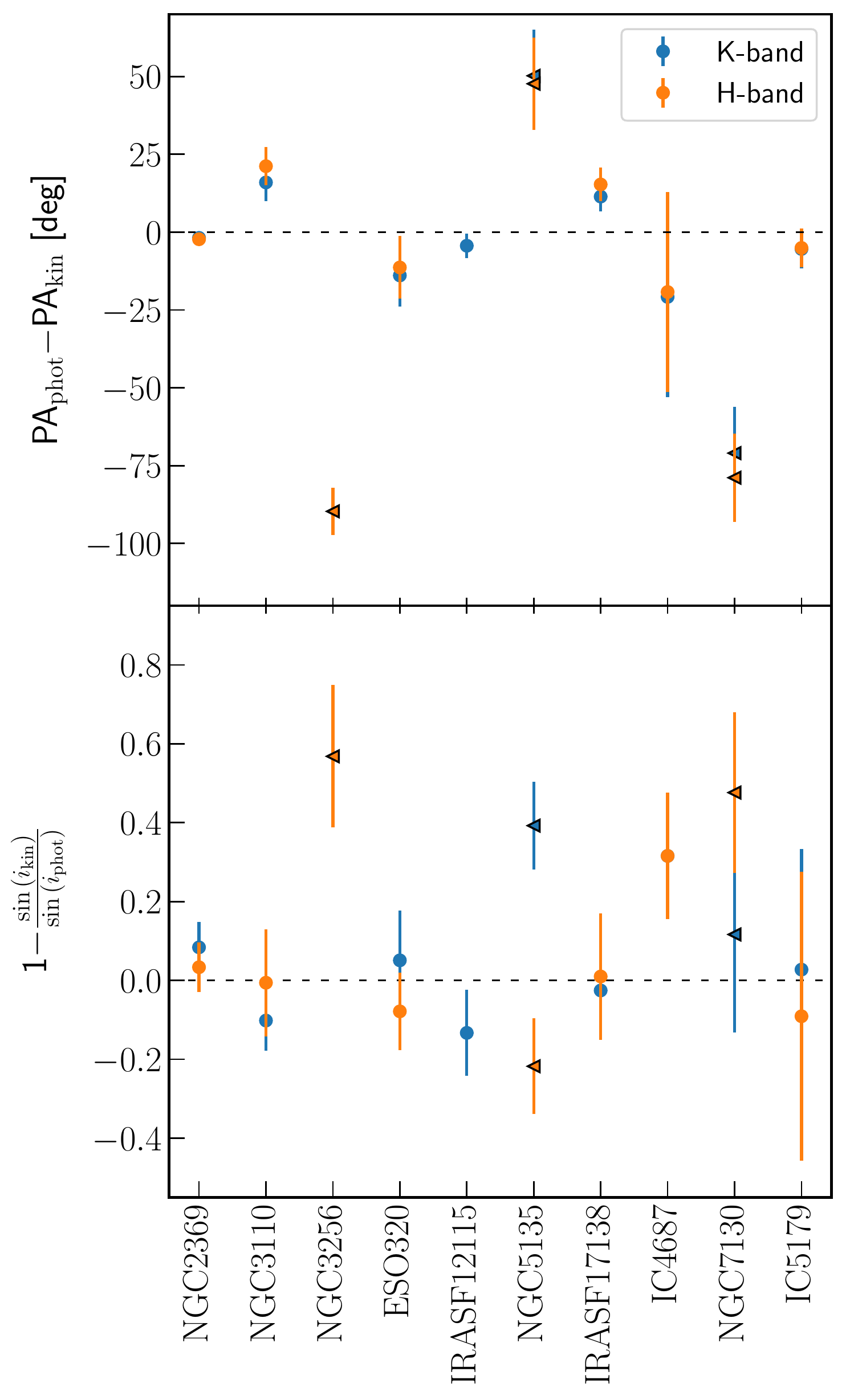}
    \caption{Comparison between the photometrically and kinematically derived parameters. The difference between the photometric (see Sect.~\ref{subsec:phot_analysis}) and the H- (in orange) and K-band (in blue) kinematic values (Sect.~\ref{subsec:results_DiskFit}) for the PA and inclination, represented as the relative difference of sin($i$), are displayed in the top and bottom panels, respectively. The galaxies represented by triangles with black borders are excluded in the subsequent analysis (see Sect.~\ref{subsec:starsvsgas})}
    \label{fig:comp_HK}
\end{figure}

\begin{table*}
\caption{Kinematic properties extracted from the \texttt{DiskFit} models for the H- and K-band velocity maps. Columns (2), (3), (7), and (8): Position of the kinematic centres (i.e. RA and Dec), as the distance from the photometric centre (see Table~\ref{tab:phot_results}). Columns (4) and (9): Position angle (PA), defined as the anti-clockwise angle from the north to the semi-major axis. Columns (5) and (10): Inclination values, computed as $\cos{i}=b/a$, where $b/a$ is the minor-to-major axis ratio. Columns (6) and (11): Orientation of the additional radial component ($\phi$=PA-$\theta_\mathrm{b}$) for those objects that showed radially structured residuals, defined as from north to east. ${\mathrm{\dagger}}$: For this object, we forced the kinematic centre to coincide with the photometric centre during the fit.} 
\centering          
\begin{tabular}{c c c c c c c c c c c }     
\hline  
NAME  &  $\Delta$RA$_{\text{H}}$ & $\Delta$Dec$_{\text{H}}$ & PA$_{\text{H}}$ & $i_{\text{H}}$ &  $\phi_{\text{H}}$  & $\Delta$RA$_{\text{K}}$ & $\Delta$Dec$_{\text{K}}$ & PA$_{\text{K}}$ & $i_{\text{K}}$ &  $\phi_{\text{K}}$  \\
   & (arcsec) & (arcsec)  &   (deg)   & (deg)   &  (deg) & (arcsec) & (arcsec)  &   (deg)   & (deg)   &  (deg)  \\
    (1)    &    (2)   &  (3)  & (4)   &  (5)  & (6) & (7) & (8) & (9) & (10)  & (11)  \\ \hline 
NGC\,2369         & -0.27$\pm$0.03 & 0.52$\pm$0.09 & 173$\pm$1 & 63$\pm$2  & ... &  -0.16$\pm$0.07 & 0.58$\pm$0.05 & 172$\pm$1  & 64$\pm$2 & ...  \\
NGC\,3110       & 0.08$\pm$0.08 &  0.25$\pm$0.08 & 168$\pm$2 & 50$\pm$7 & ...  &  -0.19$\pm$0.04 & 0.28$\pm$0.05 & 173$\pm$1 & 57$\pm$2 & ... \\ 
NGC\,3256        & -0.50$\pm$0.18  & 0.60$\pm$0.24 & 80$\pm$2 & 18$\pm$2  & ... & ...  & ... & ... & ... & ... \\
ESO\,320-G030     & -0.04$\pm$0.05 & -0.07$\pm$0.06 & 129$\pm$1 & 51$\pm$2 & 97$\pm$5  & -0.20$\pm$0.03 & -0.05$\pm$0.02 & 132$\pm$1 & 43$\pm$5 & 82$\pm$7  \\ 
IRAS\,F12115-4656 & ... &... & ... & ... & ... & -0.31$\pm$0.02 & -0.13$\pm$0.03 & 124$\pm$2 & 69$\pm$3  & ...  \\ 
NGC\,5135         & -0.94$\pm$0.04  & -0.15$\pm$0.06  & 8$\pm$3 & 39$\pm$6  & ...   & -0.72$\pm$0.15 & -0.19$\pm$0.13 & 6$\pm$2 & 18$\pm$6 & ... \\ 
IRAS\,F17138-1017  & -0.59$\pm$0.05 & -0.21$\pm$0.03 & 2$\pm$3 & 61$\pm$7  & 36$\pm$8   & -0.48$\pm$0.08 & -0.15$\pm$0.08 & 6$\pm$1 & 65$\pm$2 & 36$\pm$3 \\ 
IC\,4687 $^{\mathrm{\dagger}}$  & ... & ... & 40$\pm$3 & 18$\pm$4   & ...   & ... & ... &  41$\pm$3& 18$\pm$5 & ... \\ 
NGC\,7130      & 0.27$\pm$0.08 & -0.17$\pm$0.09 & 53$\pm$1 &  18$\pm$9  & ...   & 0.01$\pm$0.07 & 0.05$\pm$0.06 & 45$\pm$5 & 32$\pm$13 & ... \\ 
IC\,5179         & -0.55$\pm$0.37 & 0.21$\pm$0.20 & 51$\pm$2 & 70$\pm$20  & ... & -0.55$\pm$0.32 & 0.47$\pm$0.26 & 52$\pm$2 & 57$\pm$17 & ...\\ 
\hline 

\end{tabular}
\label{tab:DiskFit_HK}      
\end{table*}

We used \texttt{DiskFit} (Sect.~\ref{subsubsec:DiskFit}) to model the extracted velocity fields as rotating discs for the H and K bands in eight out of the ten targets. Although NGC\,3256 was observed in both bands, its complex kinematics and limited FoV prevent a reliable modelling using the K-band. Moreover, IRAS\,F12115-4656 was only observed, and therefore modelled, in the K-band.

Panels `f' and `h' of Fig.~\ref{fig:example_kin} show, as an example, the models obtained for ESO\,320-G030 from their interpolated velocity maps (panels `e' and `g', respectively), while the analogous figures for the rest of the sample are presented in Appendix~\ref{append:kin}. In general, the residuals between the interpolated and the modelled velocity maps (panels `i' and `k') typically show values smaller than $\sim$30\,km\,s$^{-1}$ for the H-band ($\sim$40\,km\,s$^{-1}$ for the K-band) in the innermost regions (i.e. r<0.6-1.2\,kpc), indicating that the stellar component of these objects can be well represented by a rotating disc. However, small regions with larger residuals can be found at the outskirts of the velocity maps for some of the objects, likely produced by the patchy observed velocity map due to the large bins and/or deviations from co-planarity.

Table~\ref{tab:DiskFit_HK} presents the PA, inclination and kinematic centre extracted from the \texttt{DiskFit} models for each galaxy and near-IR band. There, the orientation of the additional non-circular component (i.e. $\phi$=PA-$\theta_\mathrm{b}$) is also listed. Although the kinematically extracted parameters present small uncertainties for most of the sample, indicating a robust modelling, it is worth highlighting the larger uncertainties in IC\,5179, NGC\,5135, and NGC\,7130. The disc models of these objects are affected by the pointing strategy considered during the observations, which yields FoVs that do not symmetrically cover their velocity distribution. These effects lead to less restricted parameters and therefore larger uncertainties.

We extracted the velocity profiles from these best-fit velocity models and compared them with those obtained from the observed maps (panels `j' and `l', Fig.~\ref{fig:example_kin}). The modelled rotation curves lie within the 1\text{$\sigma$} confidence level from the original curves, shown as grey bands, for most of the sample. For four out of the ten objects, the velocity curves seem to keep growing at the outskirts of our FoV, indicating that these galaxies may have not reached the maximum rotation velocity at r$\sim$1\,kpc. Although ESO\,320-G030, IRAS\,F12115-4656, and NGC\,3256 have flatter rotation curves at r>0.5\,kpc, we cannot rule out that these values are local maxima produced by the transition between the bulge and the disc.
More specifically, these local maxima are clearly visible for IC\,5179 and NGC\,3110 as abrupt changes in the slope of their velocity curve (see Figs.~\ref{fig:kin_5179} and \ref{fig:kin_3110}). For these galaxies, one can observe steeper curves at the inner regions (r<0.5\,kpc) that are spatially correlated with elongated bulges (see Appendix~\ref{append:morph}). These results suggest that these nuclear regions are not classical bulges (i.e. dispersion-dominated) but pseudo-bulges. Moreover, flat velocity dispersion and double-humped velocity profiles (i.e. displaying a local peak) have been associated with nuclear discs and boxy/peanut (B/P) bulges in simulated and observed spiral galaxies \citep{Kormendy+04,Bureau+05,deLorenzo-Caceres+12,Fabricius+12,Mendez-Abreu+14}.

The agreement observed between the H- and K-band velocity maps was also found when we compared the geometrical parameters extracted from their models (see Fig.~\ref{fig:comp_HK}). The position angle (PA) shows differences smaller than 10$\degr$ for the whole sample with a mean value of $\langle|\mathrm{PA_H-PA_K}|\rangle$=3$\pm$2$\degr$. The inclination seems to be more affected by the limitations of the FoV. A good kinematically derived inclination relies on having a velocity map that is wide enough to characterise the isovelocity contours. For those cases with ambiguous isovelocity lines, \texttt{DiskFit} tends to model face-on velocity fields, assuming the maximum $b/a$ allowed by the code (i.e. $b/a$=0.95, $i$=18\degr). Despite these limitations, we still find good agreement between the inclination values extracted in both bands ($\langle|i_\mathrm{H}-i_\mathrm{K}|\rangle$=8$\pm$7\degr). The kinematic centre determination, although likely to be biased by the FoV, shows differences similar to the spaxel size of our IFS data ($\langle|$C$_\mathrm{H}-\mathrm{C}_\mathrm{K}|\rangle$=0.13$\pm$0.07\arcsec), and smaller than the spatial resolution (i.e. $\sim$0.63\arcsec). For IC\,4687, we forced the kinematic centre to coincide with the photometric one, as the peculiar kinematics produced by the spiral arm in the north-west region was altering the fit and yielding incorrect centre determinations. Based on this general agreement, in the subsequent analysis, we adopted the H-band as a tracer of the stellar component for simplicity.

These kinematically derived parameters are also compared with those obtained from the photometric analysis (extracted at $R_{\mathrm{r}}$, see Sect.~\ref{subsec:phot_analysis}). Figure~\ref{fig:comp_HK} shows the difference in these values for the PA and the inclination, displayed as function of sin($i$) to quantify its possible effect on the rotation velocity.

We find that, for most of the sample, there is good agreement between the values derived from the photometry and the kinematics, with differences smaller than 20$\degr$ in their PA (top panel) and <20\% in $\sin(i)$ (bottom panel). The large differences in some objects are likely produced by the complex kinematic and photometric structure of the merger (NGC\,3256) and by the difficulty in measuring reliable PA and inclinations in the most face-on galaxies (NGC\,5135 and NGC\,7130). In addition, IC\,4687 presents a large discrepancy in the inclination, produced by the low values of inclination extracted from the \texttt{DiskFit} models. The isovelocity lines of this object are almost parallel to the minor-axis in the nuclear regions and therefore the modelling is insensitive to the inclination, yielding a velocity model created with the lowest inclination allowed (i.e. 18$\degr$).

For this object, we create a grid of \texttt{DiskFit} models with the ellipticity ranging from 0.05 (i.e.$\sim$18\degr) to 0.6 (i.e.$\sim$67\degr) to study the degeneration in the inclination value. From these models, we obtain similar residuals, with differences of <5 km/s, using ellipticities from 0.05 (i.e. 18º) to 0.3 (i.e. 46º). This range of compatible inclinations is in agreement with the photometrically derived values derived in Sect.~\ref{subsec:phot_analysis} and by \citet{Bellocchi+13} (Table~\ref{tab:phot_results}).
 
Hereafter, we consider the kinematically derived inclination to correct the observed velocities, except for IC\,4687, for which we adopted the photometric value (i.e. 27$\pm$7$\degr$) as it is more constrained than the kinematically derived range of values. We refer the reader to Appendix~\ref{append:sources_notes} for more detailed descriptions of the individual results.

\subsection{Comparison with interstellar gas kinematics}
\label{subsec:starsvsgas}

\begin{figure}
\centering
    \includegraphics[width=\columnwidth]{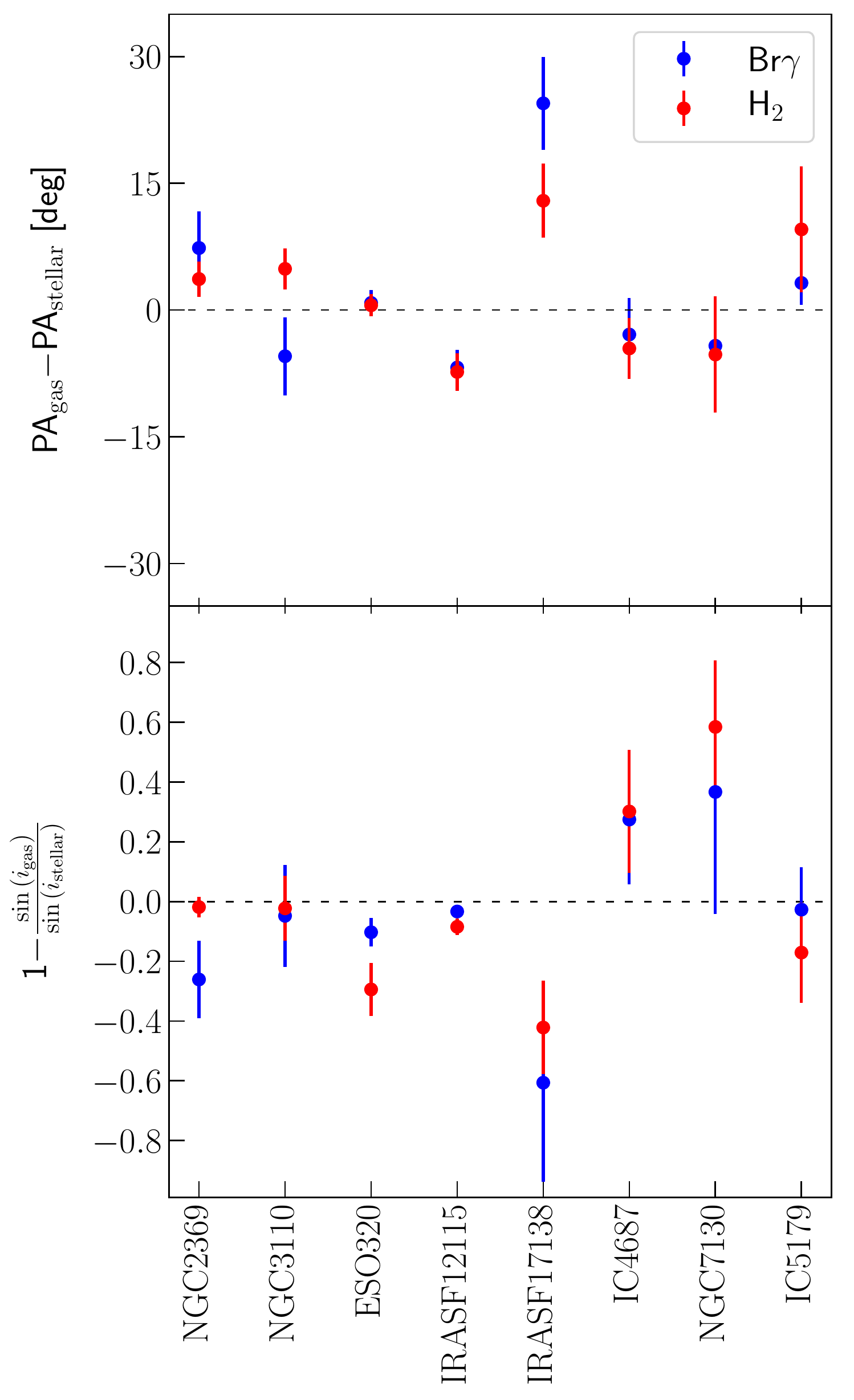}
    \caption{Comparison between the gas and stellar kinematically derived parameters. Difference between the PA and sin($i$) from the \texttt{DiskFit} models of stellar and Br$\gamma$ and H$_2$1-0\,S(1) velocity maps, in blue and red, respectively.  }
    \label{fig:PA_gas}
\end{figure}

\begin{figure*}
\centering
    \includegraphics[width=0.88\linewidth]{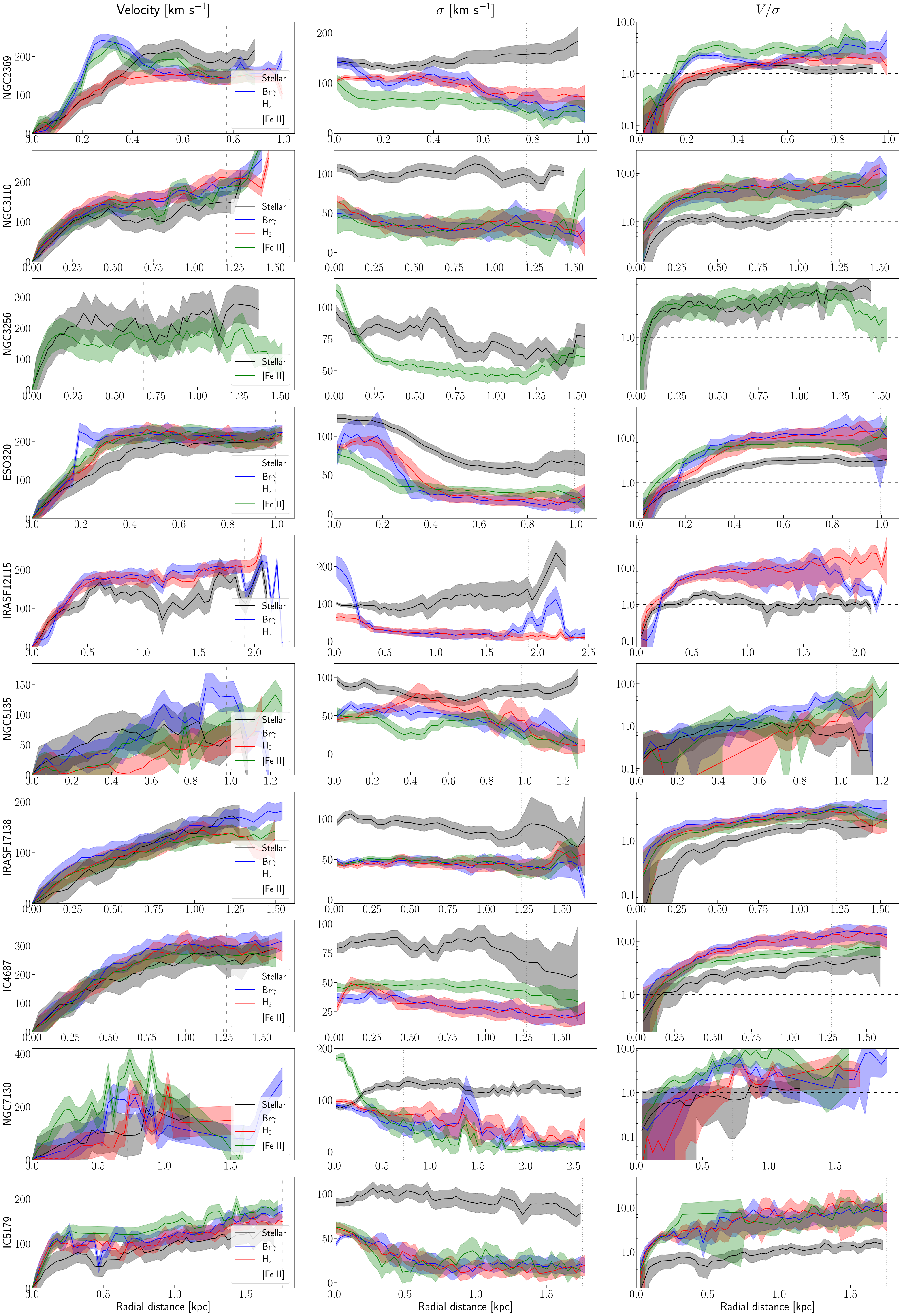}
    \caption{Gas and stellar kinematics comparison. \textit{First column}: Radial profile of the inclination-corrected velocity amplitude for the stellar, ionised (Br$\gamma$), partially ionised ([\ion{Fe}{II}]), and hot molecular gas (H$_2$1-0\,S(1)) phases in black, blue, green, and red, respectively. \textit{Second and third columns}: Ring-averaged radial profiles for the velocity dispersion and the $V/\sigma$ ratio. Dotted vertical lines mark the distance (i.e. $R_{\mathrm{r}}$) adopted to extract the kinematic values from Table~\ref{tab:dynamics_gas}. Dashed horizontal lines in the right panels mark the $V/\sigma$=1 value, which is commonly adopted to discriminate between rotation- and dispersion-supported galaxies.}
    \label{fig:gas_kin}
\end{figure*}

\begin{figure}
\centering
    \includegraphics[width=\columnwidth]{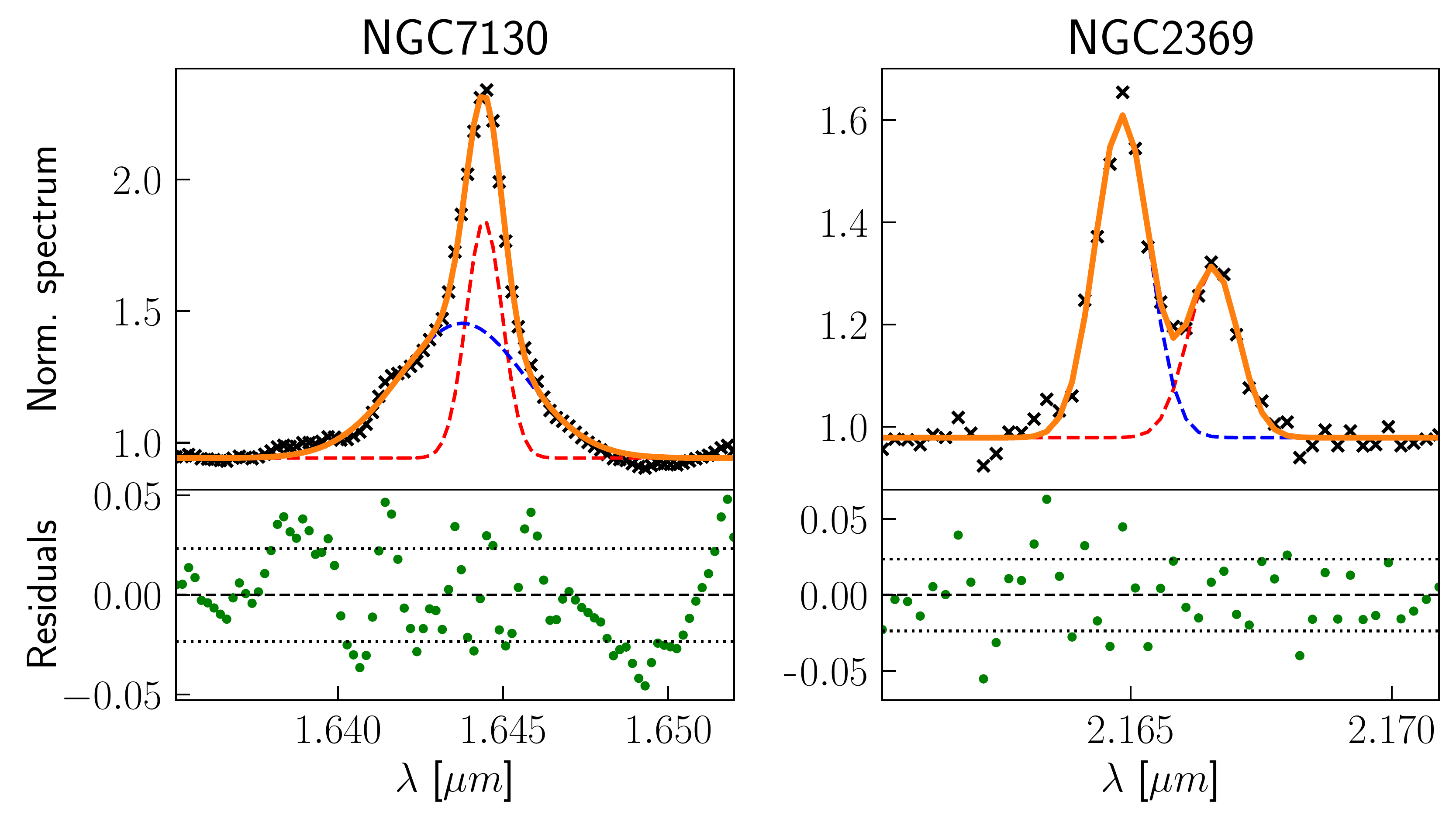}
    \caption{Double-Gaussian profile fits. Line decomposition using a double-Gaussian profile for [\ion{Fe}{II}] in NGC\,7130 (left panel) and Br$\gamma$ in NGC\,2369 (right panel). Normalised spectra are shown as black crosses, while the fitted components and their sum are displayed in blue, red, and orange, respectively. Residuals are shown as green dots in the bottom panels. Dotted horizontal lines represent the 1$\sigma$ value of the residuals.}
    \label{fig:doble_fit}
\end{figure}
 
\begin{figure}
\centering
    \includegraphics[width=\linewidth]{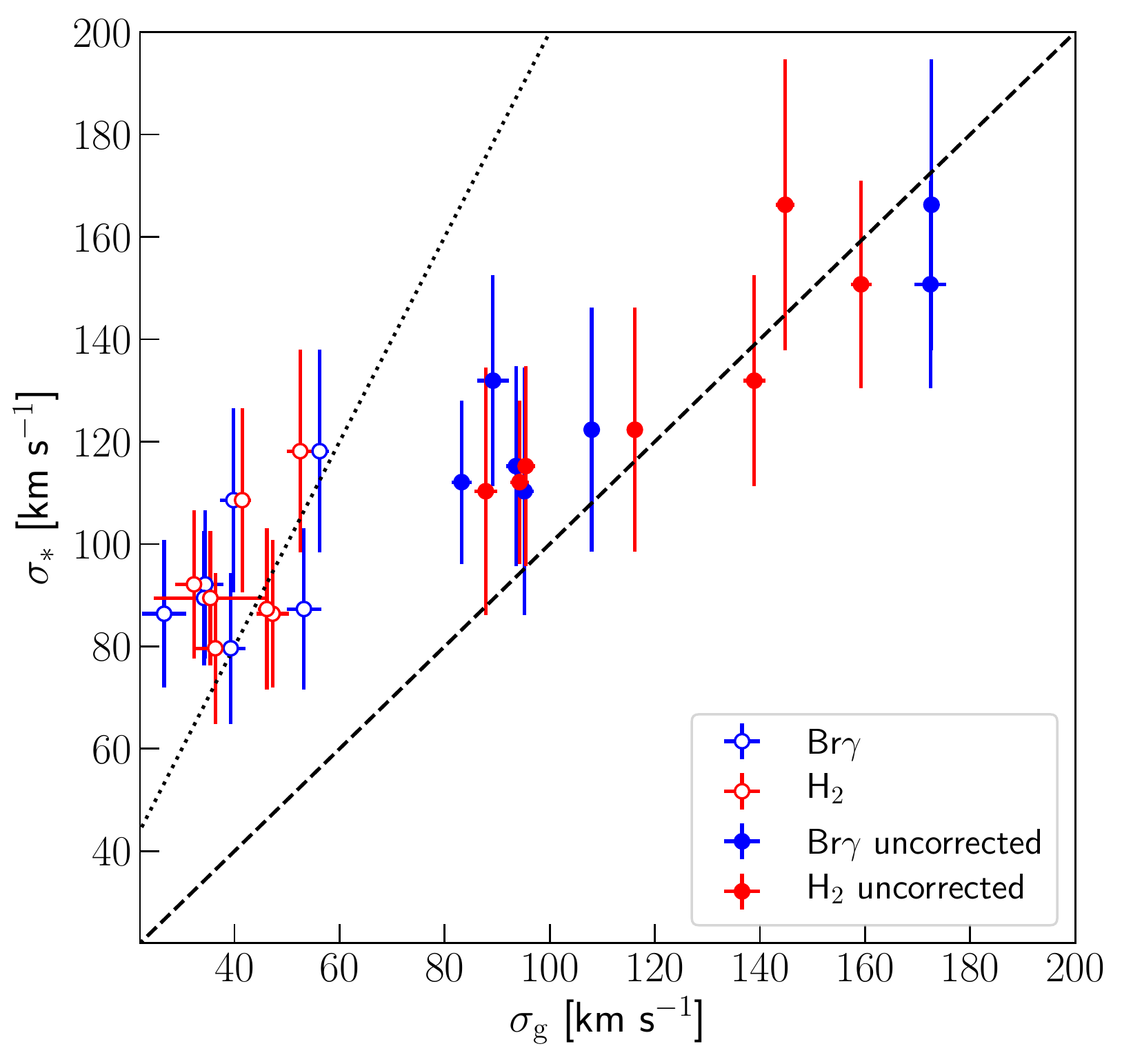}
    \caption{Effect of the aperture-smearing correction. Velocity dispersion values for the stellar and ionised and hot molecular gas phases (blue and red, respectively). Empty circles represent the $\sigma$ values obtained after correcting all the spectra within the $R_\mathrm{r}$, whereas the filled circles are the integrated $\sigma$ values without this correction.}
    \label{fig:beam_smearing}
\end{figure}

\begin{figure}
\centering
    \includegraphics[width=\columnwidth]{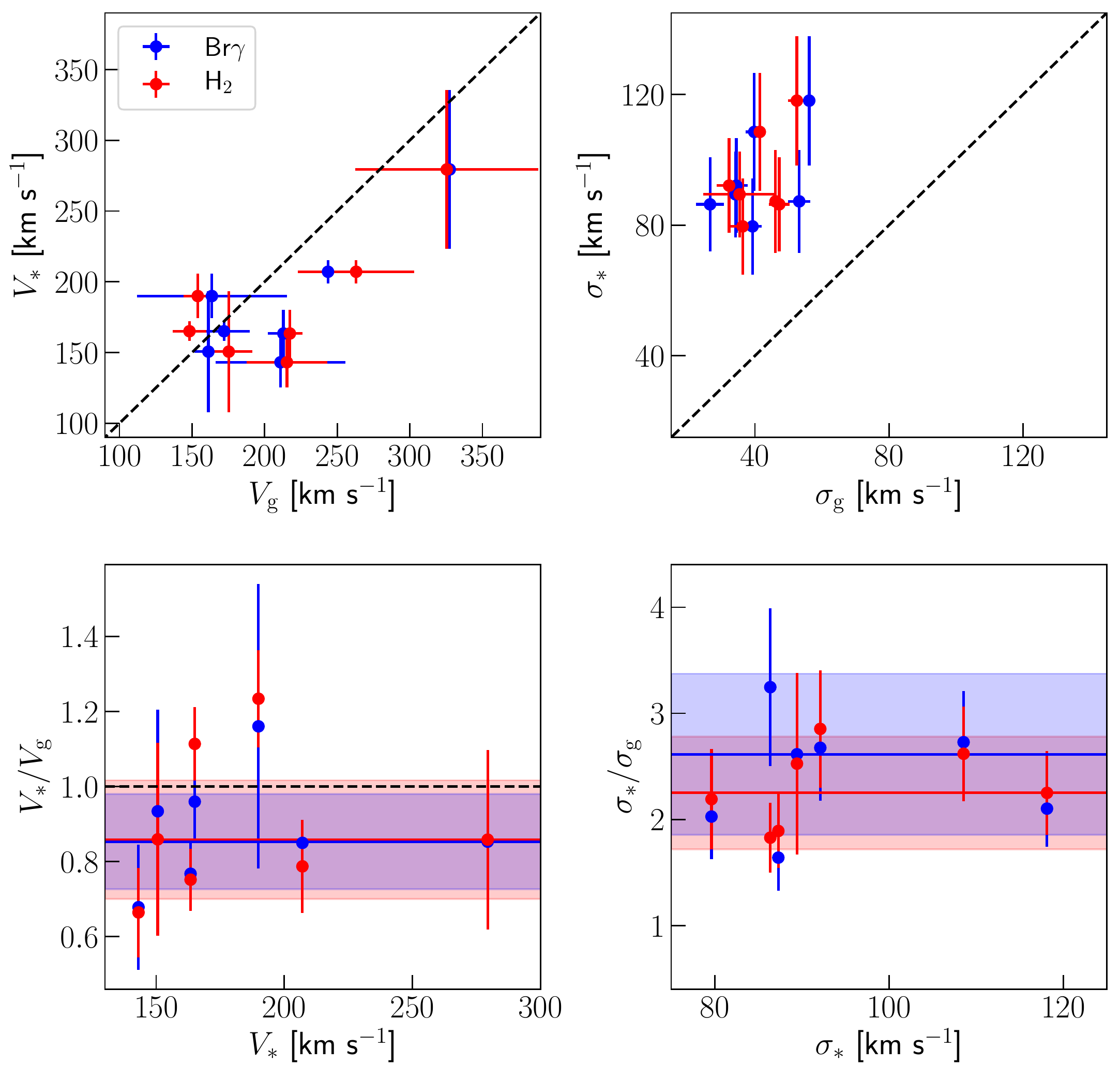}
    \caption{Gas and stellar kinematic results. Comparison between the stellar and gas (ionised and molecular in blue and red, respectively) inclination-corrected velocity amplitude (upper left panel) and velocity dispersion (upper right panel) extracted in Sect.~\ref{subsec:starsvsgas}. Dashed lines represent the 1:1 relation. Lower left and right panel shows the ratio between the stellar and gas velocity and velocity dispersion as a function of their stellar values. Red and blue horizontal lines represent the mean values for the ionised and hot molecular phases, whereas the coloured areas represent the standard deviation of the values.}
    \label{fig:AD}
\end{figure}
\begin{figure*}
\centering
    \includegraphics[width=\linewidth]{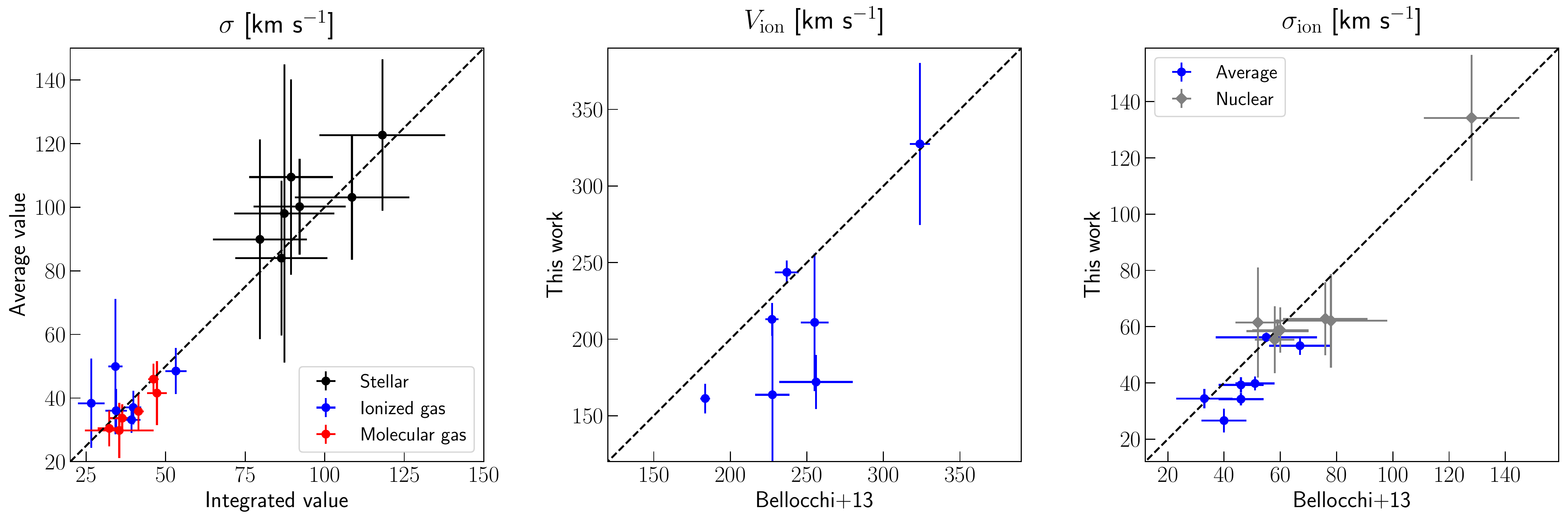}
    \caption{\textit{Left panel}: Uniformly weighted average and integrated aperture-smearing corrected velocity dispersion values obtained from the gas and stellar phases. For this comparison, we omitted the gas results of NGC\,2369 because its $\sigma$ values were not computed using the integrated spectra (see Table~\ref{tab:dynamics_gas}). \textit{Middle}: Inclination-corrected velocity amplitude for the ionised gas extracted in this work and in \cite{Bellocchi+13}. We assumed the kinematically derived inclination extracted in this work (Table~\ref{tab:DiskFit_HK}) to correct both the optical and near-IR observed velocities. \textit{Right panel}: Comparison between the average (blue) and nuclear (grey) velocity dispersion values extracted from optical and near-IR emission lines. Dashed lines represent the 1:1 relation.}
    \label{fig:Vbell_comp}
\end{figure*}

\begin{table*}
\caption{Stellar and gas kinematics results, extracted within 1$R_{\mathrm{r}}$. Columns (2), (5), and (8): Inclination-corrected velocity amplitude, defined as the value of the rotation curve at $R_\mathrm{r}$. Columns (3), (6), and (9): Aperture-smearing corrected velocity dispersion. Columns (4), (7), and (10): $V/\sigma$ for the stellar, ionised, and molecular gas phases, respectively. Uncertainties were computed with Monte Carlo simulations. The ongoing merger (i.e. NGC\,3256) was discarded from this analysis due to its limited FoV and peculiar kinematics. We also dismissed NGC\,5135 and NGC\,7130 due to their irregular velocity profiles, produced by their face-on orientations.  (${\mathrm{\dagger}}$) The emission lines of NGC\,2369 were fitted using double-Gaussian profiles (see Sect.~\ref{subsec:starsvsgas} and App.~\ref{append:2369}) and its $\sigma$ values were assumed to be the average of both components across the FoV. However, its complex kinematics does not allow us to extract a velocity amplitude and therefore we assume its value from the single-Gaussian map. } 
\centering      
\begin{tabular}{c c c c c c c c c c}     
\hline  
 NAME  & $V_{*}$ & $\sigma_{*}$ & $(V/\sigma)_{*}$ & $V_{\mathrm{ion}}$ & $\sigma_{\mathrm{ion}}$ & $(V/\sigma)_{\mathrm{ion}}$ & $V_{\mathrm{mol}}$ & $\sigma_{\mathrm{mol}}$ & $(V/\sigma)_{\mathrm{mol}}$  \\
      &  (km s$^{-1}$)    &   (km s$^{-1}$)   &   &  (km s$^{-1}$)   &   (km s$^{-1}$)     &   &   (km s$^{-1}$)    &   (km s$^{-1}$)     &    \\ 
 (1) &  (2) & (3) & (4) & (5) & (6) &  (7)  & (8) & (9) & (10)  \\ \hline 
NGC\,2369$^{\mathrm{\dagger}}$& 190$\pm$16 & 118$\pm$18 & 1.61$\pm$0.30 & 164$\pm$48 & 56$\pm$2 &  2.91$\pm$0.63 &  154$\pm$10 & 53$\pm$3 &  2.93$\pm$0.24 \\
NGC\,3110        &  143$\pm$18 & 108$\pm$18 &  1.32$\pm$0.27 &  211$\pm$44 & 40$\pm$2 &  5.30$\pm$0.87 &  216$\pm$28 & 41$\pm$2 &  5.20$\pm$0.70 \\ 
ESO\,320-G030    &  207$\pm$13 & 86$\pm$14 &  2.40$\pm$0.41 &  244$\pm$13 & 27$\pm$4 &  9.16$\pm$1.47 &  263$\pm$40 & 47$\pm$3 &  5.57$\pm$0.92  \\
IRAS\,F12115-4656 &  163$\pm$17 & 89$\pm$13 &  1.83$\pm$0.33 &  213$\pm$12 & 34$\pm$2 &  6.22$\pm$0.51 &   217$\pm$11 & 35$\pm$11 &  6.14$\pm$1.88 \\
IRAS\,F17138-1017 &  165$\pm$10 & 87$\pm$16 &  1.89$\pm$0.35 &  172$\pm$18 & 53$\pm$3 &  3.23$\pm$0.39 &  148$\pm$12 & 46$\pm$2 &  3.21$\pm$0.28 \\
IC\,4687          &  279$\pm$40 & 79$\pm$15 &  3.51$\pm$0.96 &  328$\pm$53 & 39$\pm$3 &  8.34$\pm$1.37 &  325$\pm$62 & 36$\pm$4 &  8.96$\pm$1.84  \\
IC\,5179         &   151$\pm$33 & 92$\pm$14    &  1.64$\pm$0.41 &  161$\pm$11 & 34$\pm$4 &  4.68$\pm$0.55 &   175$\pm$16 & 32$\pm$4 & 5.43$\pm$0.79  \\
\hline           
\end{tabular}
\label{tab:dynamics_gas}      
\end{table*}

In contrast to the stellar component, whose kinematics are driven by the gravitational potential, the gas phases are more easily disturbed by different mechanisms (e.g. AGN, star-formation). In this section, we compare the gas and stellar kinematics to identify possible differences in the inner regions of our LIRGs, where large SF rates and/or AGNs have already been observed in some cases. We used the kinematic maps derived for the different gas components (i.e. partially ionised, ionised, and hot molecular gas) extracted by \citet{PiquerasLopez+12} from the same SINFONI dataset.

Firstly, we used \texttt{DiskFit} to model the velocity maps for the ionised and hot molecular gas phases (traced by the Br$\gamma$ and H$_2$1-0\,S(1) lines) in eight out of the ten galaxies from the sample. From these models, we extracted the PA and inclination to probe whether the stellar and gas phases are aligned or not (see Fig~\ref{fig:PA_gas}). We did not consider the merging system (i.e. NGC\,2356) during this comparison as its southern nucleus, observed with the K-band, does not show a stellar or gas kinematics compatible with rotation. In addition, the NGC\,5135 gas phases present a highly disturbed kinematics, which prevents us from obtaining reliable rotating-disc models. 

Figure~\ref{fig:PA_gas} shows that there is good agreement between the PAs of the stellar and gas phases, with differences smaller than 10$\degr$ for most of the sample. However, the differences in sin($i$) are larger, with typical values of 20\% that can be up to 60\%. However, it should be noted that the gas and stellar kinematics not only present different intrinsic structure at small scales but also different S/Ns, yielding different sampled areas for both phases. Although these differences affect the inclination determination during the modelling, in general the results suggest that the stellar and gas kinematics are close to being aligned in the galaxies of the sample.

This alignment allowed us to confidently compare the kinematics along the major axis for both the stellar and gas components. The comparison between the inclination-corrected velocity profiles (first column, Fig.~\ref{fig:gas_kin}) reveals good agreement between the gas and stellar velocity profiles for eight out of the ten objects. Only NGC\,5135 and NGC\,7130, with low inclinations (i.e. small observed velocity amplitudes), show significant discrepancies in their velocity curves, with differences larger than 100\,km\,s$^{-1}$. In addition, we observed that in five objects the stellar velocity displays slightly lower velocity amplitudes at all radii. As observed in the stellar kinematics, the gas velocity curves of ESO\,320-G030, IRAS\,F12115-4656, and NGC\,3256 flatten at $\sim$0.5\,kpc. It is remarkable that, except for NGC\,2369, the different gas phases (ionised, partially ionised, and hot molecular) have similar rotation curves, suggesting a co-rotation of all the gas phases.

On the contrary, the velocity dispersion presents significant differences between the stellar and gas phases. In general, the stellar $\sigma$ values are approximately two times larger than the gaseous ones at every radius, although we obtained similar flat or slightly decreasing profiles (second column of Fig.~\ref{fig:gas_kin}) for both the stellar and gas phases. Consequently, the ring-averaged radial distribution of the $V/\sigma$ ratio (third column of Fig.~\ref{fig:gas_kin}) shows, for most of the sample, larger values for the gas phases at all radii. In general, the gas phases reach values associated with fast rotation (i.e. $V/\sigma$>3) in the inner $\sim$0.5\,kpc, whereas the stellar phase presents slower rotation (i.e. $V/\sigma$\,$\sim$1.5) at these radii. The stellar $V/\sigma$ ratio is almost constant until the outskirts of our FoV (i.e. r$\sim$1.5\,kpc), whereas the gas ratio increases up to $V/\sigma$\,$\sim$10. These dynamical ratios can be interpreted as a different spatial distribution of the two components, with the gas being confined to flat or thin discs and the stars having warmer orbits and populating thicker discs. The agreement between the velocity and velocity dispersion curves (and consequently the $V/\sigma$ profile) between all the gas phases could indicate the existence of a unique gaseous disc, with no significant dynamical segregation between the ionised, partially ionised, and hot molecular components. 

Despite the global agreement in the gas phases, distinctive local differences can be observed at small physical scales (i.e. $\sim$200\,pc). We observed large nuclear $\sigma_{\mathrm{g}}$ peaks in IRAS\,F12115-4656, NGC\,3256, and NGC\,7130, which have been reported to host nuclear outflows \citep{Arribas+14,Davies+14,Sakamoto+14,Knapen+19}. In these cases, the use of a single Gaussian to fit the line profile may yield larger $\sigma_{\mathrm{g}}$ values. Therefore, we performed a double Gaussian fit analysis on the nuclear spectra of these galaxies to confirm this explanation. Figure~\ref{fig:doble_fit} (left panel) shows, as an example, a double-Gaussian profile fit to the nuclear spectrum of NGC\,7130. We obtained a similar secondary, bluer, and wider component in the nuclei of these objects, likely linked to their outflows. Despite the presence of these secondary components, which can contribute to broadening the fitted line modelled with a single Gaussian, the global kinematics of these galaxies show regular rotation-dominated velocity patterns (see Appendix~\ref{append:kin}).

In addition to the objects presenting nuclear outflows, NGC\,2369 also shows distinctive structures in its velocity and $\sigma$ maps and therefore we also performed a double-Gaussian fit for this object (right panel of Fig.~\ref{fig:doble_fit}). The double-Gaussian analysis of Br$\gamma$ could be extended across the entire FoV, revealing two distinctive kinematic components (see Appendix~\ref{append:2369} for further details and velocity maps of each component). This double-line profile is misleading the extraction of the kinematics maps, especially from the gas phases. Indeed, the $V_\mathrm{rot}$ and $\sigma$ values for this object display large differences among all the phases (Table~\ref{tab:dynamics_gas}).

In addition to the comparison between the velocity curves along the major axis, we extracted the modelled velocity amplitude as well as the light-weighted average velocity dispersion for the gas and stars at $R_{\mathrm{r}}$. These values allow us to directly compare the kinematics of the different phases at apertures that grant a reliable analysis. The velocity amplitude was defined as the value of the modelled velocity curve evaluated at $R_\mathrm{r}$. We derived the $\sigma$ using \texttt{pPXF} and single-Gaussian fits (convolved with the instrumental LSF) on the $R_{\mathrm{r}}$-aperture integrated spectrum, minimising the possible effect of spurious values from low-S/N regions at the outskirts of the FoV. Each single spectrum within the aperture was previously set to rest frame using the velocity field to avoid the beam smearing produced by the aperture integration, hereafter referred to as `aperture-smearing'. The importance of this correction can be observed in Fig.~\ref{fig:beam_smearing}, where the comparison of the $\sigma$ values with and without this correction is displayed for the stellar and gas phases. This figure shows how the velocity field can increase the integrated $\sigma$ by factors of up to $\sim$1.5 and $\sim$3 in the stellar and gas phases, respectively, if we do not correct for aperture smearing, yielding compatible velocity dispersion values. 

The velocity amplitude and integrated $\sigma$ of the stellar, ionised, and hot molecular phases are listed in Table~\ref{tab:dynamics_gas} along with their associated uncertainties. We considered the associated error of the modelled velocity curve as the uncertainty on the velocity amplitude, whereas for the integrated $\sigma$, the error was derived from Monte Carlo simulations. In these simulations the flux of each spectra was varied with a Gaussian distribution of $\sigma$ equal to the flux noise. In addition, the effect of the spectral sampling of the LSF in the velocity dispersion was also considered during these Monte Carlo simulations, varying its sampling with a normal distribution of half a spectral channel. At this point, we discarded the objects NGC\,5135, NGC\,7130, and NGC\,3256 as we are not able to obtain a reliable rotating-disc model for either their gas or stellar phase. In addition, the two first objects present significant differences between the stellar and gas velocity profiles obtained from the observed maps (see Fig.~\ref{fig:gas_kin}).

Figure~\ref{fig:AD} shows the comparison between the velocity amplitude and velocity dispersion for the gas and stellar phases. We observed that the gas phases show larger amplitudes (i.e. median $ V_{\mathrm{*}}/V_{\mathrm{g}}$=0.85$\pm$0.13 and 0.86$\pm$0.16 for ionised and molecular, respectively), whereas the velocity dispersion, as observed in the radial profiles, shows larger values for the stellar component than for the gas phases (i.e. median $ \sigma_{\mathrm{*}}/\sigma_{\mathrm{g}}$=2.61$\pm$0.76 and 2.25$\pm$0.53 for the ionised and molecular gas, respectively). The $V/\sigma$ ratio shows that despite the fact that both the stellar and gas components of all the objects are rotation-supported (i.e. $V/\sigma$>1), we obtain low stellar ratios (1<$V/\sigma$<3) due to the larger stellar $\sigma$, whereas the gas is largely rotation-supported (i.e. $V/\sigma$>3). As suggested above, the difference between the stellar and gas $V/\sigma$ can be understood if the stars are rotating in thick discs while the gas phases are confined to dynamically cooler (i.e. thinner) rotating discs.

The ratios between the gas and stellar velocity and velocity dispersion obtained here are in agreement within the uncertainties with those obtained for a subsample of late-type galaxies from the SAMI survey by \citet{Barat+19} (i.e. $V_{\mathrm{*}}/V_{\mathrm{g}}$=0.77$\pm$0.19 and $\sigma_{\mathrm{*}}/\sigma_{\mathrm{g}}$=1.58$\pm$0.59). These stellar and gas kinematics were obtained from the optical continuum and main emission lines (e.g. H$\alpha$, [\ion{O}{III}], etc.), respectively, for a sample of 1559 local (z<0.11) galaxies. However, these kinematic differences between the gas and stars were not observed in the nuclear regions of local (z<0.03) interacting U/LIRGs presented in \citet{Medling+14}, based on AO-assisted OSIRIS data (i.e. $V_{\mathrm{*}}/V_{\mathrm{g}}$=0.98$\pm$0.37 and $\sigma_{\mathrm{*}}/\sigma_{\mathrm{g}}$=1.21$\pm$0.42). In this latter work, the authors extracted the stellar and gas kinematics fitting the CO (2-0) and (3-1) bands and Br$\gamma$ using \texttt{pPXF} and a single Gaussian function, respectively. From these results, we can conclude that, in terms of their dynamics, our LIRGs more closely resemble the isolated late-type galaxies (LTGs) than the interacting U/LIRGs, where the strong gravitational interaction prevents the gas phases from residing in dynamically cool orbits.

In summary, the comparison carried out here shows that the gas presents smaller velocity dispersion values and slightly larger velocity amplitudes than the stars in our LIRGs. Similar results, although unprecedented in LIRGs, were reported in the Milky Way \citep{Stromberg+46,Aumer&Binney+09} and nearby disc galaxies \citep{Herrmann+09,Martinsson+13,Beasley+15,Dorman+15,Quirk+19}, and were explained in terms of the dissipative nature of the gas, which allows it to maintain more uniform orbits and therefore not experience asymmetric drift (AD). This asymmetric drift is defined as a velocity lag between the expected circular velocity produced by the gravitational potential and the actual rotational velocity of the stars. The lag is produced as the stars populate warm orbits (due to their non-negligible velocity dispersion), where they have smaller angular momentum and therefore slower rotations than the surrounding material from cooler orbits. The increasing trend of $\sigma$ with stellar age creates the so-called age--velocity dispersion relation (i.e. AVR, \citealt{wielen+77}), where older stars have lived enough to populate more off-plane orbits, with larger velocity dispersions than the new-born ones.

In addition, numerical simulations show that the thicker stellar discs can also be produced by the presence of disturbed gas at the birth-time of the stars \citep[see][]{Brook+12,Bird+13,Ma+17,Navarro+18,Pillepich+19}. This disturbed gas would experience a process of dynamical cooling, generating stars with decreasing velocity dispersion with time. This gas settling could also explain the fact that the stellar component is in a warmer state compared with the gas phases, as the stars inherit the dynamical state of the gas they were born from.

\subsection{Optical versus near-IR emission lines: impact on the kinematics}
\label{subsec:optvsnir}

We compared our kinematic results from the ionised gas (traced by Br$\gamma$ emission) with those obtained in \citet{Bellocchi+13} using H$\alpha$ for the same objects in order to investigate the impact of using near-IR lines, which are less affected by the dust extinction, to extract the gas kinematics. In \citet{Bellocchi+13}, the H$\alpha$ line was observed with VLT/VIMOS, which provides a FoV$\sim$30$\arcsec$x30$\arcsec$ covering an area that is approximately ten times larger than our SINFONI dataset. These VIMOS data are of similar spectral resolution (i.e. $\sigma$\,$\sim$40\,km\,s$^{-1}$) but two times lower angular resolution (FWHM$\sim$1.3$\arcsec$) than those from SINFONI.

The velocity dispersion definition used in this latter work is different from the one considered here, as their values were computed as the uniformly weighted average value across the FoV. Figure~\ref{fig:Vbell_comp} (left panel) displays the $\sigma$ values obtained from our dataset considering both definitions. The differences found (i.e. <15\,km\,s$^{-1}$) are negligible and therefore we can compare the near-IR and optical $\sigma$ regardless of the definition used. We adopted the photometrically derived inclinations from this work to correct the observed optical velocities, preventing possible differences produced by assuming the inclinations from \citet{Bellocchi+13} (see Sect.~\ref{subsubsec:evolution_morph}). 

The comparison between the inclination-corrected velocities amplitudes (middle panel, Fig.~\ref{fig:Vbell_comp}) revealed that the values obtained with Br$\gamma$ are similar to or smaller than those extracted with H$\alpha$, with differences of up to 70\,km\,s$^{-1}$. These results can be explained by the fact that the velocity curves have not reached their maximum within the SINFONI FoV.

We compared the optical uniformly weighted average $\sigma$ with our near-IR integrated $\sigma$ (blue points in right panel, Fig.~\ref{fig:Vbell_comp}), obtaining similar results with the optical values being slightly larger (up to 15\,km\,s$^{-1}$). These small differences are likely due to our better angular resolution and smaller spaxel size, which minimise the effect of the beam smearing and yield smaller $\sigma$ values. 

To assess this possible explanation, we spatially degraded and re-binned our SINFONI dataset, matching the angular resolution and spaxel size of the VIMOS data. We then extracted the mean near-IR $\sigma$ value of the central region (i.e. 2.7$\arcsec$x2.7$\arcsec$) to compare with the $\sigma_\mathrm{c}$ value presented by \citet{Bellocchi+13}. This comparison at the same resolution and FoV (grey points in right panel of Fig.~\ref{fig:Vbell_comp}) revealed that most of the sample presents very similar optical and near-IR nuclear $\sigma$, whereas only two of the objects maintain the differences found in the integrated values (i.e. $\sim$15\,km\,s$^{-1}$).

From this comparison between the optical and near-IR kinematics, we conclude that the extinction at the optical wavelengths seems not to significantly affect the extraction of the ionised gas kinematics, as the nebular lines used at each wavelength range (i.e. H$\alpha$ and Br$\gamma$) yield relatively similar velocity amplitude and velocity dispersion.

\subsection{Impact of the tracer on the dynamical mass estimation}
\label{subsec:Mdyn}

\begin{table*}
\caption{Dynamical and stellar mass estimations. Columns (2), (3), and (4): Dynamical masses within 1$R_\mathrm{r}$ based on the stellar, ionised, and hot molecular gas kinematics. Column (5): Values of the asymmetric-drift correction factor, as defined in Eq.~\ref{eq:C_AD}. Columns (6) and (7): Stellar masses estimated using a constant IRAC/3.6$\mu m$ M/L derived from the \citet{U+12} and \citet{Pereira-Santaella+15} results, respectively.}
\centering      
\begin{tabular}{ c c c c c | c c }     
\hline  
 NAME  & $M^{\mathrm{stellar}}_{\mathrm{dyn}}($<$R_\mathrm{r})$  & $M^{\mathrm{ion}}_{\mathrm{dyn}}($<$R_\mathrm{r})$   & $M^{\mathrm{mol}}_{\mathrm{dyn}}($<$R_\mathrm{r})$ & $C_\mathrm{AD}$ & $M^{\mathrm{U+12}}_{\mathrm{*}}($<$R_\mathrm{r})$  & $M^{\mathrm{PS+15}}_{\mathrm{*}}($<$R_\mathrm{r})$ \\
     &  (10$^{10}\mathrm{M_{\odot}}$) &  (10$^{10}\mathrm{M_{\odot}}$) &  (10$^{10}\mathrm{M_{\odot}}$) &  &  (10$^{10}\mathrm{M_{\odot}}$) &   (10$^{10}\mathrm{M_{\odot}}$)  \\
 (1) &  (2) & (3) & (4) & (5) & (6) & (7)  \\ \hline 
NGC\,2369       &  1.0$\pm$0.3  &  0.6$\pm$0.1  &  0.5$\pm$0.1 &  1.6$\pm$0.7 & 0.7$\pm$0.2 & 1.7$\pm$0.5   \\
NGC\,3110        &   0.9$\pm$0.3 &    1.2$\pm$0.5 &   1.3$\pm$0.3 &  1.0$\pm$0.8 & 0.7$\pm$0.2 & 2.1$\pm$0.6   \\
ESO\,320-G030   &   1.3$\pm$0.2   &   1.4$\pm$0.1 &    1.7$\pm$0.5 &  1.9$\pm$0.7 & 0.8$\pm$0.2 & 2.7$\pm$0.8  \\
IRAS\,F12115-4656  &   1.7$\pm$0.5 &  2.1$\pm$0.2 &  2.2$\pm$0.2 &  1.6$\pm$1.1 & ... & ... \\
IRAS\,F17138-1017  &  1.2$\pm$0.3   &  1.0$\pm$0.2 &   0.8$\pm$0.2 &  2.1$\pm$0.8   & 0.8$\pm$0.2 & 3.2$\pm$0.9 \\
IC\,4687           &   2.6$\pm$0.9 &  3.2$\pm$1.0 &   3.2$\pm$1.2 &  1.4$\pm$0.8 & 0.8$\pm$0.2 & 2.8$\pm$0.9  \\
IC\,5179       &  1.3$\pm$0.6 &  1.1$\pm$0.2 &    1.3$\pm$0.2 &  1.1$\pm$1.1 & 0.8$\pm$0.2 & 3.9$\pm$1.2  \\
\hline           
\end{tabular}
\label{tab:Mdyn}      
\end{table*}

\begin{figure}
\centering
    \includegraphics[width=0.95\linewidth]{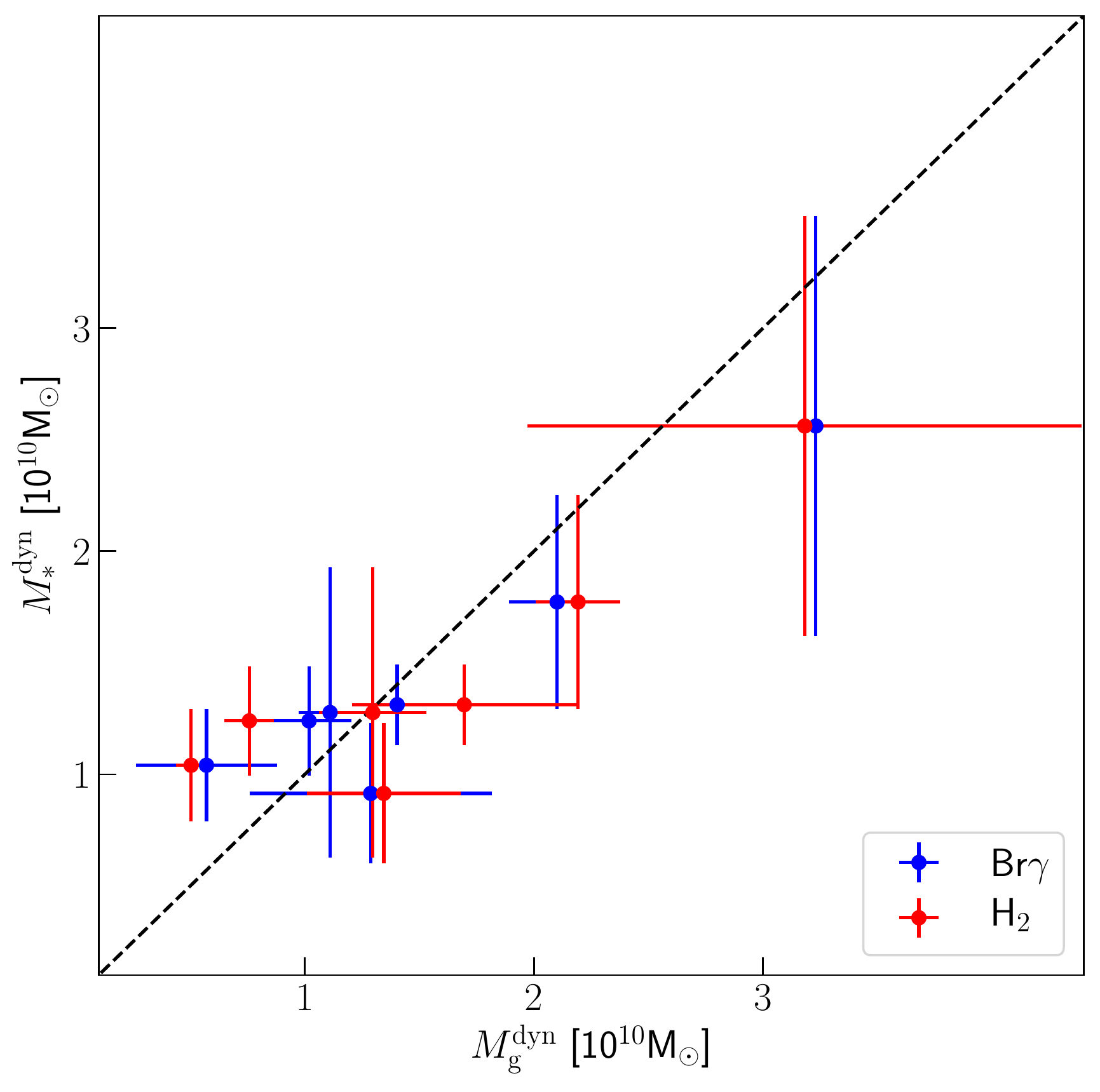}
    \caption{Dynamical mass. Comparison of dynamical masses estimated using the stellar and gas (ionised and hot molecular in blue and red, respectively) kinematics. The error bars account for the propagated uncertainties, in which the velocity uncertainties do not include the systematic errors.}
    \label{fig:Mdyn}
\end{figure}

\begin{figure}
\centering
    \includegraphics[width=0.9\linewidth]{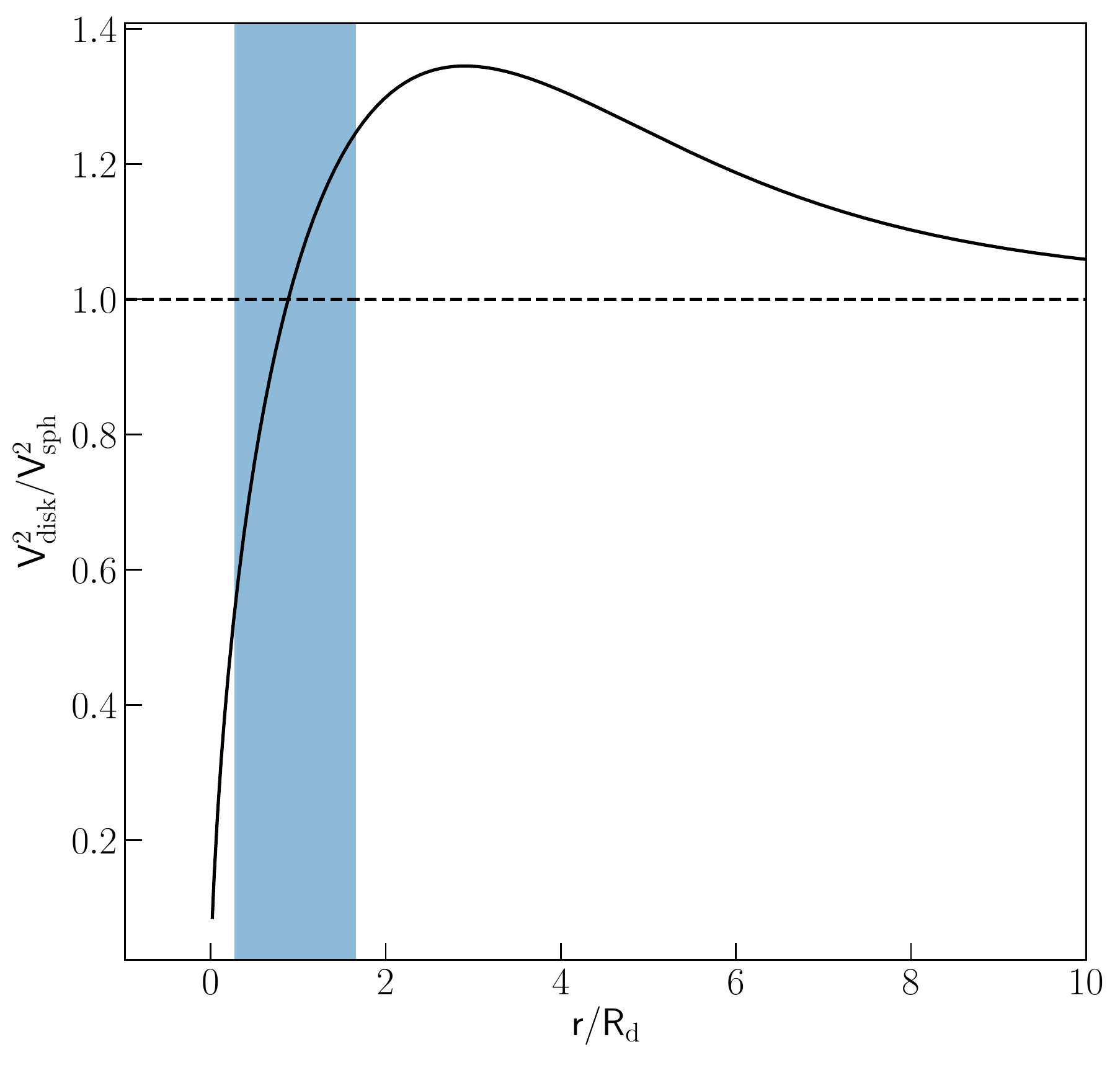}
    \caption{Circular velocity comparison. Ratio between the circular velocities expected assuming a mass distribution following an exponential disc and its equivalent-mass spherical model as a function of the scale length. The blue shadow represents the area covered by the reference radius ($R_\mathrm{r}$) from our sample.}
    \label{fig:Vdisc}
\end{figure}

In the previous section, we studied the velocity and velocity dispersion differences produced when using different tracers during the kinematic extraction. Here we examine the impact of these differences on the dynamical mass estimations. 

The limitations of our SINFONI dataset, which are restricted to a small FoV (r<1-2\,kpc) and a S/N that does not allow us to extract meaningful $h_3$ and $h_4$ Gauss-Hermite moment maps, along with the complex light structure in the innermost regions of some targets, prevent us from creating detailed kinematic models to determine the dynamical mass. However, we can still perform a rotation-curve-based analysis as a first-order approximation, because the galaxies of the sample are clearly rotation-dominated. This approximation allow us to study the `relative differences' when estimating the dynamical mass with the stellar and gas kinematics.

Assuming that the source of the gravitational potential is spherically distributed, we can estimate the dynamical mass within 1$R_\mathrm{r}$ as:
\begin{equation}
M_{\mathrm{dyn}}(R_\mathrm{r})= \frac{R_\mathrm{r}V_{\mathrm{c}}^2(R_r)}{G}
\label{eq:mdyn}
,\end{equation}
where $G$ is the gravitational constant and $V_\mathrm{c}$ is the circular velocity at $R_\mathrm{r}$. This $V_\mathrm{c}$ should account for the asymmetric drift produced by pressure forces and quantified by the velocity dispersion. To perform this asymmetric drift correction, we used the equation derived in \citet{Binney&Tremaine+87}:
\begin{equation}
\label{eq:Vc_BT}
V_\mathrm{c}^2=V_\mathrm{\phi}^2-\sigma_R^2 \left[ \frac{\partial\ln \Sigma_R}{\partial\ln r}+\frac{\partial\ln \sigma_R^2}{\partial\ln r}  + 1 -\frac{\sigma_\phi^2}{\sigma_R^2} + \frac{r}{\sigma_R^2} \frac{\partial v_R v_z}{\partial z}\right]
,\end{equation}

where $\Sigma_R$ and $\sigma_R$ are the radial profile of the stellar mass surface density and the radial component of the velocity dispersion, respectively. Assuming a cylindrical coordinate system aligned with the stellar velocity ellipsoid (i.e. SVE) and an isotropic velocity dispersion ($\sigma_\phi$=$\sigma_R$=$\sigma$), we can neglect the last three terms in Eq.~\ref{eq:Vc_BT}:

\begin{equation}
V_\mathrm{c}^2=V_\mathrm{rot}^2-\sigma^2 \left[ \frac{\partial\ln \Sigma_R}{\partial\ln r}+\frac{\partial\ln \sigma^2}{\partial\ln r} \right] \equiv V_\mathrm{rot}^2+C_\mathrm{AD}\sigma^2
\label{eq:C_AD}
.\end{equation}

Therefore, we assumed this formula to estimate the dynamical masses based on the stellar and gas kinematics (Table~\ref{tab:Mdyn}) using the inclination-corrected velocity amplitude $V_{\mathrm{rot}}$ and the integrated $\sigma$ listed in Table~\ref{tab:dynamics_gas}. In addition, we used the IRAC/3.6$\mu m$ images to derive the stellar mass surface density, which in turn we used, along with the $\sigma$ radial profiles from Fig.~\ref{fig:gas_kin}, to compute the values of $C_\mathrm{AD}$ at $R_\mathrm{r}$.

We analysed the contribution of both terms to the $C_\mathrm{AD}$ factor. As the IRAC/3.6$\micron$ PSF is significantly smaller than $R_\mathrm{r}$ for all the galaxies, it is possible to resolve the photometric structure at these scales. At $R_\mathrm{r}$, the $\Sigma_R$ variations are then due to the different morphology of each object, and the contributions to the $C_\mathrm{AD}$ factor follow differences in the photometric structure of each galaxy. Therefore, we measured individual values for this first term of the $C_\mathrm{AD}$ factor for each galaxy. For the second term, we analysed the $\sigma$ contribution to the $C_\mathrm{AD}$ factor from the kinematic data. In this term, the $R_\mathrm{r}$ radius coincides with the outskirts of the SINFONI FoVs and the contribution of this term is dominated by the scatter produced by the noise of the maps. As the $\sigma$ radial profiles are almost constant in all the galaxies, and in order to reduce the uncertainties produced by the scatter, we considered an average value for all galaxies in the sample (i.e. 0.01$\pm$0.67). The values of $C_\mathrm{AD}$ along with their associated uncertainties are presented in Table~\ref{tab:Mdyn}.

Figure~\ref{fig:Mdyn} displays the individual $M_\mathrm{dyn}$ obtained from the different phases.
For most of the sample, the dynamical masses estimated from the gas kinematics are in good agreement with those extracted from the stellar phase, with mean $M_\mathrm{dyn}^\mathrm{gas}/M_\mathrm{dyn}^\mathrm{stellar}$=1.0$\pm$0.3 and 1.1$\pm$0.3 for the ionised and molecular gas phases, respectively. Despite finding smaller stellar velocity amplitudes (by a factor of $\sim$0.8) than the gas ones, the correction of the asymmetric drift counteracts these differences, yielding similar circular velocities.

In Equation~\ref{eq:mdyn}, we assume that the source of the gravitational potential is spherically distributed. However, this approximation may not be adequate in general, as we are analysing the innermost regions of rotation-dominated galaxy discs. For this reason, we analysed the differences in $V_\mathrm{c}^2$ expected for an exponential disc and a spherically symmetric model with the same total mass. This difference is represented in Fig.~\ref{fig:Vdisc} as a function of the radial distance (normalised by the scale length). These scale lengths along with their associated uncertainties, presented in Table~\ref{tab:Reff}, were obtained by fitting exponential disc models to the IRAC surface brightness radial profiles. At our reference radii, these differences are, on average, $V_\mathrm{disc}/V_\mathrm{sph}$=0.80$\pm$0.21 with values ranging from 0.5 up to 1.3. These typical errors of 20\% are within the uncertainties calculated for the dynamical masses, suggesting that the approximation of a spherical distribution of the source of the gravitational potential is still valid at these scales.

In this section we present our estimations of the dynamical masses within $R_{\mathrm{r}}$ (i.e. $\sim$1-2\,kpc) as our kinematics results do not cover wider FoVs. At these distances, the dark-matter contributions of local LIRGs and z$\sim$2 SFGs, which are considered as the high-z counterpart of the local U/LIRG population, should be small \citep{Courteau+15,Lee+17,Genzel+17,Ubler+18}. Therefore, we can assume that our $M_\mathrm{dyn}$ values are dominated by the stellar mass. This assumption is conceptually similar to the `maximum disc' hypothesis, in which the luminous disc is responsible for the totality of the circular velocity in the inner region of the velocity curve \citep{vanAlbada+86}.

In this context, we used the IRAC/3.6$\mu m$ images to estimate non-dynamical measurements of the stellar mass at $R_\mathrm{r}$ and test this approximation. We used the stellar masses and integrated fluxes derived by \citet{Pereira-Santaella+15} and \citet{U+12} to calculate constant mass-to-light (M/L) ratios, which were then used to estimate the stellar mass within $R_\mathrm{r}$. In these latter works, the stellar masses for two samples of LIRGs were computed via spectral energy distribution (SED) fitting assuming Kroupa \citep{Kroupa+01} and Chabrier \citep{Chabrier+03} initial mass functions (IMFs), respectively. At $R_\mathrm{r}$, we obtained M/L$_\mathrm{3.6\mu m}$ values of 0.43$\pm$0.11 and 0.14$\pm$0.05 for \citet{Pereira-Santaella+15} and \citet{U+12}, respectively. These values yield stellar mass estimations that are compatible with our results, with typical values ranging from 7$\times$10$^9$\,$\mathrm{M_\odot}$ to 4$\times$10$^{10}$\,$\mathrm{M_\odot}$ (see Table~\ref{tab:Mdyn}). However, the results from \citet{U+12} should be formally preferred as the dynamical mass cannot be smaller than the stellar one. It should be noted that our stellar mass estimations are obtained assuming a constant M/L, which was derived using integrated fluxes from the whole galaxy. In the nuclear regions, the M/L might differ because of the presence of older stellar populations and larger extinction values. In addition, there are large uncertainties on the total stellar masses obtained in these previous works which are attributable to the different techniques used to fit the SED, the choice of IMF, and the star formation history (SFH), metallicities, and extinctions assumed during the fitting. These effects as well as the assumptions on the PAHs and/or hot dust emission may have an effect on the M/L differences.

In summary, we find that the differences between the stellar and the gas kinematics do not strongly affect the dynamical mass estimation, providing values that are in good agreement. These dynamical masses can be assumed to be reliable estimations of the stellar mass, as at $R_\mathrm{r}$ (i.e.$\sim$1-2\,kpc) the dark matter halo and the gas component should have a minor contribution.

\subsection{Local LIRGs as counterparts of SFGs at z$\sim$2}
\label{subsub:high-z}

High-z studies have shown that the tight correlation between the stellar mass and the star-formation rate in SFGs, defined as the star-forming `main sequence' (i.e. SFMS; \citealt{Rodighiero+11}), can be found up to z$\sim$6 \citep[see][and references therein]{Speagle+14}. In general, low-z LIRGs build up their stellar mass at rates that are 10-100 times higher than the SFMS at z$\sim$0 but at similar rates to those MS galaxies at z$\sim$1-2, with SFR$\sim$20–200\,M$_\mathrm{\odot}$\,yr$^{-1}$ and specific SFR (i.e. sSFR=SFR/M$_\mathrm{*}$) of $\sim$1–2\,Gyr$^{-1}$ \citep{Speagle+14}. Moreover, kinematic IFS studies of z>1 SFGs demonstrate that most of the main sequence galaxies are large rotating discs \citep{Forster-Schreiber+09,Forster-Schreiber+18,Wisnioski+19}, a characteristic also observed in local LIRGs \citep{Bellocchi+13}, which is one of the reasons why LIRGs are usually considered to be the local counterpart of the star-forming population at z>1.

Although a detailed comparison of local and high-z samples is always difficult because of the different observational conditions (e.g. spatial resolution, sensitivity, rest-frame wavelength range), the recent AO-assisted SINFONI study by \citet{Forster-Schreiber+18} offers a possibility to contrast the main structural and kinematic properties of their sample of 36 SFGs at z$\sim$2 with the present sample of nearby LIRGs. However, we note that despite the large angular resolution achieved on the AO-assisted data (i.e. mean FWHM=0.19$\arcsec$), there are still significant differences in linear resolution between both samples ($\sim$200\,pc for the nearby LIRGs and $\sim$1.5\,kpc for SINS/zC-SINF SFGs at z$\sim$2). In addition, the kinematic analysis of the SF z$\sim$2 galaxies is extended up to $\sim$10\,kpc, whereas our data only cover the inner $\sim$1.5\,kpc. These differences preclude a direct comparison between our local LIRGs and the z$\sim$2 SFGs, and therefore our goal is simply to contrast at first order the main structural and kinematic properties of both samples.

The fundamental structural properties of both populations show good agreement, with typical sizes and masses obtained for the LIRGs (i.e. $R_\mathrm{eff}$\,$\sim$1-5\,kpc and $M_\mathrm{dyn}$>4$\times$10$^{9}-3\times$10$^{10}$\,M$_\mathrm{\odot}$) within the range of values found in the SFGs at z$\sim$2 (i.e. $R_\mathrm{eff}$\,$\sim$1-9\,kpc and $M_\mathrm{dyn}$\,$\sim$5$\times$10$^{9}-3\times$10$^{11}$\,M$_\mathrm{\odot}$). The local LIRGs display SFRs ($\sim$20-80\,M$_\mathrm{\odot}$\,yr$^{-1}$, \citealt{Rodriguez-Zaurin+11}) also within the range of values for the z$\sim$2 SFGs (i.e. SFR$\sim10-150$\,M$_\mathrm{\odot}$\,yr$^{-1}$), which includes 14 galaxies with SFRs in the range of the ULIRGs (i.e. SFR>100). 
Kinematically, \citet{Forster-Schreiber+18} obtained integrated (r<10\,kpc) $\sigma$ values of H$\alpha$ ranging between $\sim$70\,km\,s$^{-1}$ and $\sim$200\,km\,s$^{-1}$. Overall, we find good agreement between these values and our aperture-smearing uncorrected $\sigma$ values from Br$\gamma$ (i.e. $\sim$80$-$180\,km\,s$^{-1}$; see Fig.~\ref{fig:beam_smearing}), although our integrated values were obtained at smaller apertures (i.e. r<1.5\,kpc, see Sect.~\ref{subsec:starsvsgas}). Further, the aperture-smearing corrected $\sigma$ values extracted for the LIRGs (i.e. 28-56\,km\,s$^{-1}$, see Table~\ref{tab:dynamics_gas}) are also well within the range of the intrinsic $\sigma$ for the SFGs at z$\sim$2 (i.e.$\sim$25-70\,km\,s$^{-1}$), which are assumed to be constant across the galaxy and therefore not influenced by the difference in the aperture size. The possible impact of a larger extinction on the optical wavelengths is discussed in Sect.~\ref{subsec:optvsnir}, yielding small differences in $\sigma$ (i.e. <15\,km\,s$^{-1}$) that might have a minor impact on this comparison.

The agreement in the uncorrected velocity dispersion can be understood if the central regions dominate the light-weighted integrated measurements of $\sigma$, minimising the impact of not covering the outskirts of the galaxy. Dynamically, the nearby LIRGs studied in this work are rotationally supported systems (with $V/\sigma$\,$\sim$3-8), in agreement with the results of the SFG sample (i.e. $V/\sigma$\,$\sim$1-10, whereas $\sim$70\% of them have $V/\sigma$\,$\gtrsim$2). 

In summary, the first-order comparison between our nearby LIRGs and the z$\sim$2 SFG sample \citep{Forster-Schreiber+18} reveals that both populations present similar structural and dynamic properties, supporting the commonly adopted assumption that local LIRGs are counterparts of the MS star-formation galaxies at z>1. These findings suggest that, following our results in Sect.~\ref{subsec:Mdyn}, nebular lines are good tracers of dynamical mass in SFGs at z$\sim$2 and in rotation-dominated galaxies at high-z in general. To date, evidence of rotation-dominated systems has been found until z$\sim$7 \citep{Smit+18}. This result will be useful when the next generation of near- and mid-IR IFS (e.g. HARMONI/ELT, NIRSpec/JWST, and MIRI/JWST) multiplies the number of studies of z>2 galaxies. Nevertheless, obtaining enough S/N at the continuum to extract the stellar kinematics will still be extremely time-consuming.

\section{Summary and conclusions}
\label{sec:conclusions}

In this work we use SINFONI IFS data to study the stellar kinematics from a sample of nearby LIRGs and compare it with their gas kinematics. The sample comprises ten local (mean z=0.014) LIRGs within a factor of two in distance (i.e. 40-80\,Mpc). In general, these LIRGs display low IR luminosities (i.e. 9 out of the 10 are 11.1<log($L_\mathrm{IR}/\mathrm{L_\odot}$)<11.5) and different interaction stages (5 isolated, 3 interacting, 1 ongoing merger, and 1 post-coalescence object) and spectroscopic types (i.e. 2 Seyfert, 6 \ion{H}{II,} and 2 `composite' objects). We extracted the stellar velocity and velocity dispersion maps in both H- and K-bands from the near-IR stellar continuum and compared them with the kinematics of the different gas phases (ionised, partially ionised, and hot molecular) extracted by \citet{PiquerasLopez+12} from the same dataset. Our main results can be summarised as follows.

\begin{enumerate}

    \item We do not observe significant differences in the kinematic maps extracted from the two near-IR bands (SINFONI H- and K-band). However, the larger number of absorption bands in the H-band along with the wider wavelength range grant higher S/N velocity dispersion distributions and smaller uncertainties than the K-band.
        
    \item Most of the galaxies from the sample, nine out of ten, have stellar velocity distributions that can be modelled as rotating discs. Only the kinematics of NGC\,3256, as result of its ongoing interaction, shows a complex behaviour that cannot be represented by a rotating disc. In general, the velocity dispersion shows flat or slightly-decreasing radial profiles, compatible with values found for late-type galaxies.  
    
    \item Our comparison of parameters derived from photometry (extracted from HST/NICMOS images) and kinematics shows that they provide similar results, with differences of <20$\degr$ for the PA and <20\%, in general, for sin($i$).
   
    \item We modelled the ionised and hot molecular gas velocity maps as a rotating disc for eight out of the ten galaxies. Our comparison of the PA and inclination derived from the stellar and gas phase models shows that there is good alignment between these phases. We also find good agreement between the inclination-corrected velocity profiles of the stellar and gas phases in most of the sample (i.e. 8 out of 10 objects).
    
    \item We find that, while the different gas phases studied in this work (ionised, partially ionised, hot molecular) present similar kinematics, the stellar component shows a different behaviour. In particular, we observe that the stellar component rotates slower than the gas phases (i.e. median $V_{\mathrm{*}}/V_{\mathrm{g}}$=0.85$\pm$0.13 and 0.86$\pm$0.16, for the ionised and hot molecular phases, respectively) and presents larger velocity dispersion values (i.e. median $\sigma_{\mathrm{*}}/\sigma_{\mathrm{g}}$=2.61$\pm$0.76 and 2.25$\pm$0.53, for the ionised and hot molecular phases, respectively). 
    
    \item The radial profiles of the dynamical ratio $V/\sigma$ show values compatible with rotation-supported systems (i.e. $V/\sigma$>1) at the inner $\sim$0.5\,kpc for the gas and stellar components. At the outskirts of the FoV, the gas phases reach values of $V/\sigma$\,$\sim$10, whereas the stellar ratio does not increase significantly. These results indicate that the gas is rotating in cooler (i.e. thinner) discs, whereas the stars populate warmer discs. This dynamical structure could be explained by the dissipative nature of the gas, which favours its cooling. The gas therefore experiences less asymmetric drift (AD) but the stars migrate to more disturbed orbits.
    
    \item The differences found between the $V$ and $\sigma$ values of the stellar and gas phases do not significantly affect the dynamical mass estimations in galaxies that can be modelled as rotating discs in both phases (i.e. 7 out of 10). On average, the differences between the values obtained from the gas and stellar components are $M_\mathrm{dyn}^\mathrm{gas}/M_\mathrm{dyn}^\mathrm{stellar}$=1.0$\pm$0.3 and 1.1$\pm$0.3 for the ionised and molecular gas phases, respectively.
   
    \item The comparison of our local LIRGs with the sample of MS SFGs at z$\sim$2 from \citet{Forster-Schreiber+18}, despite the different spatial scales ($\sim$1.5\,kpc and $\sim$10\,kpc, respectively) and resolutions ($\sim$0.2\,kpc and $\sim$1.5\,kpc, respectively), reveals good agreement between different structural and kinematic parameters (i.e. $R_\mathrm{eff}$, $M_\mathrm{dyn}$, SFRs, $\sigma,$ and $V/\sigma$), supporting the assumption that nearby LIRGs could be considered as local counterparts of the SF MS population at z>1. This agreement suggests that, at first order, it could be possible to use nebular lines to estimate the dynamical masses of rotating galaxies at these redshifts, where extracting the stellar continuum would be extremely time-consuming.
    
   \end{enumerate}

The H- and K-band SINFONI observations analysed in this work provide new insights into the nuclear stellar kinematics of local LIRGs. As we show above, these objects present rotation-dominated stellar discs that are globally insensitive to the presence of outflowing material expelled by AGNs or starburst events. The comparison with the gas phases reveals that these stellar discs, though aligned with the gaseous discs, are dynamically cooler. We find that the small differences in the kinematics produced by being in different dynamical states do not significantly affect the dynamical mass estimation, and support the use of nebular lines to probe the kinematics of SFGs at high-z.

\begin{acknowledgements}
    
We thank the anonymous referee for his/her useful suggestions and comments that helped to improve the final content of this work. ACG, SA, LC and BRP acknowledge support from the Spanish Ministerio de Econom\'ia y Competitividad through the grants BES-2016-078214, ESP2015-68964-P, ESP2017-83197 and PID2019-106280GB-I00. JPL acknowledges support from the Spanish Ministerio de Econom\'ia y Competitividad through grant AYA2017-85170-R. MPS acknowledges support from the Comunidad de Madrid through the Atracci\'on de Talento Investigador Grant 2018-T1/TIC-11035 and PID2019-105423GA-I00 (MCIU/AEI/FEDER,UE). This work is based on observations collected at the European Organisation for Astronomical Research in the Southern Hemisphere, Chile, programmes 077.B-0151A, 078.B-0066A, 081.B-0042A and 103.B-0867A. This paper also makes use of the following ALMA data: ADS/JAO.ALMA\#2013.1.00243.S, ADS/JAO.ALMA\#2013.1.00271.S, ADS/JAO.ALMA\#2017.1.00255.S. ALMA is a partnership of ESO (representing its member states), NSF (USA) and NINS (Japan), together with NRC (Canada) and NSC and ASIAA (Taiwan), in cooperation with the Republic of Chile. The Joint ALMA Observatory is operated by ESO, AUI/NRAO and NAOJ. This work made use of Astropy, a community-developed
core Python package for Astronomy \citep{Astropy+18}. This research has made use of the NASA/IPAC Extragalactic Database (NED) which is operated by the Jet Propulsion Laboratory, California Institute of Technology, under contract with the National Aeronautics and Space Administration. Some of the data presented in this paper were obtained from the Mikulski Archive for Space Telescopes (MAST). STScI is operated by the Association of Universities for Research in Astronomy, Inc., under NASA contract NAS5-26555. 

\end{acknowledgements}

\bibliographystyle{aa} 
\bibliography{biblio_paper.bib} 

\appendix

\section{Individual sources}
\label{append:sources_notes}

\begin{figure*}
\includegraphics[width=\linewidth]{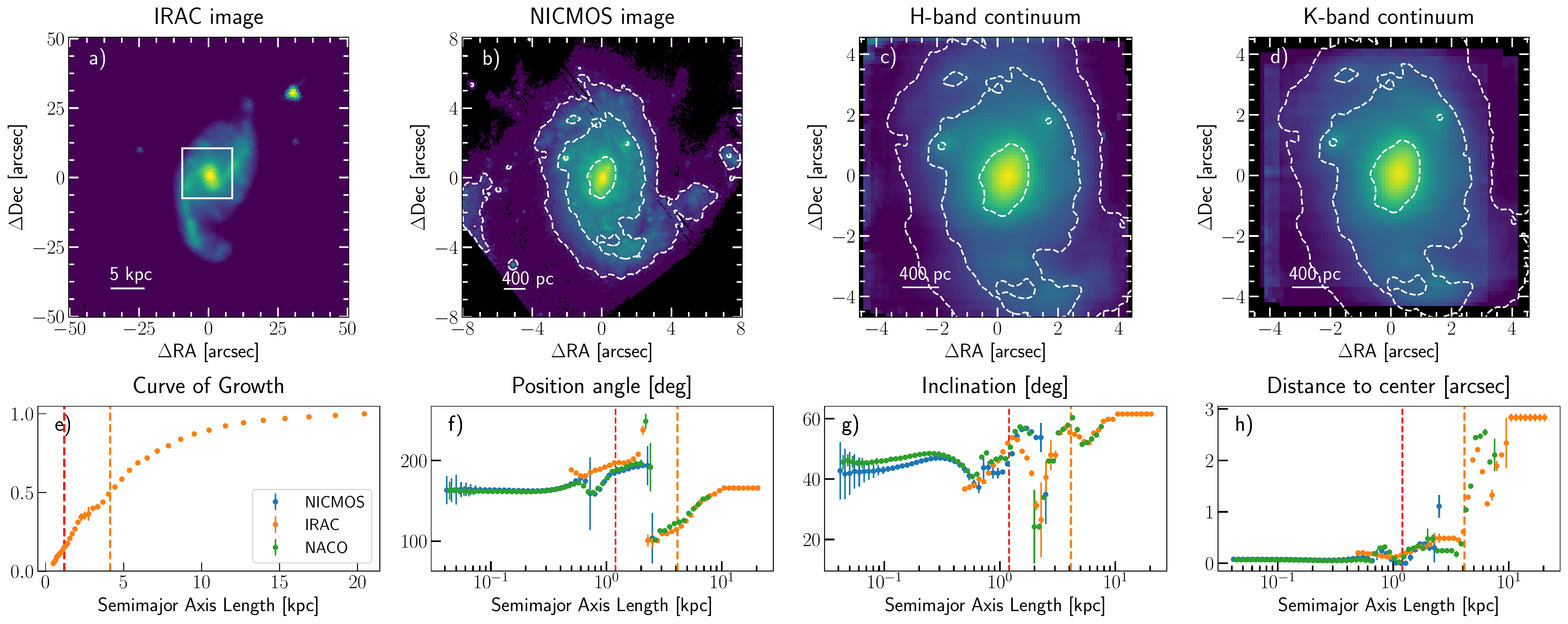} 
\caption{As in Fig.~\ref{fig:example_morph} but for NGC\,3110.}
\label{fig:morph_3110}
\end{figure*}

\begin{figure*}
\includegraphics[width=\linewidth]{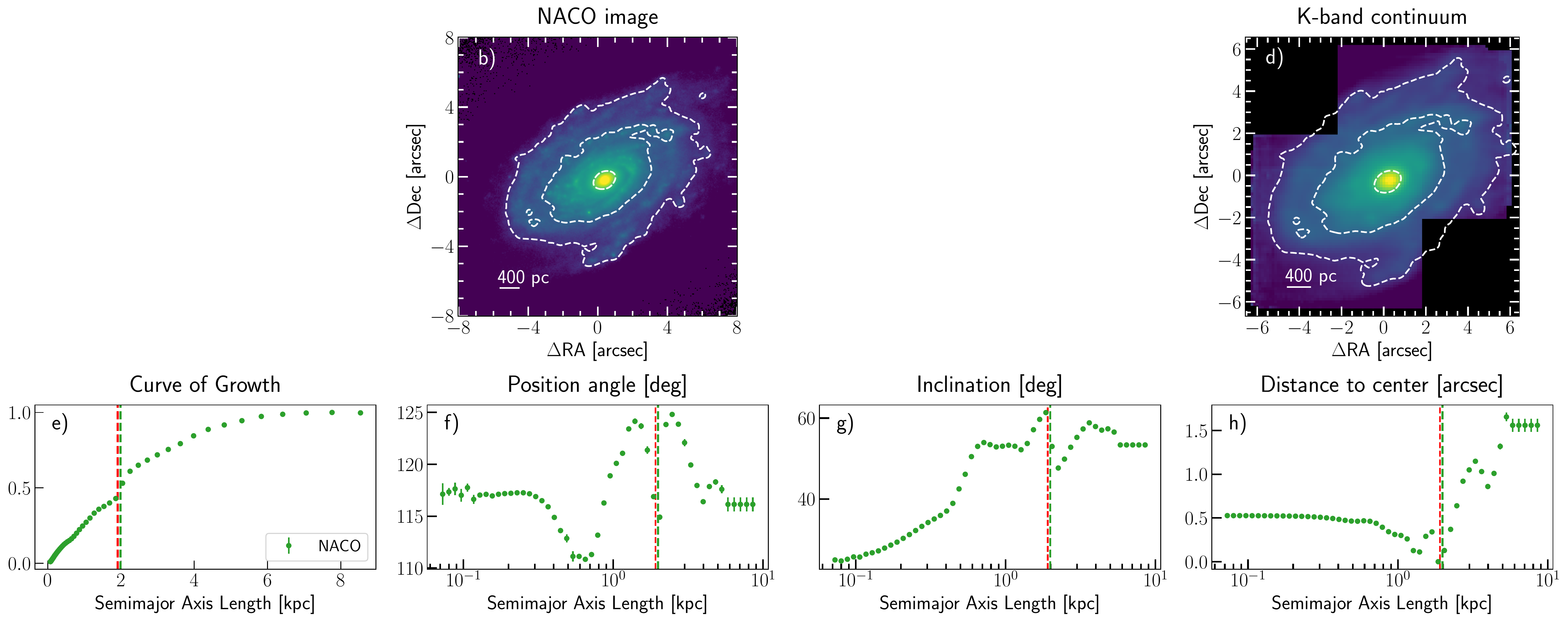} 
\caption{As in Fig.~\ref{fig:example_morph} but for IRAS\,F12115-4656. Panel `b' corresponds to the NACO image.}
\label{fig:morph_12115}
\end{figure*}

\begin{enumerate}

     \item \textbf{NGC\,2369}: This highly inclined SB(s)a galaxy presents a complex morphology with bright clumps dominating the nuclear regions along with faint arms at larger scales. Similarly to IRAS\,F17138-1017, this galaxy shows aligned ($\sim$170\degr) star-forming regions at its inner-most regions that complicate the measurement of the photometric parameters. Moreover, NGC\,2369 also presents an intense extinction filament along the star-forming regions. During the kinematic analysis, the rotating-disc modelling of the velocity maps revealed highly structured residuals (i.e. >40\,km\,s$^{-1}$). A double-Gaussian profile fit of the Br$\gamma$ line (see Sect.~\ref{subsec:gas_kin}) was performed to shed some light on the kinematics of this target. The proposed scenario (see Appendix~\ref{append:2369}) to explain the kinematics is that the light along the line of sight (LOS) comes from the different arms of the galaxy, producing two distinct kinematic components. In addition, the intense dust extinction in the innermost regions and the high inclination, precludes the detection of the emission line at the centre but produces an excess in flux spatially coinciding with local peaks of receding and approaching velocities. 

    \item \textbf{NGC\,3110}: This SB(rs)b galaxy is interacting with a minor companion located at $\sim$40\,kpc SW \citep{Yuan+10,Larson+16} and presents two bright spiral arms protruding from an elongated bright nucleus. The faint transition between the nucleus and the spiral arms yields abrupt changes in the inclination and PA determination due to loss of S/N. At larger scales, the PA and inclination show a smooth profile. As IC\,5179, this galaxy has differences between the brightness of both arms that bias the centre determination at r>5\,kpc (see Fig.~\ref{fig:morph_3110}). The kinematic analysis reveals a rotating-disc pattern orientated with PA$\sim$0\degr, aligned with the elongated bulge. In addition, the velocity curve shows a steeper profile at r<0.5\,kpc, consistent with a pseudo-bulge. We observe velocity dispersion drops located in two blobs at $\sim$0.4\,kpc NW and NE, values expected in SF regions ($\sim$60\,km\,s$^{-1}$).

    \item \textbf{NGC\,3256}: This object is a gas-rich merger where the secondary nucleus is located at $\sim$1\,kpc S. Gas outflows in both nuclei had been previously reported \citep[see][and references therein]{Emonts+14}. Despite the fact that the NICMOS image reveals a main spherical nucleus (i.e. north nucleus), the abundant, bright, spatially resolved structures of the inner regions complicate the photometric analysis within r<1\,kpc. From analysis of the outer parts, the interacting system seems to be slightly elongated (i.e. $\sim$45\degr) with PA$\sim$60$\degr$. The SINFONI data do not cover the same spatial regions in both bands, precluding the kinematic analysis. The H-band velocity map shows the rotation pattern of the main nucleus with similar orientation to the photometric PA (i.e. $\sim$80\degr) while the K-band reveals a different velocity component, likely related with the secondary nucleus. The H-band velocity curve revealed a steeper curve at r<0.2\,kpc, spatially coincident with the main nucleus. Its velocity dispersion map shows a noisy distribution, with a great diminution at $\sim$1.5\,kpc NE of the main nucleus. The fact that the SINFONI H-band FoV does not cover the secondary nucleus---because of the pointing strategy adopted--- might explain the large differences between the kinematically and photometrically derived parameters.
    
    \item \textbf{ESO\,320-G030}: This SAB(r)a galaxy has been spectroscopically classified as \ion{H}{II} \citep{PereiraSantaella+11}. Our photometric analysis reveals an inner PA of $\sim$90\degr, being spatially compatible with the presence of the nuclear bar of $\sim$1\,kpc in length discovered by \citet{Greusard+00}. The kinematic analysis reveals a radial component in the velocity map which is likely due to the presence of non-circular motions, with a similar orientation ($\phi$\,$\sim$85\degr) to the photometric PA found in the inner regions. The velocity dispersion map presents a clear centrally peaked distribution, reaching values of $\sim$130\,km\,s$^{-1}$ at the central $\sim$300\,pc.
    
    \item \textbf{IRAS\,F12115-4656}: This galaxy, classified as SA(rs)b \citep{Buta+95}, has a companion at a distance of $\sim$100\,kpc SW \citep{Arp&Madore+87}. Nuclear emission of [\ion{Si}{VI}] and [\ion{Ca}{VIII}] was detected in \citet{PiquerasLopez+12}, suggesting AGN activity. Although it was not observed with IRAC, we covered the inner $\sim$10\,kpc with the NACO/Ks image. This galaxy presents a compact spherical nucleus ($\sim$0.3\,kpc), whereas the outer kiloparsecs are dominated by disrupted spiral arms. The asymmetry generated by this arm structure strongly affected the centre determination at large apertures >4\,kpc, see Fig.~\ref{fig:morph_12115}). The stellar velocity map also reflects this structure. We observed small decrements in the velocity spatially correlated with the regions between the arms, likely because of the slight loss of continuum emission, which leads to less restricted fits where the velocity is slightly biased towards the systemic value.

    \item \textbf{NGC\,5135}: This Seyfert 2 galaxy \citep{Phillips+83}, classified as SB(s)ab, presents a nuclear region with intense SF clumps along with its AGN \citep[see][and references therein]{Busko+90,Levenson+04,Bedregal+09,Veron-Cetty+06,Colina+12}, which made the photometric analysis challenging. Although the innermost part displays low inclination values, the wider FoV covered by the imagining data reveals an inclination of $\sim$50\degr. The kinematic analysis reveals a peak in the stellar velocity dispersion map coincident with the AGN position, and a low-value ($\sigma$\,$\sim$50\,km\,s$^{-1}$) arc following the SF regions. In addition, a strong diminution in the velocity dispersion can be observed to the SW, spatially coincident with the presence of an outflow \citep{Colina+05}. The complex photometric structure of the innermost regions along with the small velocity amplitude are most likely contributing to the differences between the photometrically and kinematically derived PA and inclination.

    \item \textbf{IRAS\,F17138-1017}: This target, classified as an obscured starburst galaxy in \citet{Depoy+88}, has been suggested to be a post-coalescence merger \citep{Yuan+10,Bellocchi+16}. This galaxy shows a complex nuclear structure dominated by the presence of aligned bright star-forming regions ($\sim$30\degr). These regions bias the photometric analysis of the NICMOS image, especially the inclination and centre determination within 0.2<r<1\,kpc. The kinematic analysis reveals the presence of a radial component aligned with the SF regions ($\phi$\,$\sim$35\degr). The dispersion maps display a central peak (i.e. $\sim$110\,km\,s$^{-1}$) and regions with lower $\sigma$ (i.e. $\sim$80\,km\,s$^{-1}$) that are spatially correlated with the intense SF regions. 

    \item \textbf{IC\,4687}: This galaxy is part of an interacting group together with IC\,4686 and IC\,4689, which are at $\sim$10\,kpc and $\sim$20\,kpc SW. Its faint arms along with the presence of bright regions in the FoV yield large uncertainties in the estimation of the PA and inclination. Despite this, one can observe a well-defined spherical nucleus ($i$\,$\sim$20\degr, r<0.4\,kpc) and a slightly inclined disc ($i$\,$\sim$40\degr). The velocity dispersion map shows a flat distribution, with lower values along the arm structure likely produced by the SF regions populating the arms. The isovelocity lines in the nuclear region of this galaxy are nearly parallel to the minor axis, yielding a less restricted inclination determination, which is biased towards low values by \texttt{DiskFit}.  

    \item \textbf{NGC\,7130}: This Seyfert 2 galaxy \citep{Veron-Cetty+06} has been classified as a peculiar Sa galaxy \citep{Lu+98}, and hosts a compact circumnuclear starburst \citep{Phillips+83}. The HST image only covers the northern arm, which is the brightest in the near-IR. Thus, the determination of the photometric parameters are strongly biased by the northern region of the galaxy. This effect is clearly shown in the evolution of the centre determination, where the northern arm shifts the photocentre ($\sim$6$\arcsec$) at apertures of $\sim$8\,kpc. The low inclination of this target hinders the kinematic analysis and therefore we were only able to extract meaningful kinematic maps in the innermost region (<0.5\,kpc). The low inclination complicates the extraction of the stellar kinematics, yielding kinematically derived PA and inclinations that differ from the photometric values.
    
    \item \textbf{IC\,5179}: This SA(rs)bc galaxy, classified as starburst \citep{Veilleux+95}, presents intense \ion{H}{II} regions distributed along its spiral arms \citep{Alonso-Herrero+06}. The photometric analysis displayed a bright elongated nucleus of $\sim$0.2\,kpc and two bright inner arms. The bright difference between the SW and NE arms induced an offset of almost 4.5$\arcsec$ at apertures of $\sim$6\,kpc in the centre determination (see Fig.~\ref{fig:morph_5179}). The velocity curve shows a steeper profile in the innermost regions (i.e. r<0.3\,kpc) which is compatible with the presence of a pseudo-bulge.
    
\end{enumerate}

\section{Morphological results}
\label{append:morph}

\begin{figure*}
\includegraphics[width=\linewidth]{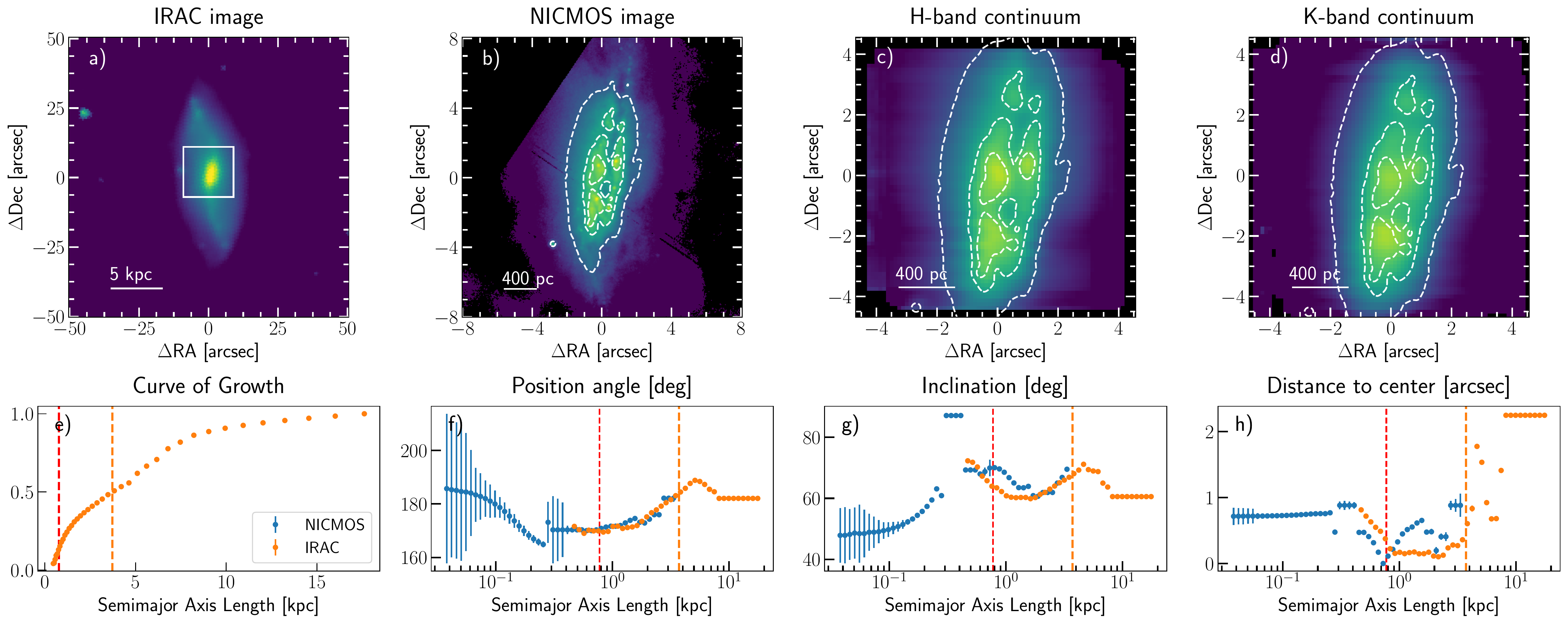} 
\caption{As in Fig.~\ref{fig:example_morph} but for NGC\,2369.}
\label{fig:morph_2369}
\end{figure*}

\begin{figure*}
\includegraphics[width=\linewidth]{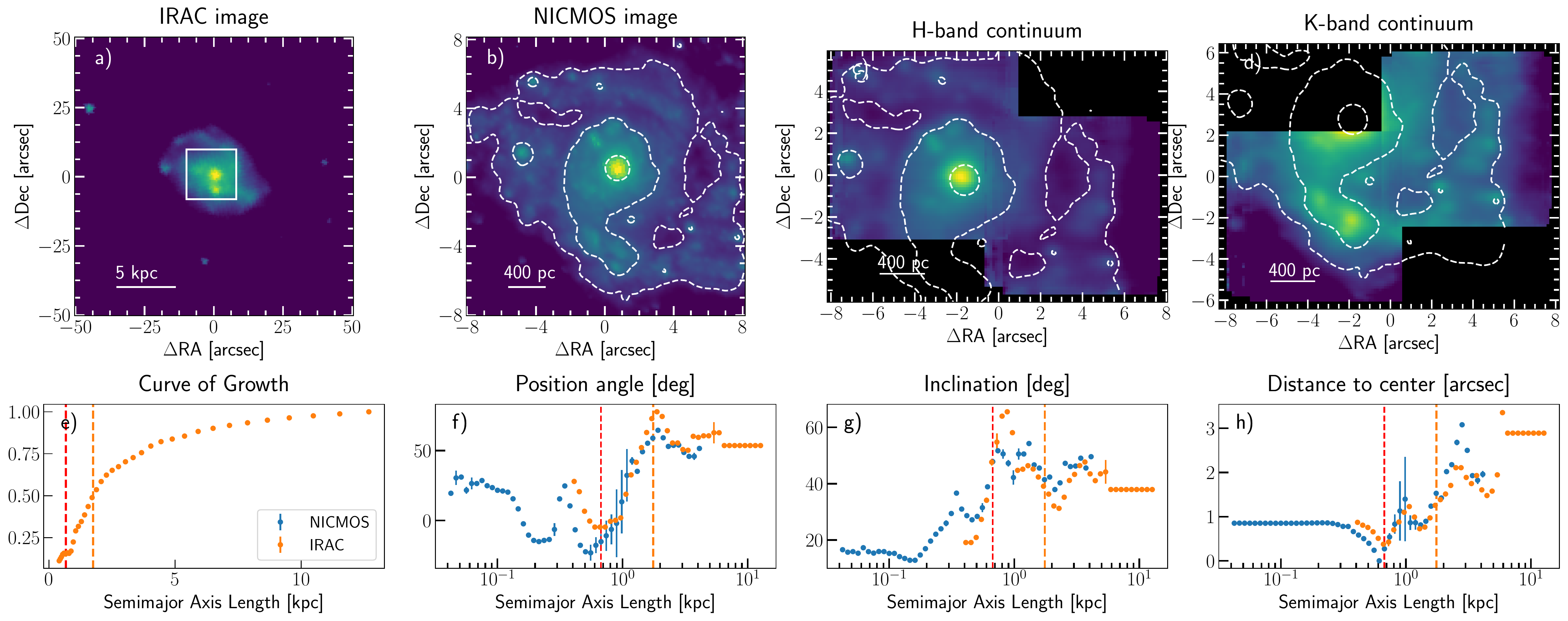} 
\caption{As in Fig.~\ref{fig:example_morph} but for NGC\,3256.}
\label{fig:morph_3256}
\end{figure*}

\begin{figure*}
\includegraphics[width=\linewidth]{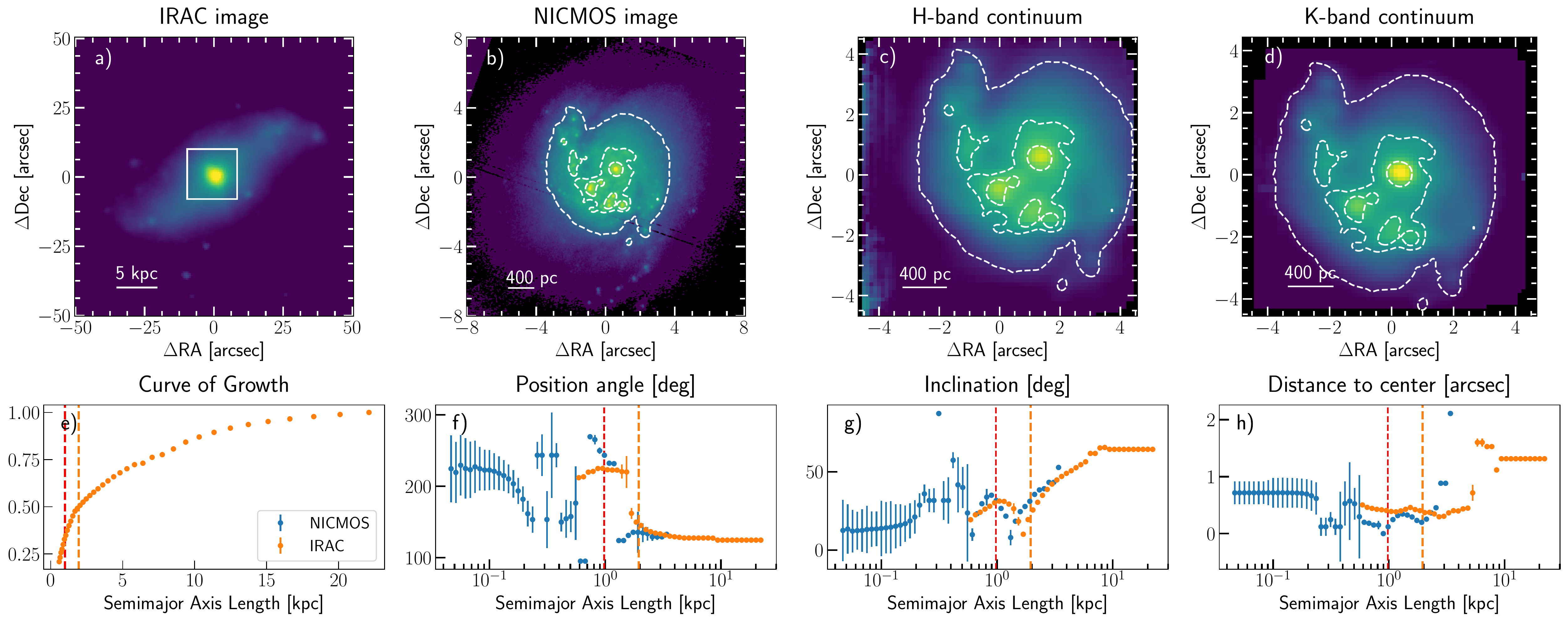} 
\caption{As in Fig.~\ref{fig:example_morph} but for NGC\,5135.}
\label{fig:morph_5135}
\end{figure*}

\begin{figure*}
\includegraphics[width=\linewidth]{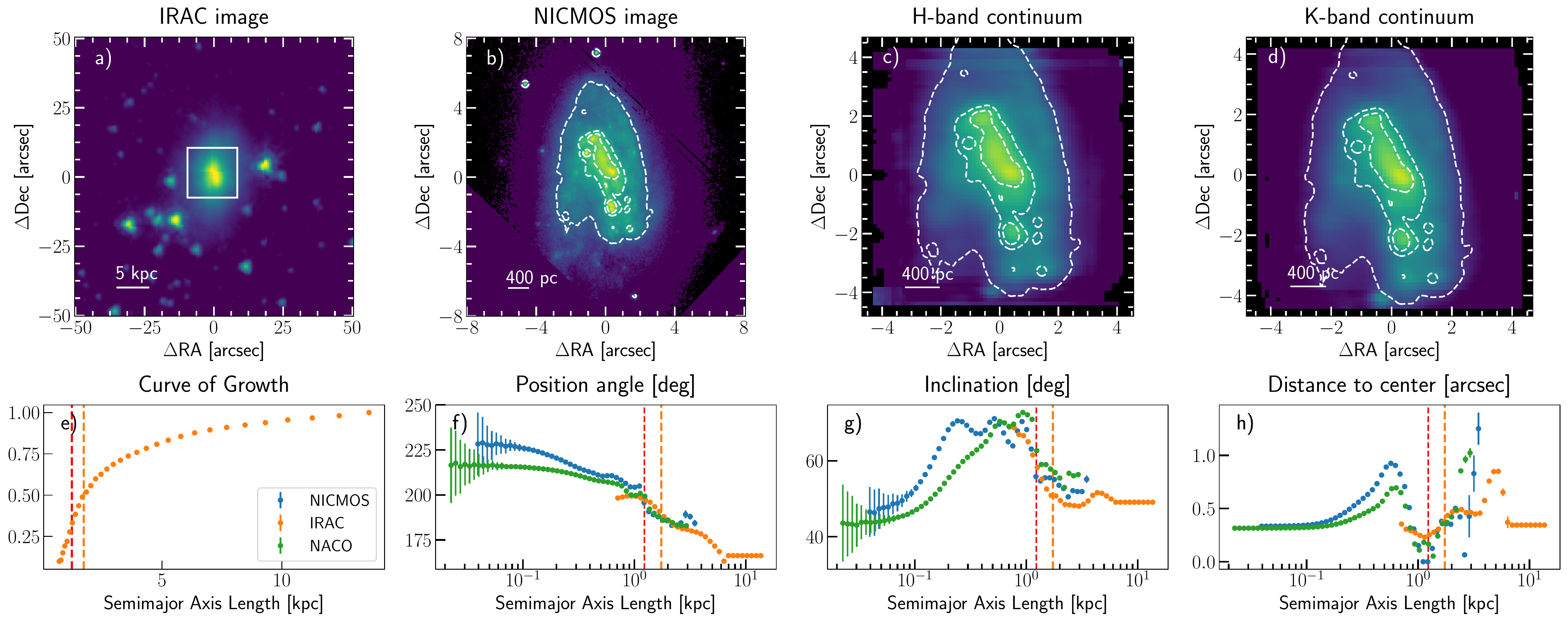} 
\caption{As in Fig.~\ref{fig:example_morph} but for IRAS\,F17138-1017.}
\label{fig:morph_17138}
\end{figure*}

\begin{figure*}
\includegraphics[width=\linewidth]{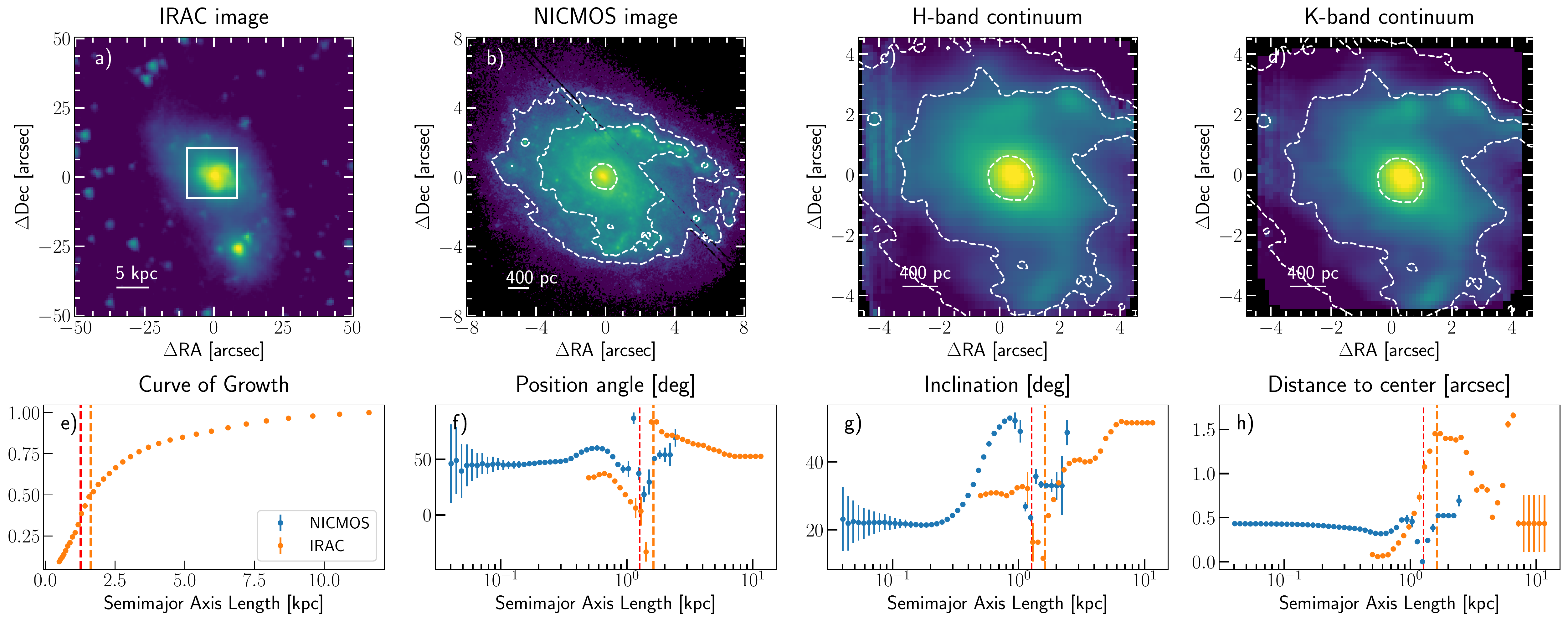} 
\caption{As in Fig.~\ref{fig:example_morph} but for IC\,4687.}
\label{fig:morph_4687}
\end{figure*}

\begin{figure*}
\includegraphics[width=\linewidth]{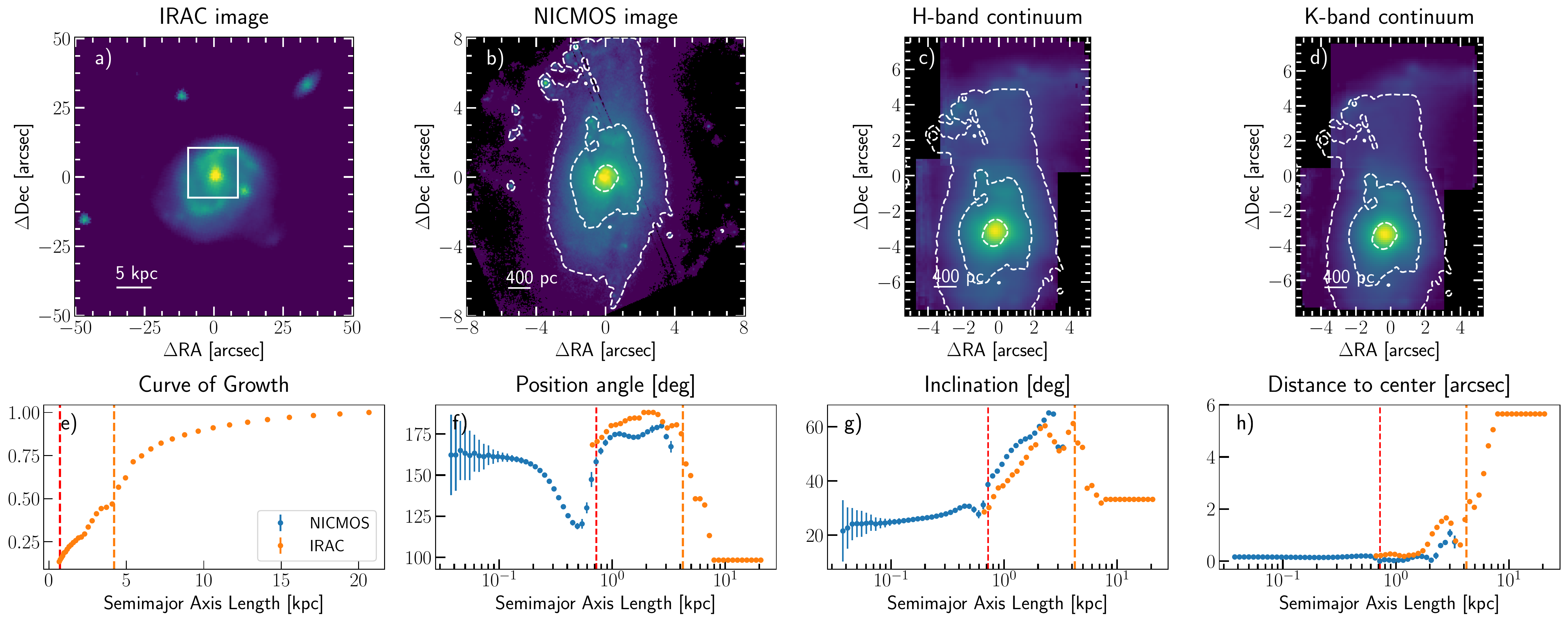} 
\caption{As in Fig.~\ref{fig:example_morph} but for NGC\,7130.}
\label{fig:morph_7130}
\end{figure*}

\begin{figure*}
\includegraphics[width=\linewidth]{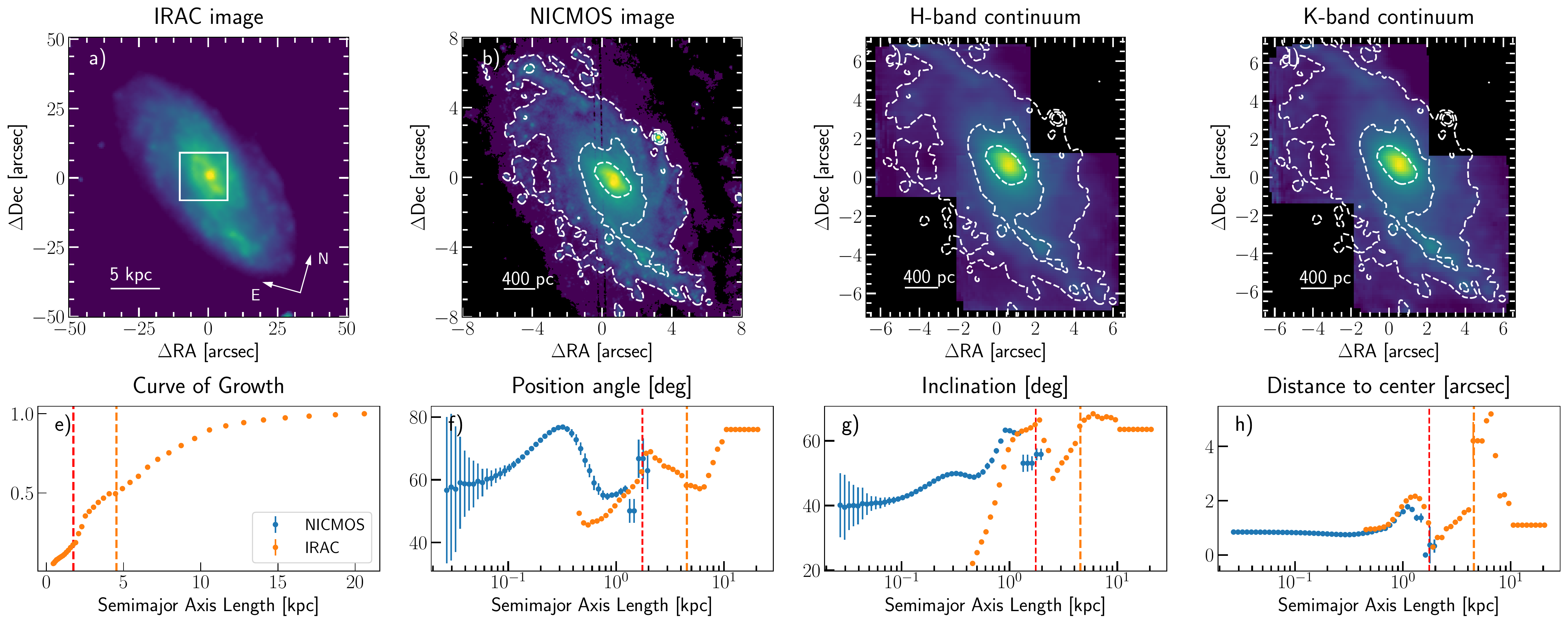} 
\caption{As in Fig.~\ref{fig:example_morph} but for IC\,5179.}
\label{fig:morph_5179}
\end{figure*}

\section{Kinematic results}
\label{append:kin}

\begin{figure*}
\includegraphics[width=\linewidth]{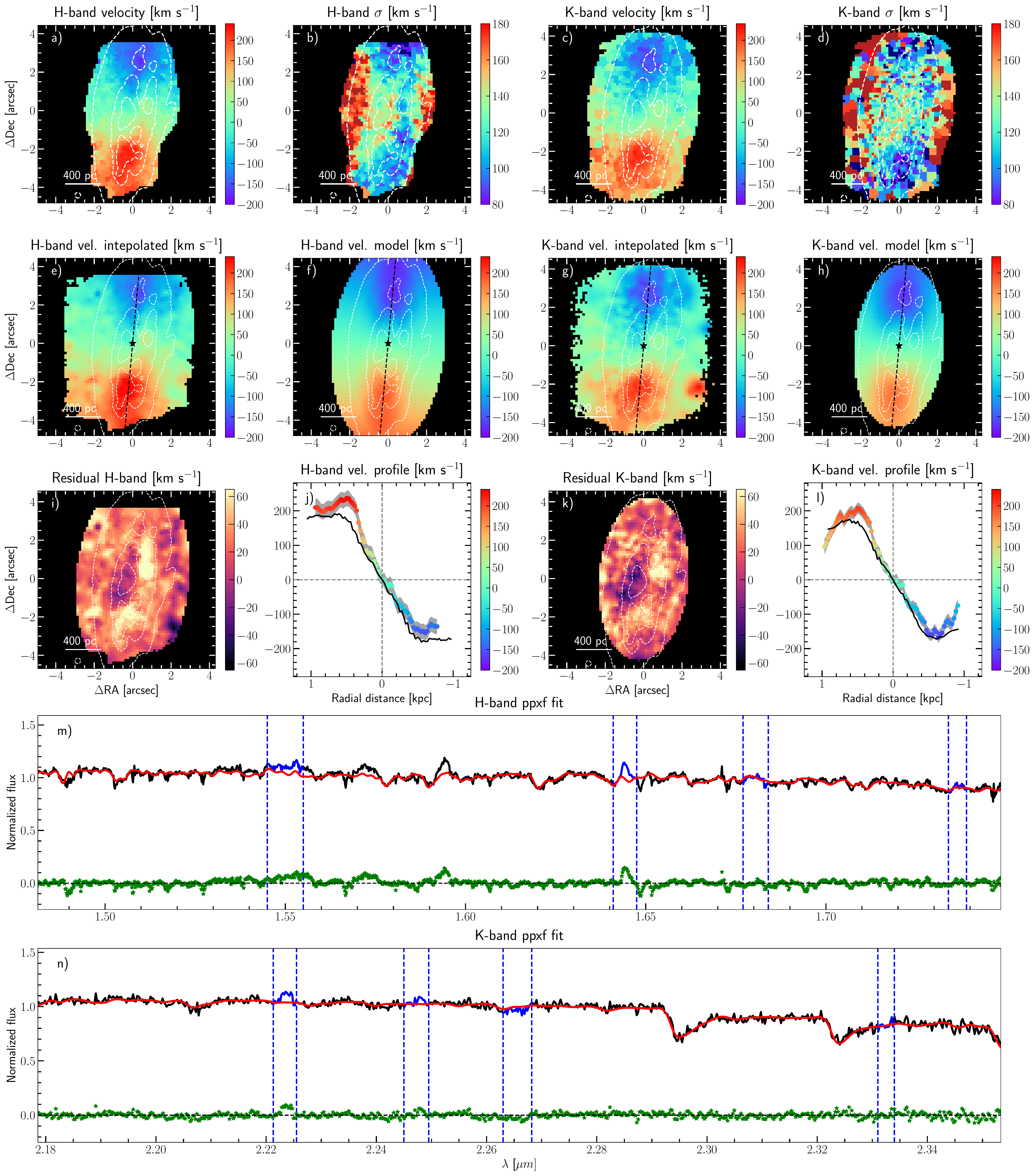} 
\caption{As in Fig.~\ref{fig:example_kin} but for NGC\,2369.}
\label{fig:kin_2369}
\end{figure*}

\begin{figure*}
\includegraphics[width=\linewidth]{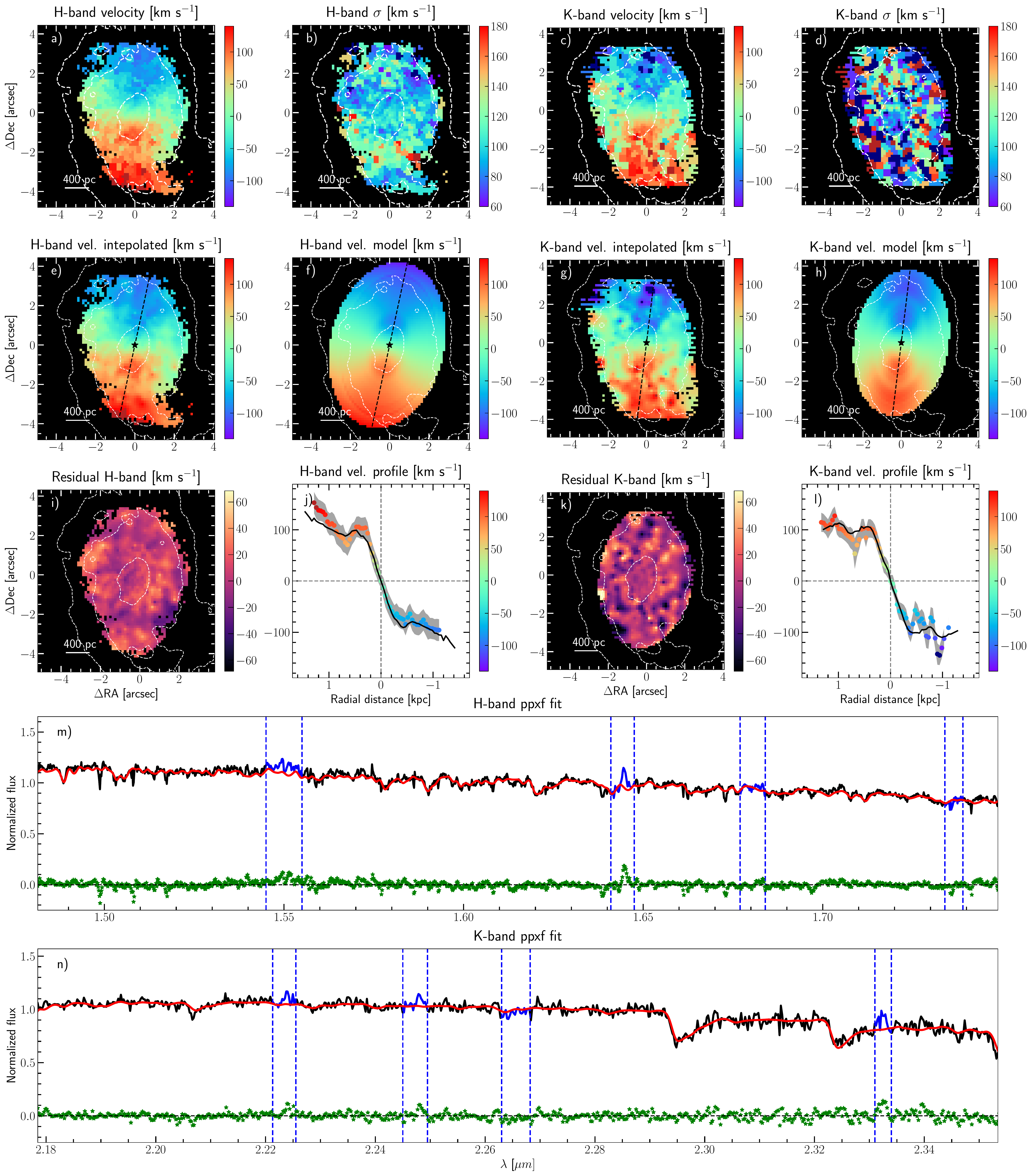} 
\caption{As in Fig.~\ref{fig:example_kin} but for NGC\,3110.}
\label{fig:kin_3110}
\end{figure*}

\begin{figure*}
\includegraphics[width=\linewidth]{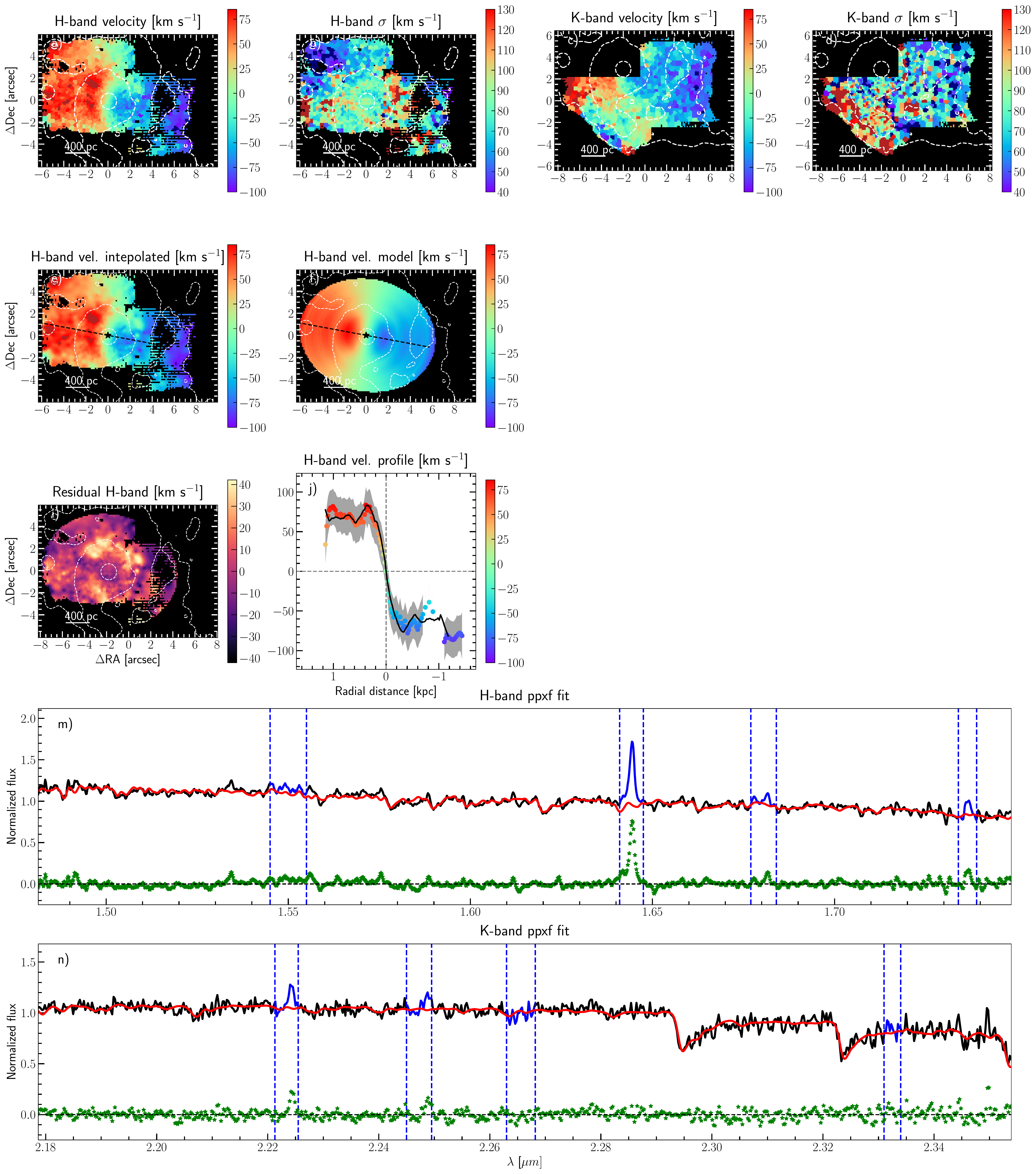} 
\caption{As in Fig.~\ref{fig:example_kin} but for NGC\,3256.}
\label{fig:kin_3256}
\end{figure*}

\begin{figure*}
\includegraphics[width=\linewidth]{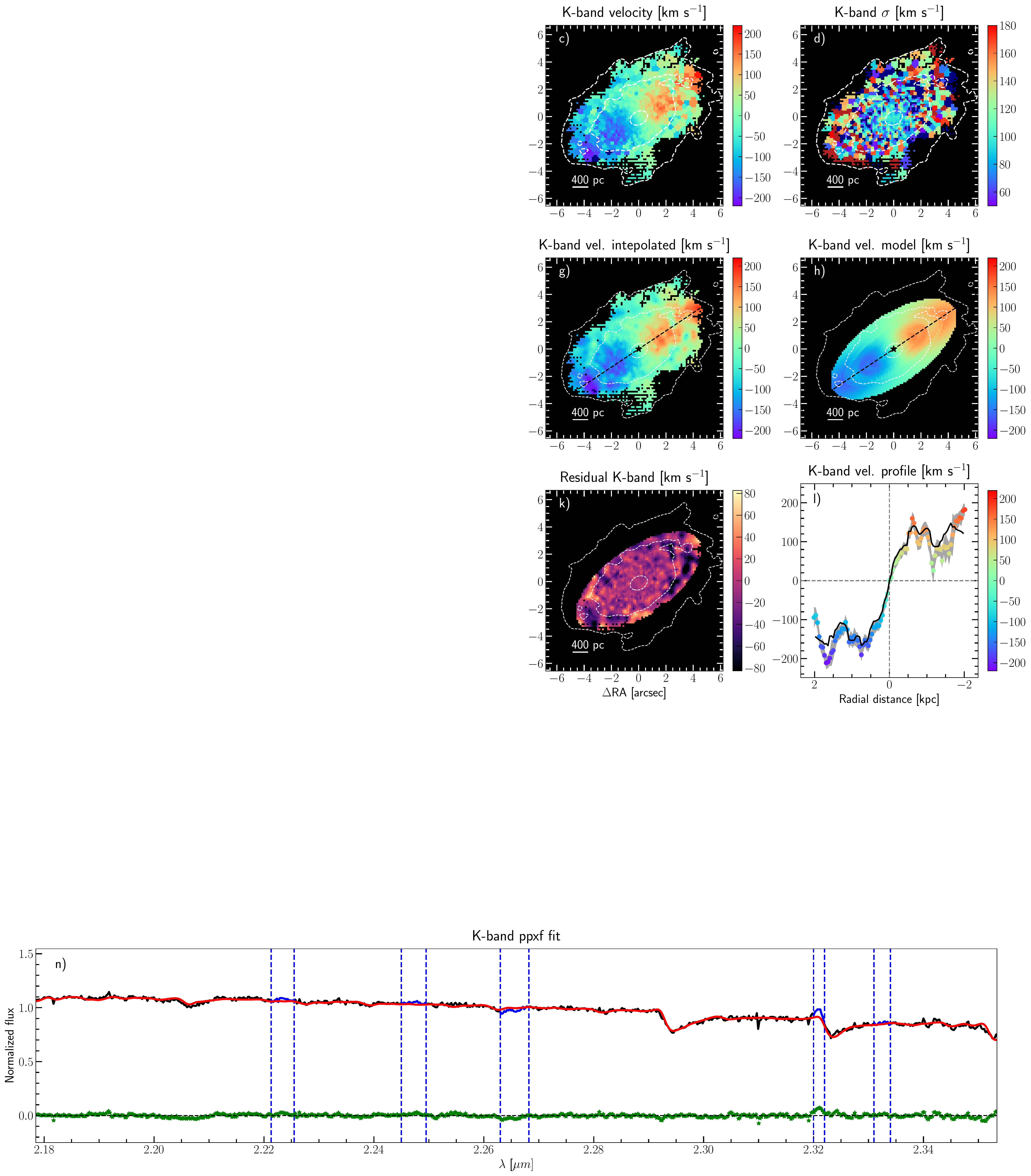} 
\caption{As in Fig.~\ref{fig:example_kin} but for IRAS\,F12115-4656.}
\label{fig:kin_12115}
\end{figure*}

\begin{figure*}
\includegraphics[width=\linewidth]{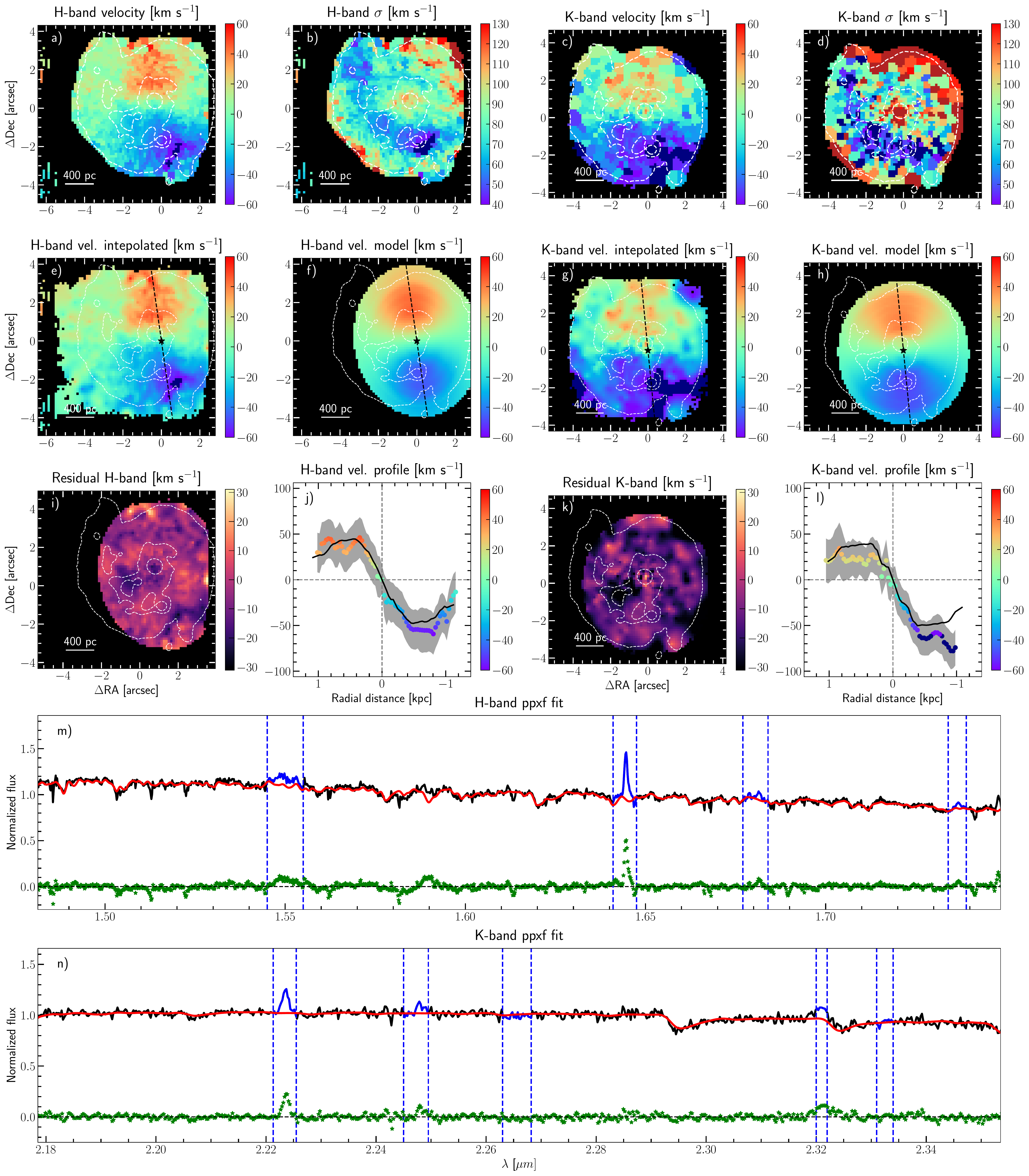} 
\caption{As in Fig.~\ref{fig:example_kin} but for NGC\,5135.}
\label{fig:kin_5135}
\end{figure*}

\begin{figure*}
\includegraphics[width=\linewidth]{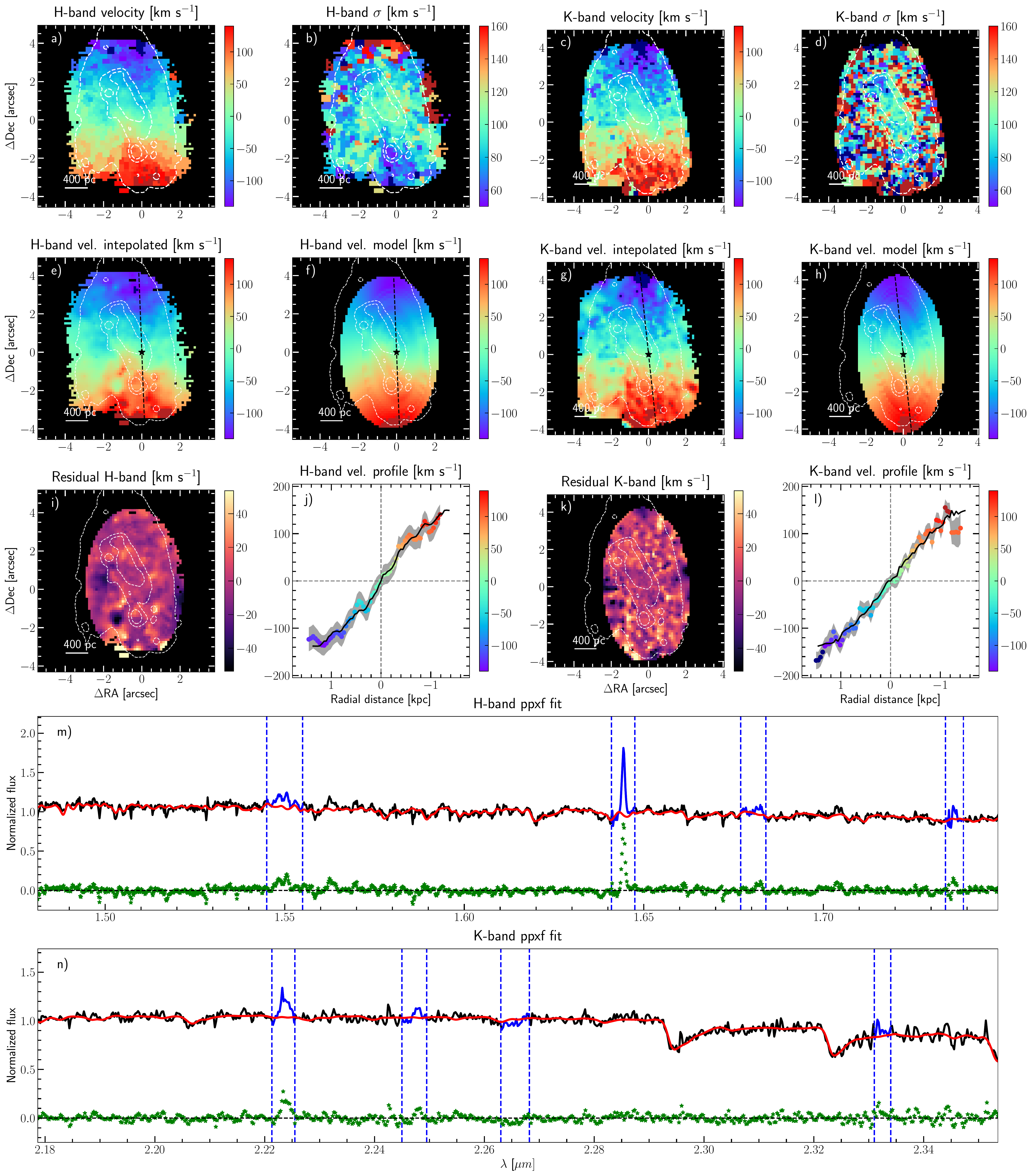} 
\caption{As in Fig.~\ref{fig:example_kin} but for IRAS\,F17138-1017.}
\label{fig:kin_17138}
\end{figure*}

\begin{figure*}
\includegraphics[width=\linewidth]{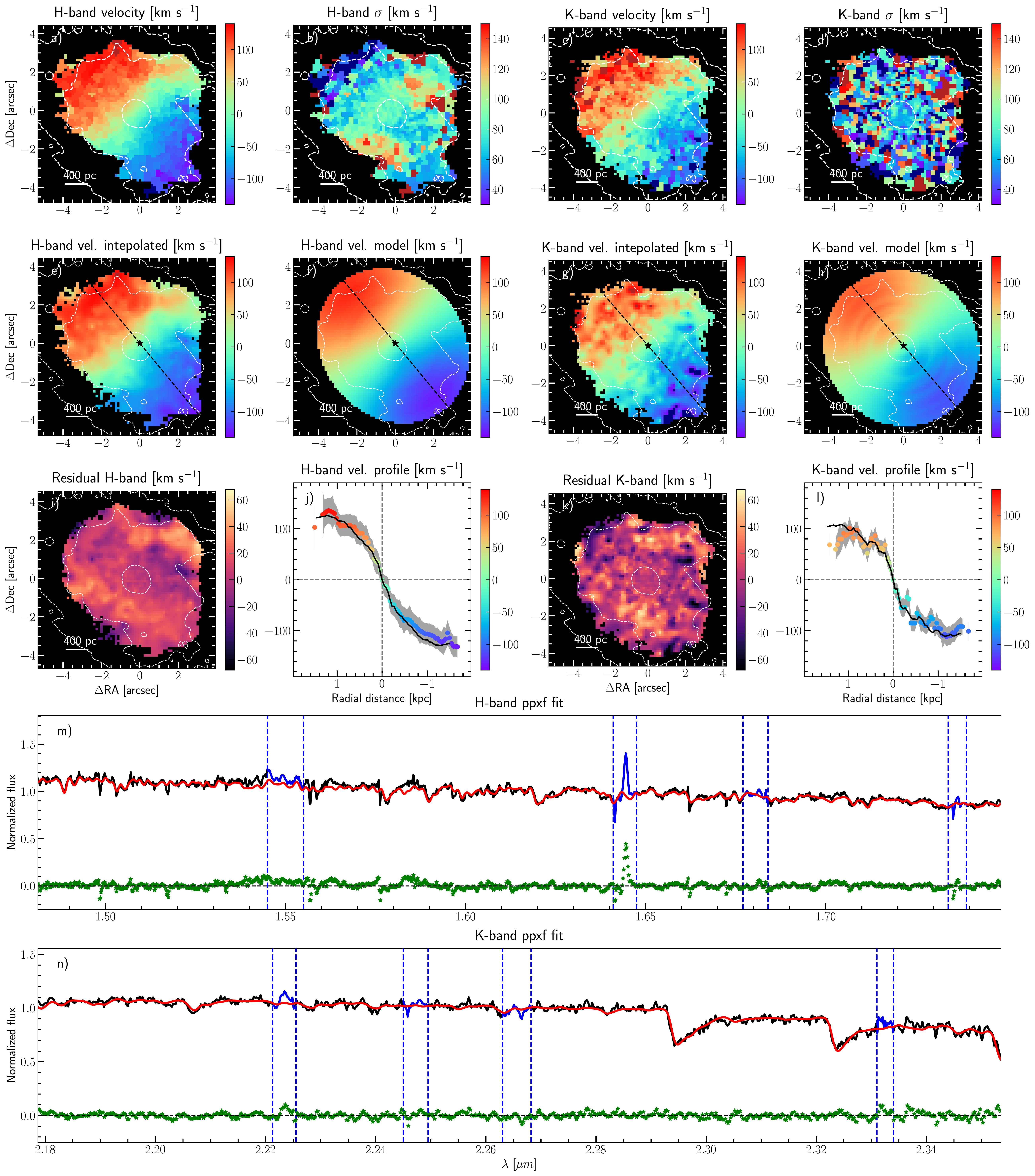} 
\caption{As in Fig.~\ref{fig:example_kin} but for IC\,4687. For this object, we adopted the morphological centre during the \texttt{DiskFit} modelling (see Sect.~\ref{subsec:results_DiskFit}).}
\label{fig:kin_4687}
\end{figure*}

\begin{figure*}
\includegraphics[width=\linewidth]{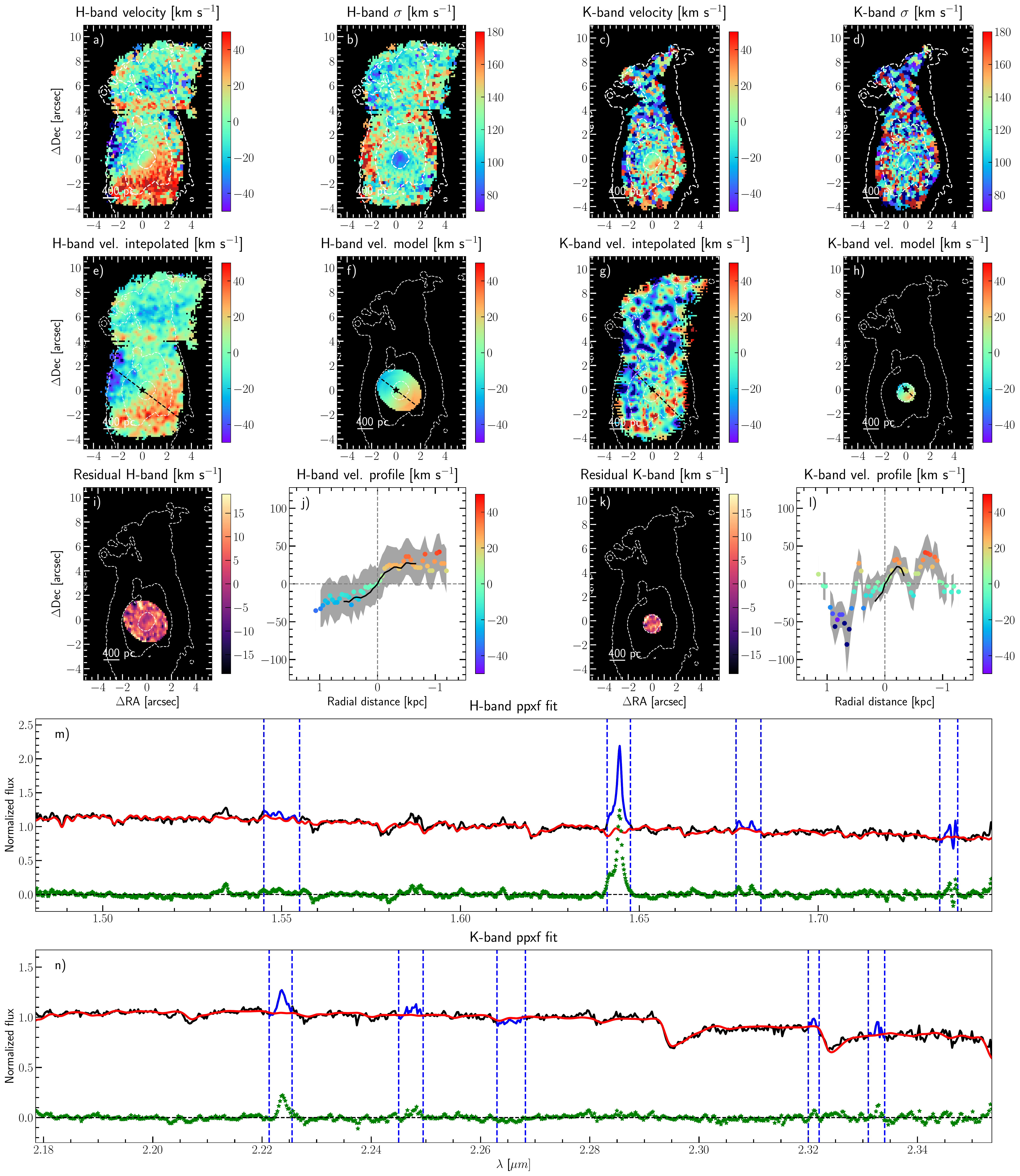} 
\caption{As in Fig.~\ref{fig:example_kin} but for NGC\,7130.}
\label{fig:kin_7130}
\end{figure*}

\begin{figure*}
\includegraphics[width=\linewidth]{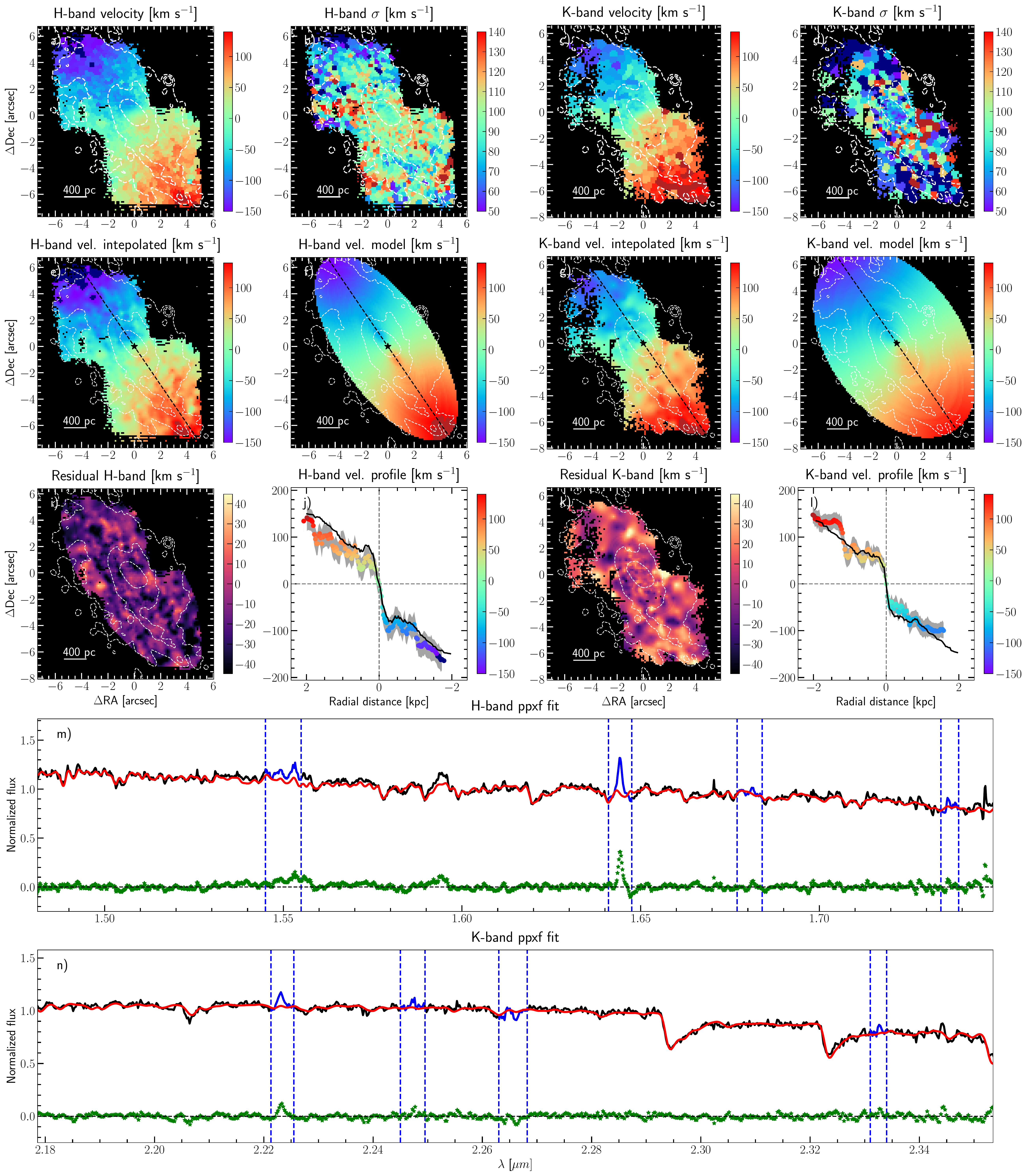} 
\caption{As in Fig.~\ref{fig:example_kin} but for IC\,5179.}
\label{fig:kin_5179}
\end{figure*}

\section{Double-Gaussian kinematic analysis of NGC\,2369}
\label{append:2369}

\begin{figure*}
\includegraphics[width=\linewidth]{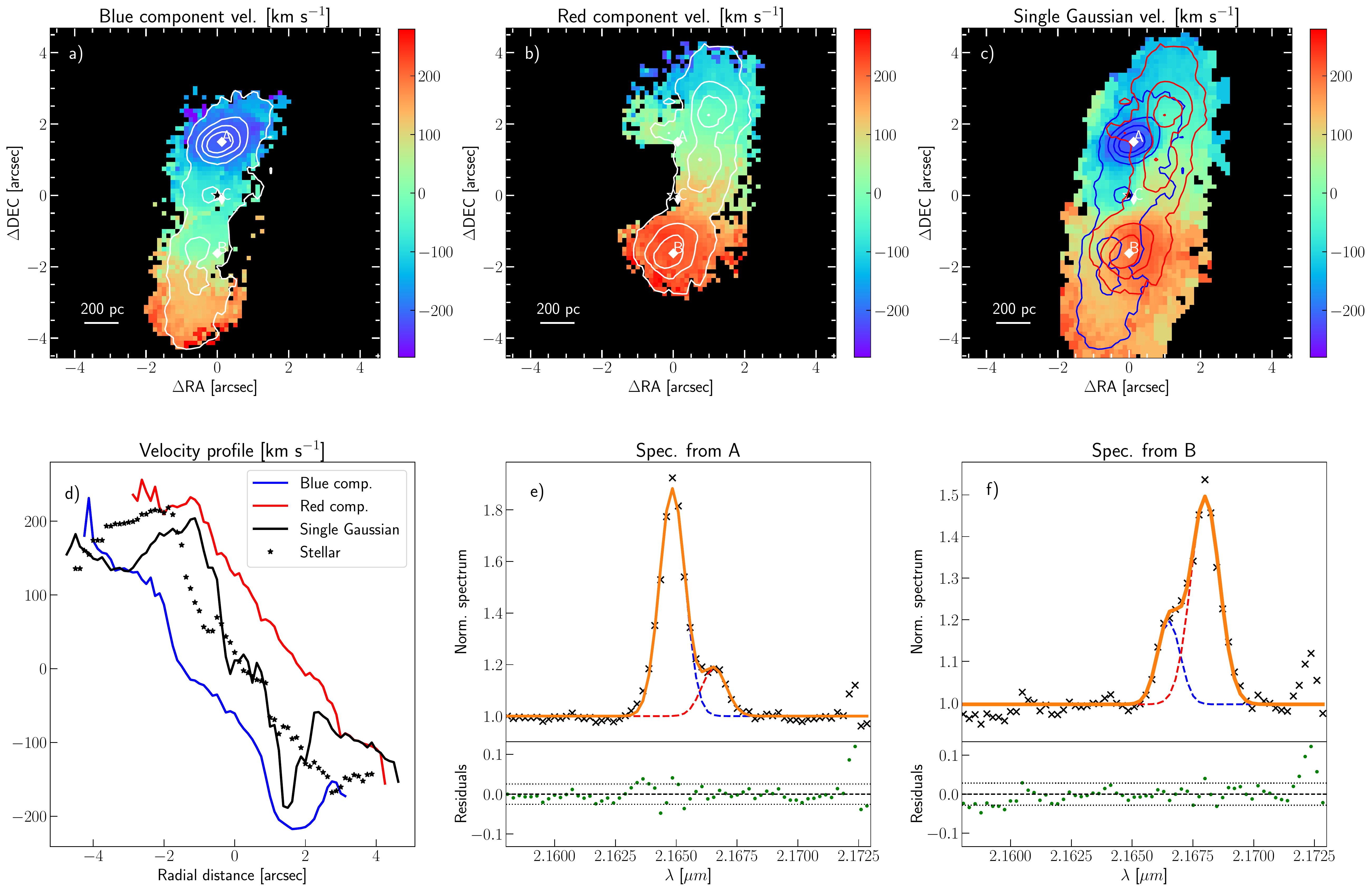} 
\caption{Double-Gaussian analysis in NGC\,2369. \textit{Panels `a' and `b':} Velocity of the blue and red components derived from the Br$\gamma$ analysis. White contours represent the flux of each component. White diamonds, labelled A, B, and C, represent the flux peak of the blue and red component, and their midpoint, respectively. Black star represents the kinematic centre derived with \texttt{DiskFit}. \textit{Panel `c'}: Velocity derived from the single-Gaussian fit carried out in \citet{PiquerasLopez+12}. Blue and red contours represent the flux contours of each component. \textit{Panel `d':} Observed velocity curves extracted using the `C' point and the PA from Table~\ref{tab:phot_results}. Red, blue, and black solid lines represent the velocity from panels `a', `b', and `c', respectively. Stars represent the stellar velocity curve extracted from the H-band and presented in Fig.~\ref{fig:kin_2369}. \textit{Panels `e' and `f'}: Double-Gaussian fits of the spectra corresponding to the flux peaks of each component (labelled A and B in the top panels, respectively). Black crosses represent the original spectra, whereas both components and the total fit are displayed by dashed blue and red, and solid orange lines, respectively. Green dots in the lower panels show the residuals of the fit.
\label{fig:NGC2369_doublegauss}}
\end{figure*}

\begin{figure}
\includegraphics[width=0.9\linewidth]{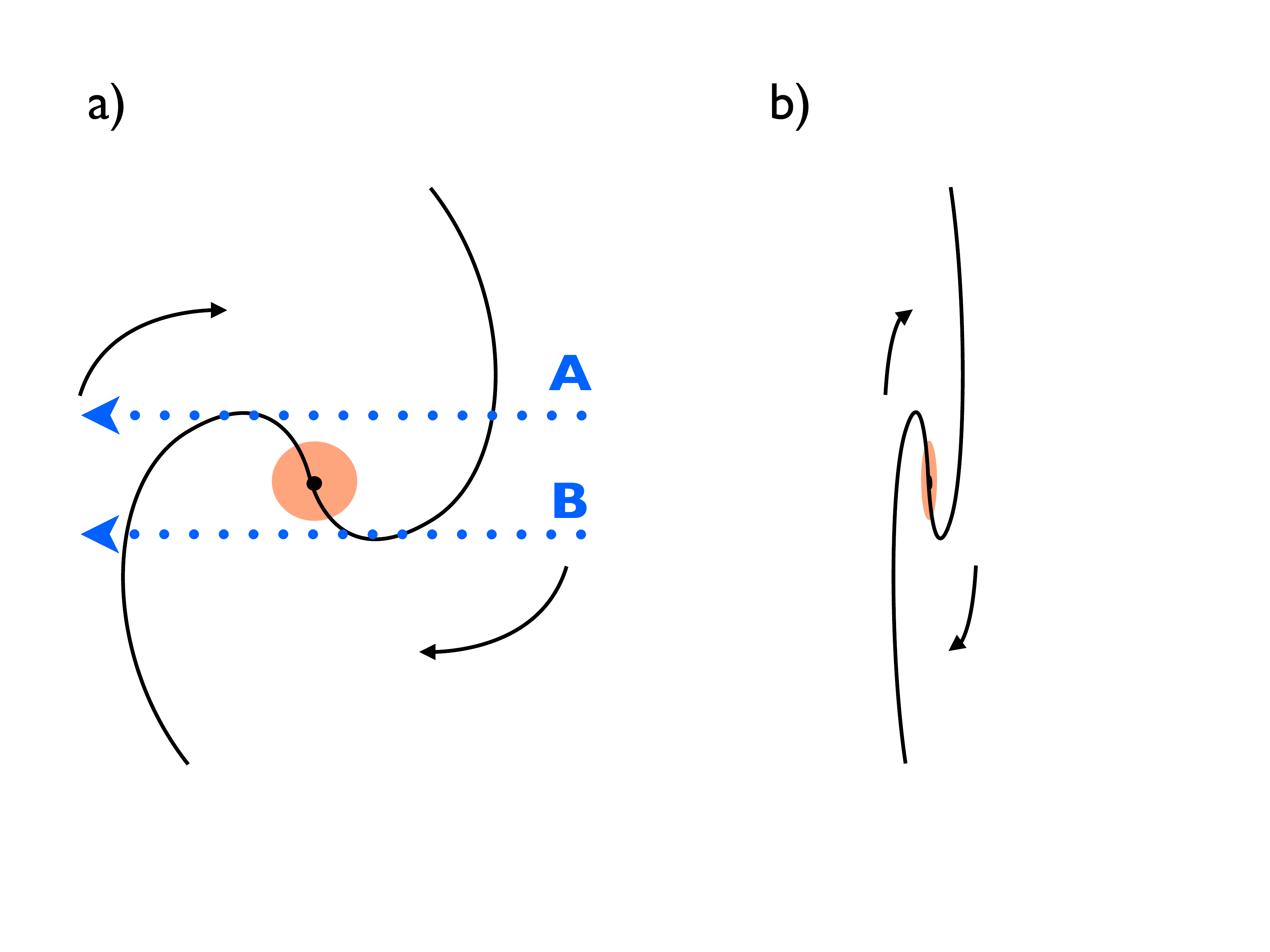} 
\caption{Sketch illustrating the scenario proposed for NGC\,2369. \textit{Panel `a'}: Face-on representation of the galaxy, where the orange circle represent the highly obscured nuclear region of this object. Blue dotted lines display the LOS of the corresponding to the regions A and B from Fig.~\ref{fig:NGC2369_doublegauss}. \textit{Panel `b'}: Representation of the galaxy orientation in the sky.}
\label{fig:NGC2369_sketch}
\end{figure}

\begin{figure}
\includegraphics[width=0.9\linewidth]{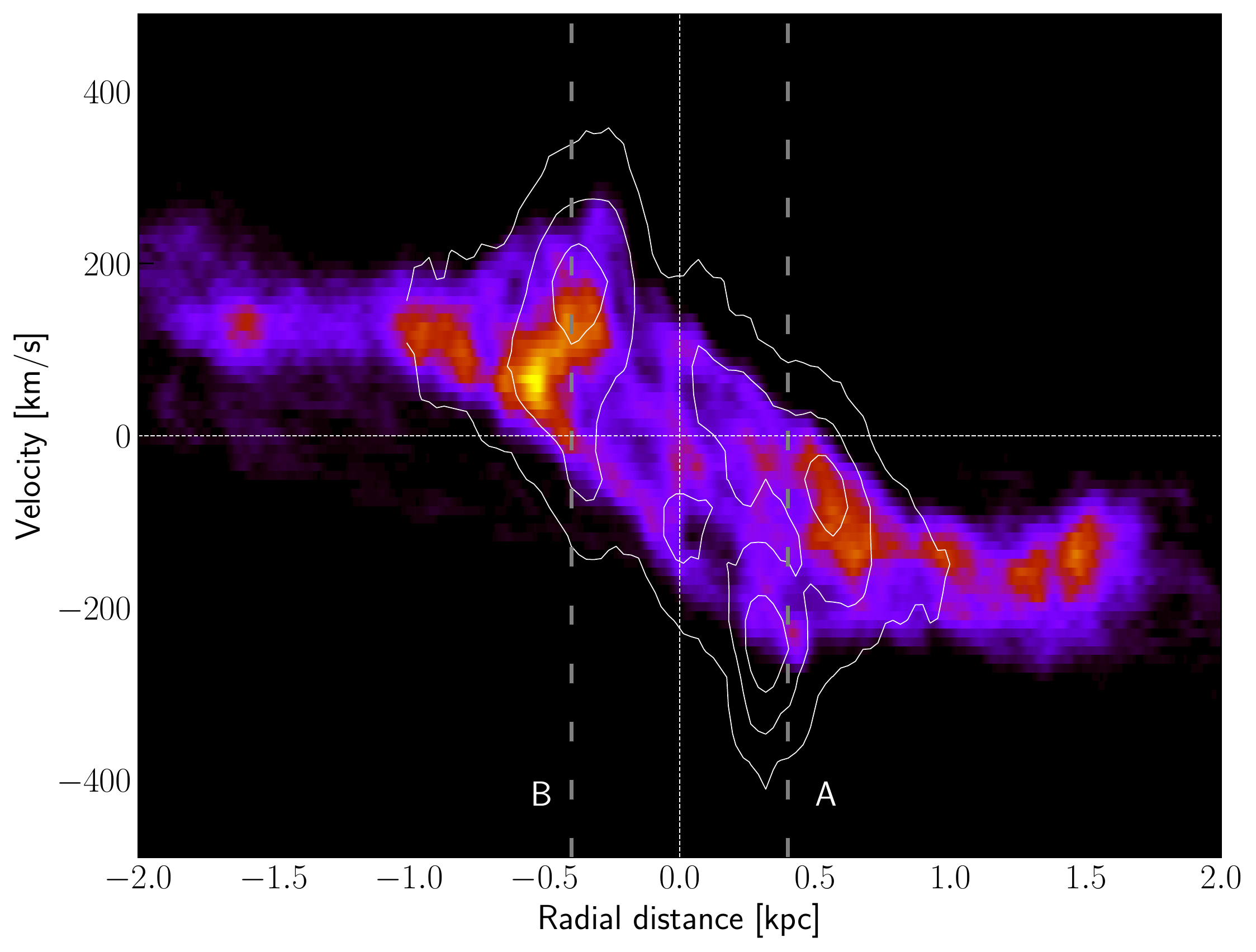} 
\caption{Position-velocity diagram of NGC\,2369. In colour scale and white contours we display the P-V diagrams along the kinematic major axis for the CO(2-1) and Br$\gamma$ observations, respectively. Vertical dashed lines represent the position of the regions A and B from Fig.~\ref{fig:NGC2369_doublegauss}.}
\label{fig:NGC2369_PV}
\end{figure}

The gas kinematics maps \citep[see][]{PiquerasLopez+12} of this object show a complex structure in all the phases, which is also observed in the stellar kinematics in this work (panels `a' and `c', Fig.~\ref{fig:kin_2369}). These structures could not be explained by the motions expected in a rotating disc or an outflow. Indeed, during the \texttt{DiskFit} modelling, these structures could not be reproduced by adding a bar-like velocity component, and its impact can be observed in the highly structured residuals (panels `i' and `k'), and in the difference between the modelled and observed velocity curves (panels `j' and `l'). 

We performed a double-Gaussian fit to the Br$\gamma$ line ---as this is the line that shows the impact of these peculiar motions most clearly--- across the whole FoV to obtain a better picture of the kinematics of this object. This analysis reveals two kinematic components showing rotating-disc patterns (panels `a' and `b', Fig.~\ref{fig:NGC2369_doublegauss}). Although both components display similar $\sigma$ values, we were able to separate them, as one is always redshifted with respect to the other. The flux contours of each component spatially coincide with the excesses in the approaching and receding velocities from the single-Gaussian velocity map (panel `c', Fig.~\ref{fig:NGC2369_doublegauss}), indicating that this velocity map is most likely the result of the kinematic analysis of the flux-weighted sum of each component. Indeed, panel `d' shows how the single-Gaussian velocity profile is compatible with this scenario, being similar to the red and blue components at each extreme, whereas the inner regions present intermediate values. We were not able to repeat this double-component analysis to the stellar component because of the limited S/N, but one can observe that the stellar velocity maps are similar to the gas phase one. 

We propose a scenario, illustrated in Fig.~\ref{fig:NGC2369_sketch}, where intense trailing spiral arms, a highly obscured nuclear region (represented by an orange circle), and the high inclination of the galaxy explain the kinematics of each component as well as the velocity map obtained with a single-Gaussian fit. In this scenario, Br$\gamma$ traces the bright emission from both arms along the LOS within the inner 1\,kpc, meaning that each kinematic component is produced by a different spiral arm.

The available CO(2-1) data (see Sect.~\ref{sec:sample}) allowed us to shed some light on the kinematics of this region, taking advantage of its better spatial and spectral resolution, and removing the effect of extinction. Figure~\ref{fig:NGC2369_PV} shows the position--velocity (i.e. P-V) diagram for the SINFONI Br$\gamma$ and ALMA CO(2-1) emission lines. The CO(2-1) diagram shows, up to r$\sim$2\,kpc, a clear velocity profile (`intrinsic' velocity) along with another two velocity components, only in the inner kiloparsec, that run almost parallel to the `intrinsic' profile. These `extra' components are the same ones observed in Br$\gamma$, where the nuclear extinction prevents us from observing the `intrinsic' velocity curve. This emission originates from both spiral arms rotating at the same velocity. In addition, we observe that, outside the inner 1\,kpc, the Br$\gamma$ velocity coincides with the intrinsic velocity, as expected in a rotating disc, traced by the CO(2-1) beyond the SINFONI FoV. These P-V diagrams are not compatible with a scenario where two dynamical structures are rotating at different velocities, as one would expect to see these components crossing each other in the centre.

\section{Stellar kinematic results of IRAS\,13120-5453}
\label{append:13120}

As part of the IFS study of U/LIRGs introduced in Sect.~\ref{sec:intro}, we recently obtained 24\,h with the near-IR IFS SINFONI (proposal 103.B-0867(A)) to extend the sample used in this work to more luminous objects. At their distances, the SINFONI K-band do not cover the CO absorption bands and, thus, these observations present an opportunity to extend the analysis of the stellar kinematics to the ULIRG regime, which was previously unexplored.

However, SINFONI was de-commissioned during the 103.B semester and the observations were not finished (only 55\% of the requested time was obtained, distributed along all the objects). Here we present, as an example, the kinematics results of the ULIRG IRAS\,13120-5453 (z=0.031). These results were not taken into account along this work as this object cannot be compared with the LIRG sample in terms of integration time, luminosity, or distance (i.e. spatial resolution).

\begin{figure*}
\includegraphics[width=\linewidth]{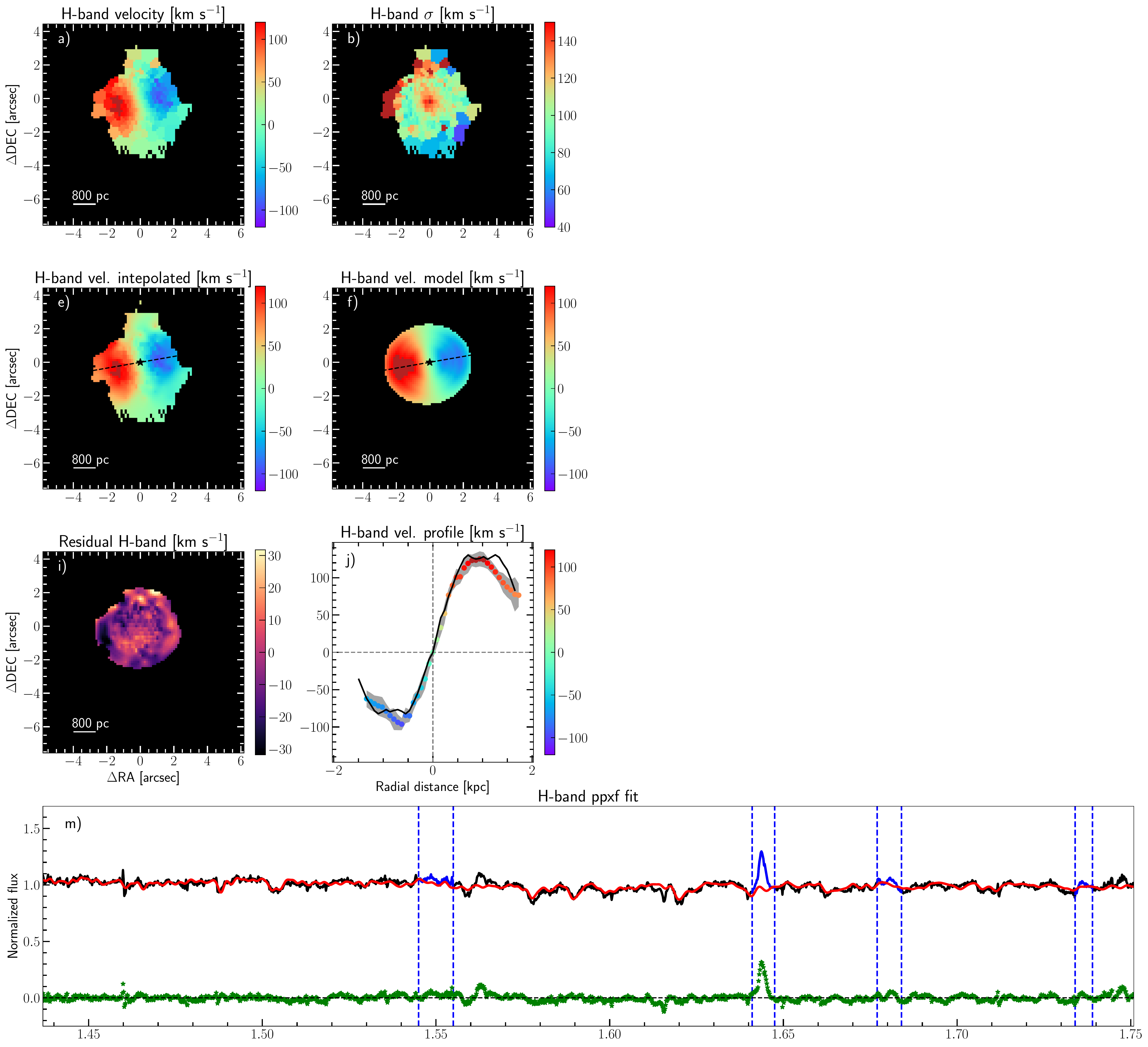} 
\caption{As in Fig.~\ref{fig:example_kin} but for IRAS\,13120-5453. The kinematic maps do not show flux contours as the photometric analysis was not carried out for this object. }
\label{fig:kin_IRAS13120}
\end{figure*}

The observations (1800\,s on target) were carried out in the H-band and seeing-limited mode (FWHM$\sim$0.85\arcsec, $\sim$0.5\,kpc at its redshift), with a plate scale of 0.125$\times$0.250\,\arcsec\,spaxel$^{-1}$. The dithering pattern adopted yielded a FoV$\sim$11\arcsec$\times$11\arcsec. The reduction and calibration, and kinematic analysis were performed as in Sect.~\ref{subsec:ifs} and ~\ref{subsec:kin_analysis}, respectively. This ULIRG presents a velocity map compatible with a rotating-disc galaxy with an amplitude of $\sim$100\,km\,s$^{-1}$ at r<1\,kpc, whereas the velocity dispersion map shows a central peak of $\sim$140\,km\,s$^{-1}$ (Fig.~\ref{fig:kin_IRAS13120}). The loss of SN at r>1\,kpc generates decrements in the measured velocities, as can be noticed in the velocity profile. These results support the advantage of obtaining the stellar kinematics using H-band observations for those ULIRGs in which the K-band does not cover the CO\,(2-0) and CO\,(3-1) bands.

\end{document}